\input{rangenii.def}
\documentstyle[12pt]{article}\oddsidemargin0cm
\title{\sf Harmonic analysis of random number generators and multiplicative 
groups of residue class rings} 
\author{{\sc Oliver Schnetz}\thanks{Institut f{\"u}r theoretische 
Physik III, Staudtstra{\ss}e 7, 91058 Erlangen, Germany,\newline 
e-mail: schnetz@pest.physik.uni-erlangen.de\newline 
Supported in parts by the DFG Graduiertenkolleg 'Starke Wechselwirkung' 
and the BMBF.\newline 
FAU-TP3-96/13}}\date{September 26, 1996}
\begin{document}\renewcommand{\arraystretch}{1.2} 
\maketitle 
\begin{abstract} 
The spectral test of random number generators (R.R. Coveyou and R.D. 
McPherson, 1967) is generalized. The sequence of random numbers 
is analyzed explicitly, not just via their $n$-tupel distributions. 
The generalized analysis of many generators becomes possible due 
to a theorem on the harmonic analysis of multiplicative groups of 
residue class rings. We find that the mixed multiplicative generator 
with power of two modulus does not pass the extended test with an 
ideal result. Best qualities has a new generator with the recursion 
formula $ X_{k+1}=aX_{k}+c{\rm\hspace{.38ex}int}(k/2){\rm\hspace
{.38ex}mod\hspace{.38ex}}2^d$. We discuss the choice of the parameters 
$a$, $c$ for very large moduli $ 2^d$ and present an implementation 
of the suggested generator with $ d=256$, $ a=2^{128}+2^{64}+2^{
32}+62181$, $ c=(2^{160}+1)\cdot 11463$.
\end{abstract} 
\tableofcontents 
\section{Introduction} 
The spectral test was proposed by R.R. Coveyou and R.D. McPherson 
in 1967 \cite{Cov}. The advantage of this test is to present an 
algebraic criterion for the quality of the generator. For the mixed 
multiplicative generator $ X_{k+1}=aX_{k}+c{\rm\hspace{.38ex}mod\hspace{.38ex}}M
$ 
\begin{equation}
\label{03}{\rm\hspace{.38ex}min\hspace{.38ex}}\{|{\bf s}|=\sqrt{s
_{1}^{2}+{\ldots}+s_{n}^{2}{}} {\rm  \ with \ }s_{a}=s_{1}+as_{
2}+{\ldots}+a^{n-1}s_{n}=0{\rm\hspace{.38ex}mod\hspace{.38ex}}M
\}
\end{equation} 
should be as large as possible \cite{Cov,Knu}. The criterion is that 
simple since the $n$-tupels of random numbers form an $n$-dimensional 
lattice (cf.\ e.g.\ Fig.\ 4). A good generator has uniformly distributed 
$n$-tupels which refers to an almost cubic lattice \cite{M,B,Nie}.

The lattice is a consequence of the (affine) linear dependence of 
$X$$_{k+1}$ on $X$$_{k}$. From the figures on the left (type I) 
we see that the relation between $k$ and $X$$_{k}$ is much more 
complicated. This is however one of the most fundamental aspects 
of randomness. In order to judge whether a sequence $X$$_{k}$ takes 
random values one would first plot the sequence itself and then 
maybe $X$$_{k+1}$ over $X$$_{k}$.

Of course, the correlation between $k$ and $X$$_{k}$ is not independent 
from the distribution of pairs $ (X_{k},X_{k+1})$. E.g., a poor 
'random' sequence $ X_{k}=ak$ lying on a line with gradient $a$ 
leads to pairs $ (X_{k},X_{k+1}=X_{k}+a)$ lying on a line with gradient 
1 shifted by $a$ off the origin. This makes it reasonable to judge 
randomness by only looking at the $n$-tupel distributions. However, 
random number generators which have identical valuation by the spectral 
test may still look quite different. The generators $ X_{k+1}=41X
_{k}+3{\rm\hspace{.38ex}mod\hspace{.38ex}}1024$ (Fig.\ 5) and $ X
_{k+1}=41X_{k}+1{\rm\hspace{.38ex}mod\hspace{.38ex}}1024$ (Fig.\ 
6), e.g., differ only by the additive constant which does not enter 
Eq.\ (\ref{03}). The lattices of pair distributions (type II in 
the figures) are similar whereas the plots of $X$$_{k}$ over $k$ 
show different behavior. The spectral test not even makes a difference 
between a prime number and a power of two modulus (cf.\ Fig.\ 1 
vs.\ Fig.\ 4).

Therefore it is desirable to include the analysis of the correlation 
between $k$ and $X$$_{k}$ into the valuation of the test. In fact 
it is possible to analyze the accumulation of random numbers along 
certain lines (which is often seen in the figures) by Fourier transformation. 
More generally we extend the spectral test by analyzing the correlation 
between $k$ and the $n$-tupel $ (X_{k},X_{k+1},{\ldots},X_{k+n-
1})$. The generators mentioned above (Figs.\ 5, 6) acquire different 
valuations. Fig.\ 6 is preferred since the random numbers spread 
more uniformly in Fig.\ 6I than in Fig.\ 5I (cf.\ Ex.\ 4.3 1.(a)).

We will develop a powerful theorem (Thm.\ 3.4, Cor.\ 3.5) on the 
harmonic analysis of multiplicative groups of residue class rings 
that allows us to algebraically perform the test for the standard 
generators. It becomes even possible to evaluate the extended test 
on more generators than the original spectral test has yet been 
applied. This has the advantage to have more freedom for the search 
after an optimum generator.

In fact we will find that the commonly used mixed multiplicative 
generator always shows correlations along certain lines if the modulus 
is not a prime number. We will improve the generator until these 
correlations will essentially disappear (from Fig.\ 4 via Fig.\ 
9 to Fig.\ 11). We finally recommend the recursion formula 
\begin{equation}
\label{02}X_{k+1}=aX_{k}+c{\rm\hspace{.38ex}int}\left( k/2
\right) {\rm\hspace{.38ex}mod\hspace{.38ex}}2^d
\end{equation} 
with the parameters 
\begin{eqnarray}
d=2^{k}d_{0}&,&a=2^{2^{k-1}d_{0}}+2^{2^{k-2}d_{0}}+{\ldots}+2^{2d
_{0}}+\left( 3~580~621~541{\rm\hspace{.38ex}mod\hspace{.38ex}}2^
{d_{0}}\right) ,\nonumber \\ 
\label{01}&&c=\left( 2^{{\rm int}\left( 2^{k+1}{} /3\right) d_{0
}}+1\right) \left( 3~370~134~727{\rm\hspace{.38ex}mod\hspace{.38ex}}2^
{d_{0}}\right) .
\end{eqnarray} 
In particular, the case $ d_{0}=16$, $ k=4$ is discussed in Ex.\ 
5.1 1.

This generator is supposed to be a good choice with respect to the 
following three criteria.

Firstly, the sequence of numbers provided by the generator should 
behave as close to a true random sequence as possible.

Secondly, the calculation of random numbers should be as fast as 
possible. The generator given in Eqs.\ (\ref{02}), (\ref{01}) is 
explicitly constructed to have best performance. It is important 
to note that this is not independent from the first criterion. It 
is possible to produce better random numbers the more effort one 
spends in calculating the numbers. Figs.\ 4 and 8 show how a simple 
doubling of the digits of the modulus improves the randomness of 
the generator. In general we can produce arbitrarily good random 
numbers with e.g.\ $ d_{0}=32$ and large $k$ in Eq.\ (\ref{01}).

Thirdly, the properties of the random numbers should be known as 
detailed as possible. It is not sufficient to use a messy, opaque 
formula. It has often been seen that this leads to numbers which 
are far from being random \cite{Knu}. As long as one is not familiar 
with the qualities of the generator one can never rely on the results 
gained with it. The full evaluation of the generalized spectral 
test is supposed to provide a profound knowledge of the generator.

\vspace{1ex}
\noindent{}We start with the development of the generalized spectral 
test in the next section. In Sec.\ \ref{rings} we derive some mathematical 
results on the Fourier transformation of residue class rings. In 
Sec.\ \ref{generators} we apply the result of the previous section, 
Thm.\ 3.4, to a series of commonly used and some new generators. 
Finally we discuss the choice of parameters in Sec.\ \ref{param}.

\section{The generalized spectral test} 
\subsection{Review of the spectral test} 
We start with a short review of the spectral test \cite{Cov,Knu} 
in which we try to stress its geometrical meaning. The idea is to 
plot all $n$-tupels of successive random numbers in an $n$-dimensional 
diagram. This is done, e.g.\ for $ n=2$ in the figures of type II.

Mathematically a figure is presented as a function $g$ which is 1 
at every dot and 0 elsewhere. If $ \NX$ is the period of the generator 
$X$, that is the smallest number with $ X_{k+\NX }=X_{k}
\hspace*{1ex}\forall k$, then 
\begin{equation}
g\left( x_{1},{\ldots},x_{n}\right) =\sum _{k=1}^\NX \delta _{x_{
1},X_{k}}\cdots \delta _{x_{n},X_{k+n-1}}\equiv \sum _{k\in 
{\ErgoBbb Z}_\NX }\delta _{{\bf x},{\bf X}_{k}}\hspace{.6ex},
\end{equation} 
where $ \delta _{a,b}=1$ if $ a=b$ and $ \delta _{a,b}=0$ if $ a
\neq b$ (for later convenience we also write the Kronecker $
\delta $ as $ \delta _{a=b}$). Moreover we have introduced the notation 
\begin{equation}
{\bf X}_{k}=\left( X_{k},X_{k+1},{\ldots},X_{k+n-1}\right) 
\hspace{.6ex},\hspace{2ex}{\bf x}=\left( x_{1},x_{2},{\ldots},x_{
n}\right) \hspace{.6ex},\hspace{2ex}{\ErgoBbb Z}_\NX =
{\ErgoBbb Z}/\NX{\ErgoBbb Z}\hspace{.6ex}.
\end{equation}

We want to check whether the dots accumulate along certain hyper-planes 
(see e.g.\ the lines in Fig.\ 1II). To this end we select a hyper-plane 
and project all the dots onto a line perpendicular to it. If points 
accumulate along the plane many dots will lie on top of each other, 
otherwise the dots are spread uniformly over the line.

The hyper-plane $H$ is determined by its Hesse normal form 
\begin{equation}
H=\{{\bf x}:s_{1}x_{1}+s_{2}x_{2}+{\ldots}+s_{n}x_{n}\equiv 
{\bf s}\cdot {\bf x}=0\}\hspace{.6ex},\hspace{2ex} |{\bf s}|
\equiv \sqrt{{\bf s}\cdot {\bf s}} \equiv \sqrt{s_{1}^{2}+
{\ldots}+s_{n}^{2}{}} \neq 0\hspace{.6ex}.
\end{equation} 
The line is stretched by the factor $ |{\bf s}|$. The position of 
a point $ {\bf X}_{k}$ on the line perpendicular to the plane is 
given by the number $ {\bf s}\cdot {\bf X}_{k}$.

Next we wind the line up to a circle so that the modulus $M$ as point 
on the line lies on top of the 0. For a suitable choice of $ 
{\bf s}$, namely $ {\bf s}=(-3,7)$ all points in Fig.\ 1II lie now 
on the point represented by the number 256.

The points are realized as complex phases on the unit circle. We 
obtain the assignment 
\begin{equation}
{\bf X}_{k}\mapsto \exp\left( \frac{2\pi i}{M} {\bf s}\cdot 
{\bf X}_{k}\right) \hspace{.6ex}.
\end{equation} 
Finally we draw arrows from the center of the circle to all the dots 
and add them. The length of the resulting vector describes how the 
dots are balanced on the circle. If the dots spread uniformly the 
arrows cancel each other and the resulting vector is small. If, 
on the other hand, all dots lie on top of each other the length 
of the arrows sums up to $ \NX$.

If we restrict ourselves to integer $s$$_{1}$, {\dots}, $s$$_{n}$ 
(accumulation of random numbers always occur along hyper-planes 
given by integer $s$$_{i}$) the resulting vector is given by the 
Fourier transform of $g$, 
\begin{equation}
\hat{g}\left( {\bf s}\right) =\frac{1}{\sqrt{\NX}} \sum _{
{\bf x}\in {\ErgoBbb Z}^{n}_\NX }g\left( {\bf x}\right) \exp
\left( \frac{2\pi i}{M} {\bf s}\cdot {\bf x}\right) =\frac{1}{
\sqrt{\NX}} \sum _{k\in {\ErgoBbb Z}_\NX }\exp\left( \frac{2
\pi i}{M} {\bf s}\cdot {\bf X}_{k}\right) \hspace{.6ex},
\end{equation} 
where we have introduced the normalization factor $ \NX^{-1/2}$. 
The information about accumulations along the hyper-plane is contained 
in $ |\hat{g}|^{2}$, the phase of $ \hat{g}$ is irrelevant.

\vspace{1ex}
\noindent{}We remember that for the mixed multiplicative generator 
the $n$-tupels form a lattice (which is displaced off the origin). 
So $ |\hat{g}|^{2}({\bf s})$ will assume the maximum value $ \NX
$ if $ {\bf s}$ lies in the dual lattice, $ {\bf s}\cdot 
{\bf X}_{k}=C+\ell M$, $ C,\ell \in {\ErgoBbb Z}$, otherwise $ 
|\hat{g}|^{2}({\bf s})$ is zero. Since $ X_{k}=c(a^{k}-1)/(a-1){\rm\hspace
{.38ex}mod\hspace{.38ex}}M$ this means, if $ {\rm\hspace
{.38ex}gcd}(c,M)=1$ and $X$ has full period, that $ \forall k:
\hspace*{1ex}(a^{k}-1)s_{a}=0{\rm\hspace{.38ex}mod\hspace{.38ex}}M
{\rm\hspace{.38ex}gcd}(a-1,M)$ from which Eq.\ (\ref{03}) follows 
(cf.\ Ex.\ 4.3 1).

\subsection{Generalization of the spectral test\label{gentest}} 
We generalize the spectral test by caring for the sequence in which 
the $n$-tupels are generated. The index $k$ is added to the $n$-tupel 
$ {\bf X}_{k}$ as zeroth component and we define $g$ as 
\begin{equation}
g\left( x_{0},{\bf x}\right) =\sum _{k\in {\ErgoBbb Z}_\NX }
\delta _{x_{0},k}\delta _{{\bf x},{\bf X}_{k}}=\delta _{{\bf x}
,{\bf X}_{x_{0}}}\hspace{.6ex}.
\end{equation} 
The geometrical interpretation remains untouched but now we consider 
also the figures of type I. With the Fourier transform of $g$ we 
introduce some notation that will be needed later, 
\begin{equation}
\label{1}\hat{g}_{{\ErgoBbb Z}_{N_{1}}}[N_{2},M,X]\left( s_{0},
{\bf s}\right) =\frac{1}{\sqrt{N_{1}{}}} \sum _{k\in 
{\ErgoBbb Z}_{N_{1}}}\exp\left( \frac{2\pi i}{N_{2}{}} s_{0}k+
\frac{2\pi i}{M} {\bf s}\cdot {\bf X}_{k}\right) \hspace{.6ex}.
\end{equation} 
Normally $ N_{1}=N_{2}=\NX$ is the period of $X$. We write $ g[N_{
2},M,X]$ if $ N_{1}=N_{2}$ and we also suppress $N$$_{2}$ if $ N
_{1}=N_{2}=\NX$. However, sometimes it is convenient to use a multiple 
$N$$_{2}$ of the period $ \NX$ instead of the period itself. Moreover 
for some values of $ (s_{0},{\bf s})$, e.g.\ $ (0,{\bf 0})$, the 
right hand side is periodic in $k$ with a smaller period $ N_{1
}|N_{2}$. In this case we may sum over $ {\ErgoBbb Z}_{N_{1}}$ only 
and use the subscript $ {\ErgoBbb Z}_{N_{1}}$. Note that $ 
\sum _{k\in {\ErgoBbb Z}_{N_{1}}}f(k)$ always implies that $f$ is 
a function on $ {\ErgoBbb Z}_{N_{1}}$ which means that $f$ is periodic, 
$ f(k)=f(k+N_{1})\hspace*{1ex}\forall k$. We also suppress $M$ and 
$X$ on the left hand if the context is clear.

The sum over $k$ is hard to evaluate since in the exponential $k$ 
is combined with $X$$_{k}$. However in fact we are interested in 
$ |\hat{g}|^{2}$ and find 
\begin{eqnarray}
|\hat{g}|^{2}\left( s_{0},{\bf s}\right) &=&\frac{1}{\NX} \sum _
{k,k'\in {\ErgoBbb Z}_\NX }\exp\left( \frac{2\pi i}{\NX} s_{0}
\left( k'-k\right) +\frac{2\pi i}{M} {\bf s}\cdot \left( 
{\bf X}_{k'}-{\bf X}_{k}\right) \right) \nonumber \\ 
\label{15}&=&\frac{1}{\NX} \sum _{\Delta k\in {\ErgoBbb Z}_\NX 
}\exp\left( \frac{2\pi i}{\NX} s_{0}\Delta k\right) \sum _{k
\in {\ErgoBbb Z}_\NX }\exp\left( \frac{2\pi i}{M} {\bf s}\cdot 
\left( {\bf X}_{k+\Delta k}-{\bf X}_{k}\right) \right) 
\hspace{.6ex}.
\end{eqnarray} 
The sum over $k$ has no linear $k$-dependence, only differences of 
random numbers occur. Like in the standard spectral test in many 
cases the sum over $k$ can be evaluated. The result is often simple 
enough to be able to evaluate the sum over $ \Delta k$ also.

If, e.g., $ X_{k}=a^{k}{\rm\hspace{.38ex}mod\hspace{.38ex}}P$ where 
$P$ is a prime number and $a$ is a primitive element of $ 
{\ErgoBbb Z}_P^\times $, the multiplicative group of $ 
{\ErgoBbb Z}_P$, we find that $ {\bf s}\cdot ({\bf X}_{k+
\Delta k}-{\bf X}_{k})=s_{a}a^{k}(a^{\Delta k}-1)=s_{a}\tilde
{k}(a^{\Delta k}-1){\rm\hspace{.38ex}mod\hspace{.38ex}}P$ for some 
$ 0\neq \tilde{k}\in {\ErgoBbb Z}_P$. If $k$ runs through the $ P
-1$ values of $ {\ErgoBbb Z}_P^\times $ then $ \tilde{k}$ sweeps 
out the whole $ {\ErgoBbb Z}_P\backslash \{0\}$. The sum over $ 
\tilde{k}$ can be evaluated yielding $ P\delta _{\Delta k=0}-1$ 
for $ s_{a}\neq 0$. Finally the sum over $ \Delta k$ gives together 
with the normalization $ |\hat{g}|^{2}=P/(P-1)-\delta _{s_{0}=0
}$ (cf.\ Eq.\ (\ref{36}) for $ d=1$).

Note that the standard spectral test corresponds to $ s_{0}=0$. We 
give some simple results on $ |\hat{g}|^{2}$ in the following lemma.

\pagebreak[3]

\noindent {\bf Lemma} 2.1. 
\begin{eqnarray}
|\hat{g}|^{2}[c_{1}X_{k+c_{2}}+c_{3}]\left( s_{0},{\bf s}
\right) &=&|\hat{g}|^{2}[X_{k}]\left( s_{0},c_{1}{\bf s}
\right) \nonumber \\ 
\label{20}&=&|\hat{g}|^{2}[X_{k}]\left( s_{0},{\bf s}\right) \hbox{ 
if }\exists c_{0}:\hspace*{1ex}c_{1}X_{k}=X_{k+c_{0}}
\hspace{.6ex},\\ 
\label{42}|\hat{g}|^{2}_{{\ErgoBbb Z}_{c_{2}\NX }}[c_{1}\NX ]
\left( s_{0},{\bf s}\right) &=&c_{2}{} \delta _{s_{0}=0{\rm\hspace
{.38ex}mod\hspace{.38ex}}c_{1}}|\hat{g}|^{2}\left( s_{0}/c_{1},
{\bf s}\right) \hspace{.6ex},\\ 
\label{29}|\hat{g}|^{2}_{{\ErgoBbb Z}_{1}}=1&,&|\hat{g}|^{2}
\left( s_{0},{\bf 0}\right) =\NX \delta _{s_{0}=0{\rm\hspace
{.38ex}mod}\NX }\hspace{.6ex},\\ 
\label{30}\sum _{s_{0}\in {\ErgoBbb Z}_\NX }|\hat{g}|^{2}\left( s
_{0},{\bf s}\right) =\NX&,&\sum _{{\bf s}\in {\ErgoBbb Z}_M^{n}
}|\hat{g}|^{2}\left( s_{0},{\bf s}\right) =M^{n}\hbox{ if }
{\bf X}_{k}={\bf X}_{k'}\Rightarrow k=k'{\rm\hspace{.38ex}mod}\NX 
.\hspace*{1cm}
\end{eqnarray}

\pagebreak[3]

\noindent {\bf Proof}. The proofs are straight forward. We show 
Eq.\ (\ref{42}) to get used to the notation: 
\begin{eqnarray*}
\hat{g}_{{\ErgoBbb Z}_{c_{2}\NX }}[c_{1}\NX ]\left( s_{0},
{\bf s}\right) &=&\sqrt{1/c_{2}\NX} \sum _{k\in {\ErgoBbb Z}_{c_{
2}\NX }}\exp\left( \frac{2\pi i}{c_{1}\NX} s_{0}k+\frac{2\pi i}
{M} {\bf s}\cdot {\bf X}_{k}\right) \\ 
&=&\sqrt{c_{2}/c_{1}^{2}\NX} \sum _{k\in {\ErgoBbb Z}_{c_{1}\NX 
}}\exp\left( \frac{2\pi i}{c_{1}\NX} s_{0}k+\frac{2\pi i}{M} 
{\bf s}\cdot {\bf X}_{k}\right) \\ 
&=&\sqrt{c_{2}/c_{1}^{2}\NX} \sum _{k_{1}\in {\ErgoBbb Z}_\NX }
\sum _{k_{2}\in {\ErgoBbb Z}_{c_{1}}}\exp\left( \frac{2\pi i}{c_{
1}\NX} s_{0}\left( k_{1}+\NX k_{2}\right) +\frac{2\pi i}{M} 
{\bf s}\cdot {\bf X}_{k_{1}}\right) \\ 
&=&\sqrt{c_{2}/\NX} \delta _{s_{0}=0{\rm\hspace{.38ex}mod\hspace{.38ex}}c
_{1}}\sum _{k_{1}\in {\ErgoBbb Z}_\NX }\exp\left( \frac{2\pi i}
{\NX} \frac{s_{0}}{c_{1}{}} k_{1}+\frac{2\pi i}{M} {\bf s}
\cdot {\bf X}_{k_{1}}\right) \hspace{.6ex}.
\end{eqnarray*}\hfill $\Box $
\pagebreak[3] 

One may also be interested in correlations between non-successive 
random numbers like $X$$_{k}$ and $X$$_{k+2}$. In general it is 
possible to study $n$-tupels $ {\bf X}_{k+\tau }\equiv (X_{k+
\tau _{1}},{\ldots},X_{k+\tau _{n}})$.  This amounts to replacing 
$ {\bf X}_{k}$ by $ {\bf X}_{k+\tau }$ and $s$$_{a}$ by $ s_{a,
\tau }\equiv s_{1}a^{\tau _{1}}+{\ldots}+s_{n}a^{\tau _{n}}$ in 
our results.

\subsection{Valuation with the generalized spectral test} 
Now we have to clarify how the calculation of $ |\hat{g}|^{2}$ leads 
to a valuation of the generator.

We can not expect that $ |\hat{g}|^{2}$ vanishes identically outside 
the origin since in this case $g$ would be constant. Eq.\ (\ref{30}) 
shows that the mean value of $ |\hat{g}|^{2}$ is 1.

What would we expect for a sum of truly random phases? Real and imaginary 
part of a random arrow with length 1 have equal variance $ 1/2$. 
For large $ \NX$ the sum of arrows is therefore normally distributed 
with density $ 1/\pi \NX\cdot \exp(-(x^{2}+y^{2})/\NX )dxdy=\exp
(-r^{2}{} /\NX )dr^{2}{} /\NX$. Thus $ z=|\hat{g}|^{2}$ has the 
density $ \exp(-z)$ for a true random sequence and the expected 
value for $ |\hat{g}|^{2}$ is 1.

This means that values of $ |\hat{g}|^{2}\le 1$ can be accepted. 
It is clear that for a given $ (s_{0},{\bf s})\neq (0,{\bf 0})$ 
the correlations are worse the higher $ |\hat{g}|^{2}(s_{0},
{\bf s})>1$ is. But what does the location of an $ (s_{0},
{\bf s})$ with $ |\hat{g}|^{2}(s_{0},{\bf s})>1$ mean for the generator?

We remember that $ (s_{0},{\bf s})$ may be seen as normal vector 
on the hyper-plane along which the accumulations occur. If e.g.\ 
$ n=1$ and $ (s_{0},s_{1})=(1,1)$ the corresponding 1-plane has 
the equation $ x_{0}+x_{1}=0$ (cf.\ e.g.\ Figs.\ 4I, 5I). If the 
$k$-axis and the $X$$_{k}$-axis are normalized to length 1 this 
line has length $ \sqrt{2}$. With the normal vector (3,1) (cf.\ 
Fig.\ 6I) one obtains the equation $ 3x_{0}+x_{1}=0{\rm\hspace
{.38ex}mod\hspace{.38ex}}1$ which intersects the unit cube three 
times and therefore has the length $ \sqrt{3^{2}+1} =\sqrt{10}$. 
Accumulations along this longer line are less important than along 
the short line. In the extreme case where the line fills the whole 
unit cube by intersecting it very often, accumulations can hardly 
be recognized. Note that in this sense the normal vectors $ (s_{
0},s_{1})$ and $ (2s_{0},2s_{1})$ do not determine the same line. 
The latter one contains e.g.\ the points $ (1/2,0)$, $ (0,1/2)$, 
$ (1,1/2)$, $ (1/2,1)$. It has twice the length of the former one 
and too large a $ |\hat{g}|^{2}$ has half the effect.

We generalize these considerations to $ n>1$ by taking the area of 
the $n$-dimensional hyper-plane with normal vector $ (s_{0},
{\bf s})$ as measure for the importance of the accumulations detected. 
The area is given by $ |(s_{0},{\bf s})|=(s_{0}^{2}+{\bf s}^{2}
)^{1/2}$, the Euclidean length of the normal vector.

We can relate both mechanisms by defining the quality parameter 
\begin{equation}
\label{55}Q_{n}\left( s_{0},{\bf s}\right) \equiv \frac{|\left( s
_{0},{\bf s}\right) |}{|\hat{g}\left( s_{0},{\bf s}\right) |^{2
}{}} \hspace{.6ex},\hspace{2ex}Q_{n}\equiv {\rm\hspace{.38ex}max\hspace{.38ex}}_
{\left( s_{0},{\bf s}\right) \in {\ErgoBbb Z}_\NX \times 
{\ErgoBbb Z}_M^{n}\backslash \{0,{\bf 0}\}}Q_{n}\left( s_{0},
{\bf s}\right) \hspace{.6ex}.
\end{equation} 
Good generators have $ Q_{1}\approx 1$. It is hard to achieve $ Q
_{n}\approx 1$ for $ n>1$ (see however Sec.\ \ref{multrec}). More 
realistic is $ Q_{n}\approx M^{1/n-1}$ (cf.\ Sec.\ \ref{param}) 
which means that the distribution of $n$-tupels deteriorates for 
higher $n$. In general small $n$ are more important than large $n$. 
Apart from the value of $Q$$_{n}$ also the number of sites $ (s_{
0},{\bf s})$ at which $ Q_{n}(s_{0},{\bf s})=Q_{n}$ is relevant 
(cf.\ Ex.\ 4.3 1.(a)).

Let us try to find an interpretation for $Q$$_{n}$. Assume the generator 
produces only multiples of $ t|M$. Then $ |\hat{g}|^{2}(0,s_{1}
=M/t,0,{\ldots},0)=\NX$, thus $ Q_{n}(0,M/t,0,{\ldots},0)=M/t\NX
$, and $ \NX Q_{n}$ determines the number of non-trivial digits. 
In general $ \NX Q_{n}$ may be larger than $M$ and therefore we 
say that $ \tilde{M}_{n}\equiv {\rm\hspace{.38ex}max}(\NX Q_{n}
,M)$ determines the number of digits we can rely on. Analogously 
$ \tilde{N}_{n}\equiv {\rm\hspace{.38ex}max}(\NX Q_{n},\NX)$ gives 
the quantity of random numbers for which the $n$-tupel distributions 
are reasonably random. Specifically $ \tilde{M}_{n}(s_{0},
{\bf s})={\rm\hspace{.38ex}max}(\NX Q_{n}(s_{0},{\bf s}),M)$ determines 
the digits and $ \tilde{N}_{n}(s_{0},{\bf s})={\rm\hspace
{.38ex}max}(\NX Q_{n}(s_{0},{\bf s}),\NX )$ the quantity of random 
numbers not affected by accumulations perpendicular to $ (s_{0}
,{\bf s})$ (cf.\ \cite[,p. 90]{Knu}).

Note however that these are only crude statements. If, e.g., the 
'period' of the generator is enlarged by simply repeating it then 
$ \NX Q_{n}(s_{0},{\bf s})$ remains unaffected only if $ s_{0}=0
$. Moreover, a high $ |\hat{g}|^{2}(s_{0},{\bf s})$ may be harmful 
even if $ |(s_{0},{\bf s})|$ is large (cf.\ Fig.\ 3I with $ Q_{
1}>1$).

Note that $Q$$_{n}$ is a relative quality parameter. Although $Q$
$_{n}$ usually does not increase with larger modulus (for $ n>1
$ is actually decreases) the quality of the generator gets better 
since $ \NX$ grows (cf.\ Fig.\ 4 vs.\ Fig.\ 8).

\section{Multiplicative groups of residue class rings\label
{rings}} 
This section provides the mathematical tools that are needed to analyze 
several random number generators. We derive Eq.\ (\ref{7}) which 
determines $ |\hat{g}|^{2}$ for $ X_{k}=a^{k}{\rm\hspace
{.38ex}mod\hspace{.38ex}}M$ for various moduli $M$ and multipliers 
$a$. There are some connections to the theory of Gau{\ss} sums (see 
e.g.\ \cite[\S 3]{Gauss}) where however the calculation of the absolute 
is almost trivial. In our case the exponent is not quadratic but 
has a linear and an exponential component (cf.\ Cor.\ 3.5, but also 
Ex.\ 4.6 4). The modulus is not restricted to a prime number. We 
are concerned with residue class rings rather than with fields.

Throughout the paper we use the following notation, 
\begin{eqnarray}
{\ErgoBbb Z}_M&\equiv &{\ErgoBbb Z}_{/{\ErgoBbb Z}M}\hbox{ integers 
}{\rm\hspace{.38ex}mod\hspace{.38ex}}M\hspace{.6ex}.\nonumber 
\\ 
{\ErgoBbb Z}_M^\times &\equiv &\hbox{\hspace{.38ex}multiplicative 
group of }{\ErgoBbb Z}_M\hspace{.6ex},\hspace{2ex} a\in 
{\ErgoBbb Z}_M^\times \Longleftrightarrow {\rm\hspace{.38ex}gcd}
\left( a,M\right) =1\hspace{.6ex}.\nonumber \\ 
\langle a\rangle _M&\equiv &\{a^{k}{\rm\hspace{.38ex}mod\hspace{.38ex}}M
,k\in {\ErgoBbb Z}\}\hbox{ , the cyclic group of }a{\rm\hspace
{.38ex}mod\hspace{.38ex}}M\hspace{.6ex}.\nonumber \\ 
\label{51}s_{a}&\equiv &\textstyle  \sum _{j=1}^{n}s_{j}a^{j-1}
\hspace{.6ex},\hspace{2ex}s_{a,M}\hspace*{1ex}\equiv 
\hspace*{1ex}{\rm\hspace{.38ex}gcd}\left( s_{a},M\right) 
\hspace{.6ex}.\\ 
N&\equiv &N_{a,M}\equiv |\langle a\rangle _M|\hbox{ , the order of 
}a{\rm\hspace{.38ex}mod\hspace{.38ex}}M\hspace{.6ex},\nonumber 
\\ 
&&\hbox{\hspace{.38ex}the subscripts $a$ and $M$ are suppressed except 
for in }s_{a}\hbox{ and }s_{a,M}.\nonumber \\ 
\delta _{a=b}&=&1\hspace*{2ex}\hbox{\hspace{.38ex}if }a=b
\hspace{.6ex},\hspace{2ex}\delta _{a=b}\hspace*{1ex}=
\hspace*{1ex}0\hspace*{2ex}\hbox{\hspace{.38ex}if }a\neq b
\hspace{.6ex}.\nonumber 
\end{eqnarray}

\pagebreak[3]

\noindent {\bf Definition} 3.1. 
Let $ M\in {\ErgoBbb N}$, $ a\in {\ErgoBbb Z}_M^\times $. Then $ m
\equiv m_{a}\equiv m_M\equiv m_{a,M}$ is the smallest positive integer 
with ($ N_{m}=|\langle a\rangle _{m}|$) 
\begin{eqnarray}
\label{21}&&m\hbox{ contains every prime factor of }M,\\ 
\label{22}&&4|m\hbox{ if }8|M,\\ 
\label{3}&&m={\rm\hspace{.38ex}gcd}\left( a^{N_{m}}-1,M\right) 
\hspace{.6ex}.
\end{eqnarray} 
Moreover 
\begin{equation}
\label{26}m_{0}\equiv {\rm\hspace{.38ex}gcd}\left( m,\Mm\right) 
\hspace{.6ex},\hspace{2ex}M_{0}\equiv M/mm_{0}=M/{\rm\hspace
{.38ex}gcd}\left( m^{2},M\right) \hspace{.6ex}.
\end{equation} 
In practice $m$ is easily determined by Eq.\ (\ref{23}). We begin 
with a technical proposition.

\pagebreak[3]

\noindent {\bf Proposition} 3.2. 
For $ M_{1}|M$, $ a\in {\ErgoBbb Z}_M^\times $ we define 
\begin{equation}
\label{24}c\equiv \frac{a^{N_{m}}-1}{m}\cdot \left\{ 
\begin{array}{ll}1&\hbox{\hspace{.38ex}if }M_{1}\hbox{ odd,\hspace{.38ex}}
\\ 
\left( 1+\frac{m}{2}+\frac{M}{4}\right) &\hbox{\hspace{.38ex}if }M
_{1}\hbox{ even, }M=4{\rm\hspace{.38ex}mod\hspace{.38ex}}8\hbox{ 
and }m=2{\rm\hspace{.38ex}mod\hspace{.38ex}}4,\\ 
\left( 1+\frac{m}{2}\right) &\hbox{\hspace{.38ex}else.\hspace{.38ex}}
\end{array}\right. 
\end{equation} 
With this definition we obtain for $ X_{k}=a^{k}$  
\begin{eqnarray}
\hspace{-5ex}\label{4}&&c\in {\ErgoBbb Z}^\times _\Mm
\hspace{.6ex},\\ 
\hspace{-5ex}\label{5}&&a^{N_{m}M_{1}k}=1+cmM_{1}k{\rm\hspace
{.38ex}mod \ gcd}\left( m^{2}M_{1},M\right) \hspace{.6ex},\\ 
\hspace{-5ex}\label{14}&&\exists c_{k}\in {\ErgoBbb Z}^\times _
\Mm \hbox{ with }a^{N_{m}k}=1+c_{k}mk{\rm\hspace{.38ex}mod\hspace{.38ex}}M
\hspace{.6ex},\\ 
\hspace{-5ex}\label{6}&&\hat{g}=\frac{\sqrt{N}}{N_{m}M_{1}{}} 
\sum _{k\in {\ErgoBbb Z}_{N_{m}M_{1}}}\hspace{-2ex}\delta _{
\frac{N_{m}M}{mN} s_{0}+s_{a}a^{k}c=0{\rm\hspace{.38ex}mod\hspace{.38ex}}
\frac{M}{mM_{1}{}} }\exp\left( \frac{2\pi i}{N} s_{0}k+\frac{2
\pi i}{M} s_{a}a^{k}\right) \!,\hbox{ if }M|m^{2}M_{1}.\\ 
\hspace{-5ex}\label{25}&&\hspace*{2ex}=\frac{\sqrt{N}}{N_{m}{}} 
\sum _{k\in {\ErgoBbb Z}_{N_{m}}}\hspace{-1.5ex}\delta _{ms_{0}
+s_{a}a^{k}\left( a^{N_{m}}-1\right) =0{\rm\hspace{.38ex}mod\hspace{.38ex}}M
}\exp\left( \frac{2\pi i}{N} s_{0}k+\frac{2\pi i}{M} s_{a}a^{k}
\right) \!,\hbox{ if }M|m^{2}.
\end{eqnarray}

\pagebreak[3]

\noindent {\bf Proof}. 
Eq.\ (\ref{4}) is an immediate consequence of Def.\ 3.1. Moreover 
\begin{eqnarray*}
&&a^{N_{m}M_{1}k}-1=\left( a^{N_{m}}-1+1\right) ^{M_{1}k}-1\\ 
&=&\left( a^{N_{m}}-1\right) M_{1}k\left( 1+\frac{M_{1}k-1}{2}
\left( a^{N_{m}}-1\right) \right) +\left( a^{N_{m}}-1\right) ^{
2}\sum _{j=3}^{M_{1}k}{M_{1}k\choose j}\left( a^{N_{m}}-1
\right) ^{j-2}.
\end{eqnarray*} 
First we show that the sum over $j$ is a multiple of $M$$_{1}$. Obviously 
$ {\ErgoBbb N}\ni {M_{1}k\choose j}={M_{1}k-1\choose j-1}\frac{M
_{1}k}{j}$. Since $j$ has at most $ j-2$ prime factors for $ j
\ge 3$ and $ a^{N_{m}}-1$ has by definition every prime factor of 
$M$$_{1}$ we obtain $ {\ErgoBbb N}\ni {M_{1}k-1\choose j-1}(a^{N
_{m}}-1)^{j-2}k/j={M_{1}k\choose j}(a^{N_{m}}-1)^{j-2}{} /M_{1}
$.

The first term equals $ (a^{N_{m}}-1)M_{1}k{\rm\hspace{.38ex}mod\hspace{.38ex}}m
^{2}M_{1}$ if $M$$_{1}$ is odd, and $ (a^{N_{m}}-1)M_{1}k(1+m/2
){\rm\hspace{.38ex}mod\hspace{.38ex}}m^{2}M_{1}$ if $M$$_{1}$ is 
even. Finally the term $ M/4$ can be added without harm if $M$$
_{1}$ is even and $ M=4{\rm\hspace{.38ex}mod\hspace{.38ex}}8$. This 
gives Eq.\ (\ref{5}).

Let $ M_{2}={\rm\hspace{.38ex}max\hspace{.38ex}}_{j}{\rm\hspace
{.38ex}gcd}(k,M^{j})$. Then $ k_{2}\equiv k/M_{2}\in 
{\ErgoBbb Z}_M^\times $ and with $ M_{1}\equiv {\rm\hspace
{.38ex}gcd}(M_{2},M)$ we have $ {\rm\hspace{.38ex}gcd}(m^{2}M_{
1},M)={\rm\hspace{.38ex}gcd}(m^{2}M_{2},M)=\ell _{1}m^{2}M_{2}+
\ell _{2}M$ for some $ \ell _{1},\ell _{2}\in {\ErgoBbb Z}$. With 
$ k_{2}^{-1}\in {\ErgoBbb Z}_M^\times $, $ \ell _{3}\in 
{\ErgoBbb Z}$ Eq.\ (\ref{5}) gives $ a^{N_{m}k}=1+cmk+\ell _{3}
\ell _{1}m^{2}M_{2}k_{2}k_{2}^{-1}{\rm\hspace{.38ex}mod\hspace{.38ex}}M
$. We find $ c_{k}\equiv c+\ell _{3}\ell _{1}mk_{2}^{-1}\in 
{\ErgoBbb Z}^\times _\Mm$ from Eqs.\ (\ref{21}), (\ref{4}).

Eq.\ (\ref{6}) is obtained if one plugs Eq.\ (\ref{5}) into the definition 
(\ref{1}) of $ \hat{g}$. We find 
\begin{eqnarray*}
\hat{g}&\stackrel{{\rm (\ref{42})\hspace{.38ex}}}{=}&\frac{1}{M_{
1}\sqrt{N}} \sum _{k\in {\ErgoBbb Z}_{NM_{1}}}\exp\left( \frac{
2\pi i}{N} s_{0}k+\frac{2\pi i}{M} s_{a}a^{k}\right) \\ 
&=&\frac{1}{M_{1}\sqrt{N}} \sum _{k\in {\ErgoBbb Z}_{N_{m}M_{1}
}}\sum _{k'\in {\ErgoBbb Z}_{N/N_{m}}}\exp\left( \frac{2\pi i}{N
} s_{0}\left( k+N_{m}M_{1}k'\right) +\frac{2\pi i}{M} s_{a}a^{k
}\left( 1+cmM_{1}k'\right) \right) \hspace{.6ex}.
\end{eqnarray*} 
The sum over $ k'$ gives the Kronecker $\delta $ and we get Eq.\ 
(\ref{6}).

Finally Eq.\ (\ref{25}) will be obtained for $ M_{1}=1$ with Eq.\ 
(\ref{8}). This equation will next be derived independently of Eq.\ 
(\ref{25}).\hfill $\Box $
\pagebreak[3] 

\pagebreak[3]

\noindent {\bf Lemma} 3.3. 
\begin{eqnarray}
\hspace{-5ex}&&\hbox{\hspace{.38ex}If }m'\hbox{ is the smallest number 
that meets Eqs.\ (\ref{21}), (\ref{22}) then }\nonumber \\ 
\hspace{-5ex}\label{23}&&\hspace*{1cm}N_{m}=N_{m'}
\hspace{.6ex},\hspace{2ex}m={\rm\hspace{.38ex}gcd}\left( a^{N_{m'}
}-1,M\right) \hspace{.6ex}.\\ 
\hspace{-5ex}&&N_{m_{2^{k}p_{1}^{k_{1}}\cdots p_{\ell }^{k_{
\ell }}}}=N_{2^{k_{0}}p_{1}\cdots p_{\ell }}={\rm\hspace
{.38ex}lcm}\left( N_{2^{k_{0}}},N_{p_{1}},{\ldots},N_{p_{\ell }
}\right) \hspace{.6ex},\nonumber \\ 
\hspace{-5ex}\label{27}&&\hspace*{1cm}\hbox{\hspace{.38ex}if }p_{
1},{\ldots},p_{\ell }\hbox{ are odd, prime and }k_{0}={\rm\hspace
{.38ex}min}\left( k,2\right) .\\ 
\hspace{-5ex}&&N_{m}=1\Leftrightarrow m={\rm\hspace{.38ex}gcd}
\left( a-1,M\right) \Leftrightarrow a-1\hbox{ meets conditions 1 
and 2 of Def.\ 3.1.\hspace{.38ex}}\\ 
\hspace{-5ex}&&M=P^d,\hspace*{1ex}2\neq P\hbox{ prime: $a$ is primitive, 
}\langle a\rangle _{P^d}={\ErgoBbb Z}_{P^d}^\times 
\Leftrightarrow m=P\wedge N_{m}=P-1\hspace{.6ex}.\\ 
\hspace{-5ex}&&M=2^d,\hspace*{1ex}d\ge 3:\hspace*{1ex}a=1{\rm\hspace
{.38ex}mod\hspace{.38ex}}4\Leftrightarrow N_{m}=1
\hspace{.6ex},\hspace{2ex}a=3{\rm\hspace{.38ex}mod\hspace{.38ex}}4
\Leftrightarrow N_{m}=2\hspace{.6ex}.\\ 
\hspace{-5ex}\label{12}&&M'|M\Rightarrow N_{m_{M'}}|N_{m}
\hspace{.6ex},\hspace{2ex}m_{M'}|m\hspace{.6ex}.\\ 
\hspace{-5ex}\label{13}&&m_m=m\hspace{.6ex}.\\ 
\hspace{-5ex}\label{11}&&\langle a\rangle _M=\langle a\rangle _{
m}+m{\ErgoBbb Z}_\Mm\hspace{.6ex}.\\ 
\hspace{-5ex}\label{8}&&\frac{N}{M} =\frac{N_{m}}{m}{} 
\hspace{.6ex}.\\ 
\hspace{-5ex}\label{9}&&m_{m_{0}}=m_\Mm\hspace{.6ex}.\\ 
\hspace{-5ex}\label{18}&&\frac{\N}{N_{m_{0}}{}} =M_{0}
\hspace{.6ex},\hspace{2ex}\frac{N}{\N} =\frac{N_{m}m_{0}}{N_{m_{
0}}{}} \hspace{.6ex}.\\ 
\hspace{-5ex}\label{10}&&N_{m_{0}}={\rm\hspace{.38ex}gcd}\left( N
_{m},\N \right) \hspace{.6ex}.\\ 
\hspace{-5ex}\label{16}&&N_{m_{a^A}}=\frac{N_{m}}{{\rm\hspace
{.38ex}gcd}\left( N_{m},A\right) } \hspace{.6ex},\hspace{2ex}m_
{a^A}=\frac{{\rm\hspace{.38ex}gcd}\left( N,A\right) m}{{\rm\hspace
{.38ex}gcd}\left( N_{m},A\right) } \hspace{.6ex}.\\ 
\hspace{-5ex}\label{19}&&N_{m_{a^\N }}=\frac{N_{m}}{N_{m_{0}}{}
} \hspace{.6ex},\hspace{2ex}m_{a^\N }=\frac{M}{m_{0}{}} 
\hspace{.6ex}.
\end{eqnarray}

\pagebreak[3]

\noindent {\bf Proof}. 
Obviously $ m'|m$ and thus $ N_{m'}|N_{m}$ since $ \langle a
\rangle _{m'}$ is a subgroup of $ \langle a\rangle _{m}$. Since 
$m$ is minimal we find $ N_{m}=N_{m'}$ and thus Eq.\ (\ref{23}). 
Eqs.\ (\ref{27})--(\ref{13}) follow directly from Eq.\ (\ref{23}).

We use Eq.\ (\ref{6}) with $ s_{0}=0$ and $ n=1$ to Fourier transform 
$ \langle a\rangle _M$. We choose $M$$_{1}$ with $ s_{1}m^{2}M_{
1}=0{\rm\hspace{.38ex}mod\hspace{.38ex}}M$ and obtain that $ 
\hat{g}$ vanishes unless $ ca^{k}s_{1}mM_{1}=0{\rm\hspace
{.38ex}mod\hspace{.38ex}}M$. Since $ a\in {\ErgoBbb Z}_M^
\times $, $ c\in {\ErgoBbb Z}^\times _\Mm$ (Eq.\ (\ref{4})) this 
implies $ s_{1}mM_{1}=0{\rm\hspace{.38ex}mod\hspace{.38ex}}M$. Thus 
for non-vanishing $ \hat{g}$ and $ M_{1}'=M_{1}/{\rm\hspace
{.38ex}gcd}(M_{1},m)$ we have $ s_{1}m^{2}M_{1}'=0{\rm\hspace
{.38ex}mod\hspace{.38ex}}M$ which enables us to use Eq.\ (\ref{6}) 
again (with $M$$_{1}$ replaced by $ M_{1}'$) yielding the stronger 
condition $ s_{1}mM_{1}'=0{\rm\hspace{.38ex}mod\hspace{.38ex}}M
$. Since $m$ contains every prime factor of $M$ and $ M_{1}|M$ we 
can continue until $ M_{1}^{\prime {\ldots}\prime }=1$ and finally 
we obtain 
\begin{equation}
\hat{g}[N]\left( 0,{\bf s}\right) =\sqrt{N} /N_{m}\delta _{s_{1}
=0{\rm\hspace{.38ex}mod}\Mm }\hat{g}[N_{m}]\left( 0,{\bf s}
\right) \hspace{.6ex}.
\end{equation} 
This fixes $ \langle a\rangle _M$ to be $ \{a^{k}+k'm{\rm\hspace
{.38ex}mod\hspace{.38ex}}M,\hspace*{1ex}k\in {\ErgoBbb Z}_{N_{m
}},k'\in {\ErgoBbb Z}_\Mm\}$ (Eq.\ (\ref{42})). This is the statement 
of Eq.\ (\ref{11}) and Eq.\ (\ref{8}) follows immediately.

From $ m_\Mm |\Mm$ and $ m_\Mm |m$ (Eq.\ (\ref{12})) we get $ m_
\Mm |m_{0}$. Eqs.\ (\ref{13}) and (\ref{12}) give $ m_\Mm =m_{m_
\Mm }|m_{m_{0}}|m_\Mm$, thus Eq.\ (\ref{9}) is verified.

Eq.\ (\ref{18}) follows from Eqs.\ (\ref{8}), (\ref{9}): 
\begin{displaymath}
\N \stackrel{{\rm (\ref{8})\hspace{.38ex}}}{=} \frac{M}{m} 
\frac{N_{m_\Mm }}{m_\Mm} \stackrel{{\rm (\ref{9})\hspace{.38ex}}
}{=} \frac{MN_{m_{m_{0}}}}{mm_{m_{0}}{}} \stackrel{{\rm (\ref{8})\hspace{.38ex}}
}{=} \frac{MN_{m_{0}}}{mm_{0}{}} \stackrel{{\rm (\ref{8})\hspace{.38ex}}
}{=} \frac{NN_{m_{0}}}{N_{m}m_{0}{}} \hspace{.6ex}.
\end{displaymath} 
To prove Eq.\ (\ref{10}) we begin with 
\begin{displaymath}
\frac{m_{0}}{N_{m_{0}}{}} {\rm\hspace{.38ex}gcd}\left( N_{m},N_\Mm 
\right)  \stackrel{{\rm (\ref{8}),(\ref{9})\hspace{.38ex}}}{=} 
\frac{m_\Mm }{N_{m_\Mm} }{} {\rm\hspace{.38ex}gcd}\left( N_{m},N_
\Mm \right)  \stackrel{{\rm (\ref{8})\hspace{.38ex}}}{=} {\rm\hspace
{.38ex}gcd}\left( \frac{N_{m}m_\Mm }{N_{m_\Mm} }{} ,\frac{M}{m}
\right) \hspace{.6ex}.
\end{displaymath} 
We will show that the right hand side equals $m$$_{0}$. From Eq.\ 
(\ref{14}) we know that there exists a $ c'\in {\ErgoBbb Z}^
\times _\frac{\Mm }{m_\Mm}$ with 
\begin{displaymath}
cm\equiv a^{N_{m}}-1=a^{N_{m_\Mm }\frac{N_{m}}{N_{m_\Mm} }{} }-1
=c'm_\Mm \frac{N_{m}}{N_{m_\Mm} }{} {\rm\hspace{.38ex}mod\hspace{.38ex}}
\frac{M}{m}\hspace{.6ex}.
\end{displaymath} 
Since $ c\in {\ErgoBbb Z}^\times _\Mm$ we have also $c$, $ c''\equiv cc'
^{-1}\in {\ErgoBbb Z}^\times _\frac{\Mm }{m_\Mm}$ and the above 
equation implies that there exists a $ k\in {\ErgoBbb Z}$ with $ N
_{m}m_\Mm /N_{m_\Mm }=kM/m+c''m$. Finally we find for the right 
hand side of the first equation ($ c''\in {\ErgoBbb Z}^\times _
\frac{\Mm }{m_{0}}$, Eq.\ (\ref{9})) 
\begin{displaymath}
{\rm\hspace{.38ex}gcd}\left( m_\Mm N_{m}/N_{m_\Mm }{} ,M/m
\right) ={\rm\hspace{.38ex}gcd}\left( k\Mm +c''m,M/m \right) =m_{
0}\hspace{.6ex}.
\end{displaymath} 
Since $ m|m_{a^A}|a^{AN_{m_{a^A}}}-1$, we get $ N_{m}|AN_{m_{a^A
}}$. On the other hand $ N_{m_{a^A}}$ is the smallest number with 
this property, so $ N_{m_{a^A}}=N_{m}/{\rm\hspace{.38ex}gcd}(N_{
m},A)$. Moreover we find 
\begin{displaymath}
m_{a^A}{} \stackrel{{\rm (\ref{8})\hspace{.38ex}}}{=} \frac{N_{m_
{a^A}}M}{N_{a^A}{}} \stackrel{{\rm (\ref{16})\hspace{.38ex}}}{=
} \frac{N_{m}M}{{\rm\hspace{.38ex}gcd}\left( N_{m},A\right) N_{a^A
}{}} =\frac{N_{m}M{\rm\hspace{.38ex}gcd}\left( N,A\right) }{{\rm\hspace
{.38ex}gcd}\left( N_{m},A\right) N} \stackrel{{\rm (\ref{8})\hspace{.38ex}}
}{=} \frac{{\rm\hspace{.38ex}gcd}\left( N,A\right) m}{{\rm\hspace
{.38ex}gcd}\left( N_{m},A\right) } \hspace{.6ex}.
\end{displaymath} 
Finally the first statement of Eq.\ (\ref{19}) follows from Eq.\ 
(\ref{16}) and Eq.\ (\ref{10}). Moreover 
\begin{displaymath}
m_{a^\N }{} \stackrel{{\rm (\ref{16})\hspace{.38ex}}}{=} \frac{{\rm\hspace
{.38ex}gcd}\left( N,\N \right) m}{{\rm\hspace{.38ex}gcd}\left( N
_{m},\N \right) } \stackrel{{\rm (\ref{10})\hspace{.38ex}}}{=} 
\frac{\N m}{N_{m_{0}}{}} \stackrel{{\rm (\ref{18}),(\ref{26})\hspace{.38ex}}
}{=} \frac{M}{m_{0}{}} \hspace{.6ex}.
\end{displaymath}\hfill $\Box $
\pagebreak[3] 

\pagebreak[3]

\noindent {\bf Theorem} 3.4. 
For $ M\in {\ErgoBbb N}$ let $ X_{k}=a^{k}{\rm\hspace{.38ex}mod\hspace{.38ex}}M
$. If $a$, $ {\bf s}\cdot {\bf X}_{0}\equiv s_{a}\equiv \sum _{
j=1}^{n}s_{j}a^{j-1}\in {\ErgoBbb Z}_M^\times $ we obtain for $ 
|\hat{g}|^{2}$ defined in Eq.\ (\ref{1}) and $m$, $m$$_{0}$ given 
by Def.\ 3.1, 
\begin{eqnarray}
\label{17}|\hat{g}|^{2}[N,X_\bullet ]\left( s_{0},{\bf s}
\right) &=&\frac{1}{N_{m}{}} \sum _{k\in {\ErgoBbb Z}_{N_{m}}}|
\hat{g}|^{2}[N/\N ,X_{k+\N \bullet }]\left( s_{0},{\bf s}\cdot 
{\bf X}_{0}\right) \\ 
\label{7}&&\hspace{-3cm}=\hspace*{1ex}\left\{ 
\begin{array}{l}\displaystyle {} \frac{m_{0}}{N_{m_{0}}{}} |\hat
{g}|^{2}_{{\ErgoBbb Z}_{N_{m}/N_{m_{0}}}}[N/\N ,X_{k+\N\bullet }
]\left( s_{0},{\bf s}\cdot {\bf X}_{0}\right) \hspace{.6ex},\\ 
\hspace*{4ex}\hbox{\hspace{.38ex}if }\exists k\in {\ErgoBbb Z}_
{N_{m_{0}}}:s_{0}+{\bf s}\!\cdot \!\left( {\bf X}_{N_{m}+k}\!-
\!{\bf X}_{k}\right) \!/\!m=\frac{m_{0}}{2}\delta _{2|\frac{M}{mm
_{0}{}} }{\rm\hspace{.38ex}mod\hspace{.38ex}}m_{0}\hspace*{1ex}
,\\ 
0\hspace*{3ex}\hbox{\hspace{.38ex}else .\hspace{.38ex}}
\end{array}\right. 
\end{eqnarray}

\pagebreak[3]

\noindent {\bf Corollary} 3.5. 
For all $s$$_{a}$ and all generators of the type 
\begin{equation}
X_k=c_{1}a^{k}+c_{2}k+c_{3}{\rm\hspace{.38ex}mod\hspace{.38ex}}M
\end{equation} 
with period $ \NX$ we have 
\begin{eqnarray}
\label{28}&&|\hat{g}|^{2}[\NX ,M,X]\left( s_{0},{\bf s}\right) =
\frac{\NX}{N'} |\hat{g}|^{2}[N',M',X]\left( \frac{N'}{\NX} s_{0
},\frac{M'}{M} {\bf s}\right) \hspace{.6ex},\hspace{2ex}\hbox
{\hspace{.38ex}with\hspace{.38ex}}\\ 
&&M'=M/{\rm\hspace{.38ex}gcd}\left( c_{1}s_{a},M\right) ,
\hspace*{1ex}N'=N_{M'}\hspace{.6ex},\nonumber 
\end{eqnarray} 
and for the right hand side of Eq.\ (\ref{28}) Eqs.\ (\ref{17}), 
(\ref{7}) apply. With $ m'=m_{M'}$, $ m'_{0}={m_{0}}_{M'}$, $ M_{
0}'=M'/m'm_{0}'$ we obtain 
\begin{equation}
\label{34}|\hat{g}|^{2}[\NX ,X_\bullet ]\left( s_{0},{\bf s}
\right) =\left\{ 
\begin{array}{l}\displaystyle {} \frac{\NX}{N_{m'}N_{m_{0}'}M_{0
}'} |\hat{g}|^{2}_{{\ErgoBbb Z}_{N_{m'}/N_{m_{0}'}}}[\frac{\NX}
{N_{m_{0}'}M_{0}'} ,X_{k+N_{m_{0}'}M_{0}'\bullet }]\left( s_{0}
,{\bf s}\cdot {\bf X}_{0}\right) \hspace{.6ex},\\ 
\hspace*{3ex}\hbox{\hspace{.38ex}if }\displaystyle {} \exists k
\!\in \!{\ErgoBbb Z}_{N_{m_{0}'}}:\frac{N_{m'}s_{0}}{\NX} +
\frac{{\bf s}\!\cdot \!\left( {\bf X}_{N_{m'}+k}\!-\!{\bf X}_{k
}\right) }{M} =\frac{\delta _{2|M_{0}'}}{2M_{0}'} {\rm\hspace
{.38ex}mod\hspace{.38ex}}\frac{1}{M_{0}'} ,\\ 
0\hspace*{2ex}\hbox{\hspace{.38ex}else .\hspace{.38ex}}
\end{array}\right. 
\end{equation} 
The denominators in the above condition are understood according 
to 
\begin{equation}
\label{56}a/b =c/d {\rm\hspace{.38ex}mod\hspace{.38ex}}1/M_{0}'\Longleftrightarrow adM
_{0}'=cbM_{0}'{\rm\hspace{.38ex}mod\hspace{.38ex}}bd
\hspace{.6ex}.
\end{equation}

\pagebreak[3]

\noindent {\bf Proof} of the theorem. 
We start with Eq.\ (\ref{15}) 
\begin{eqnarray*}
&|\hat{g}|^{2}&=\hspace*{1ex}\frac{1}{N} \sum _{k,\Delta k\in 
{\ErgoBbb Z}_N}\exp\left( \frac{2\pi i}{N} s_{0}\Delta k+\frac{
2\pi i}{M} s_{a}a^{k}\left( a^{\Delta k}-1\right) \right) \\ 
&\stackrel{{\rm (\ref{11})\hspace{.38ex}}}{=}&\frac{1}{N} \sum _
{k\in {\ErgoBbb Z}_{N_{m}}}\sum _{k'\in {\ErgoBbb Z}_\Mm }\sum _
{\Delta k\in {\ErgoBbb Z}_N}\exp\left( \frac{2\pi i}{N} s_{0}
\Delta k+\frac{2\pi i}{M} s_{a}\left( a^{k}+mk'\right) \left( a^
{\Delta k}-1\right) \right) \\ 
&=&\frac{1}{N_{m}{}} \sum _{k\in {\ErgoBbb Z}_{N_{m}}}\sum _{
\Delta k\in {\ErgoBbb Z}_N}\delta _{s_{a}\left( a^{\Delta k}-1
\right) =0{\rm\hspace{.38ex}mod}\Mm }\exp\left( \frac{2\pi i}{N
} s_{0}\Delta k+\frac{2\pi i}{M} s_{a}a^{k}\left( a^{\Delta k}-1
\right) \right) \\ 
&=&\frac{1}{N_{m}{}} \sum _{k\in {\ErgoBbb Z}_{N_{m}}}\sum _{
\Delta k\in {\ErgoBbb Z}_{N/\N }}\exp\left( \frac{2\pi i}{N} \N 
s_{0}\Delta k+\frac{2\pi i}{M} s_{a}a^{k}\left( a^{\N \Delta k}
-1\right) \right) \\ 
&&[\hbox{\hspace{.38ex}replace }\Delta k\hbox{ by }\Delta k-
\Delta k'\hbox{ and }k\hbox{ by }k+\N \Delta k']\\ 
&=&\frac{1}{N_{m}{}} \sum _{k\in {\ErgoBbb Z}_{N_{m}}}\sum _{
\Delta k\in {\ErgoBbb Z}_{N/\N }}\hspace{-2ex}\exp\left( \frac{
2\pi i}{N} \N s_{0}\left( \Delta k-\Delta k'\right) +\frac{2
\pi i}{M} s_{a}a^{k}\left( a^{\N \Delta k}-a^{\N \Delta k'}
\right) \!\!\right) \\ 
&&[\hbox{\hspace{.38ex}average over }\Delta k'\hbox{ with }\N /N
\sum \nolimits _{\Delta k'\in {\ErgoBbb Z}_{N/\N }}]\\ 
&=&\frac{1}{N_{m}{}} \sum _{k\in {\ErgoBbb Z}_{N_{m}}}|\hat{g}|^{
2}[N/\N ,X_{\tilde{k}}=a^{k+\N \tilde{k}}]\left( s_{0},s_{a}
\right) 
\end{eqnarray*} 
which proves Eq.\ (\ref{17}).

Now we split the $k$-sum according to $ k=k_{0}+{\rm\hspace
{.38ex}gcd}(N_{m},\N )k'=k_{0}+N_{m_{0}}k'$. One can replace $ k'$ 
by $ \N /N_{m_{0}}k'$ since $ k'\in {\ErgoBbb Z}_{N_{m}/N_{m_{0
}}}$ and $ \N /N_{m_{0}}\in {\ErgoBbb Z}^\times _{N_{m}/N_{m_{0
}}}$. So $ k=k_{0}+\N k'$ and the $ k'$-dependence drops, since 
it amounts to a shift $ \tilde{k}\mapsto \tilde{k}+k'$ of $ X_{
\tilde{k}}=a^{k+\N \tilde{k}}$ (cf.\ Eq.\ (\ref{20})). The sum over 
$ k'$ gives simply $ N_{m}/N_{m_{0}}$.

Since $ m_{a^\N }=0{\rm\hspace{.38ex}mod\hspace{.38ex}}m$ and $ m_
{a^\N }=0{\rm\hspace{.38ex}mod}\Mm$ we have  $ (m_{a^\N })^{2}=0
{\rm\hspace{.38ex}mod\hspace{.38ex}}M$ and Eq.\  (\ref{25}) applies 
with $a$ replaced by $ a^\N$ and $s$$_{a}$ by $ a^{k}s_{a}$. With 
Eq.\ (\ref{19}) we get 
\begin{eqnarray*}
&&\hspace{-.5cm}|\hat{g}|^{2}\hspace*{1ex}=\hspace*{1ex}\frac{1}
{N_{m_{0}}{}} \sum _{k\in {\ErgoBbb Z}_{N_{m_{0}}}}\frac{N/\N}{N
_{m}/N_{m_{0}}{}} \cdot \frac{N_{m_{0}}}{N_{m}{}} \cdot \\ 
&&\hspace{-.5cm}\left| \sum _{k'\in {\ErgoBbb Z}_\frac{N_{m}}{N_
{m_{0}}{}} }\delta _{\frac{s_{0}M}{m_{0}{}} +s_{a}a^{k+\N k'}(a^
{\N N_{m}/N_{m_{0}}}-1)=0{\rm\hspace{.38ex}mod\hspace{.38ex}}M}
\exp\left( \frac{2\pi i}{N} \N s_{0}k+\frac{2\pi i}{M} s_{a}a^{k
+\N k'}\right) \right| ^{2}.
\end{eqnarray*} 
The $ k'$-dependence drops $ {\rm\hspace{.38ex}mod\hspace{.38ex}}M
$ (Eq.\ (\ref{18})) in the argument of the Kronecker $\delta $. 
The $ k'$-sum interchanges with the Kronecker $\delta $ and we obtain 
with Eq.\ (\ref{18}) 
\begin{displaymath}
|\hat{g}|^{2}=\frac{m_{0}}{N_{m_{0}}{}} \sum _{k\in {\ErgoBbb Z}_
{N_{m_{0}}}}\delta _{\frac{M}{m_{0}{}} s_{0}+s_{a}a^{k}(a^{N_{m
}M_{0}}-1)=0{\rm\hspace{.38ex}mod\hspace{.38ex}}M}|\hat{g}|^{2}_
{{\ErgoBbb Z}_\frac{N_{m}}{N_{m_{0}{}} }}[N/\N ,X_{k+\N 
\bullet }].
\end{displaymath} 
We use Eq.\ (\ref{5}) with $ M_{1}=M_{0}$, $ k=1$ yielding $ {\rm\hspace
{.38ex}mod\hspace{.38ex}}M$ 
\begin{displaymath}
a^{N_{m}M_{0}}-1=c\frac{M}{m_{0}{}} =\frac{M}{m_{0}{}} \left( 
\frac{a^{N_{m}}-1}{m}\right) \left( 1+\frac{m}{2}\delta _{2|
\frac{M}{mm_{0}{}} }\right)  \stackrel{{\rm (\ref{14})\hspace{.38ex}}
}{=} \frac{M}{m_{0}{}} \left( \frac{a^{N_{m}}-1}{m}+\frac{m}{2}
\delta _{2|\frac{M}{mm_{0}{}} }\right) .
\end{displaymath} 
Finally we divide the argument of the Kronecker $\delta $ by $ M
/m_{0}$ and notice that, if $ 2|\frac{M}{mm_{0}}$, we have $ m/2
=m_{0}/2=-m_{0}/2{\rm\hspace{.38ex}mod\hspace{.38ex}}m_{0}$. Since 
the $k$-sum has at most one non-vanishing term we obtain the desired 
result.\hfill $\Box $
\pagebreak[3] 

\pagebreak[3]

\noindent {\bf Proof} of Cor.\ 3.5. 
With the definition $ {\bf s}'\equiv M'c_{1}{\bf s}/M$ we find $ s
_{a}'\equiv \sum _{j=1}^{n}s'_{j}a^{j-1}=c_{1}s_{a}/{\rm\hspace
{.38ex}gcd}(c_{1}s_{a},M)\in {\ErgoBbb Z}_{M'}^\times $. With $ 
\NNX \equiv {\rm\hspace{.38ex}lcm}(\NX ,M)$ we obtain 
\begin{eqnarray*}
|\hat{g}|^{2}\left( s_{0},{\bf s}\right) &=&\frac{1}{\NX} 
\left| \sum _{k\in {\ErgoBbb Z}_\NX }\exp\left( \frac{2\pi i}{\NX
} s_{0}k+\frac{2\pi i}{M} \left( c_{1}s_{a}a^{k}+c_{2}\sum _{j=
1}^{n}s_{j}\left( k+j-1+c_{3}\right) \right) \right) \right| ^{
2}{}\\ 
&=&\frac{1}{\NX} \left| \sum _{k\in {\ErgoBbb Z}_\NX }\exp
\left( \frac{2\pi i}{\NNX} \left( \frac{\NNX}{\NX} s_{0}+\frac{
\NNX}{M} c_{2}\sum _{j=1}^{n}s_{j}\right) k+\frac{2\pi i}{M'} s_{
a}'a^{k}\right) \right| ^{2}.
\end{eqnarray*} 
We can replace $ \sum _{k\in {\ErgoBbb Z}_\NX }$ by $ \NX\!/\!\NNX
\!\sum _{k\in {\ErgoBbb Z}_{\NNX }}$. This sum splits into $ 
\sum _{k_{0}\in {\ErgoBbb Z}_{N'}}\sum _{k'\in {\ErgoBbb Z}_{\NNX 
/N'}}$ with $ k=k_{0}+N'k'$. The exponential depends linearly on 
$ k'$, thus the sum over $ k'$ gives a Kronecker $\delta $, 
\begin{displaymath}
\frac{1}{\NX} \left| \sum _{k\in {\ErgoBbb Z}_\NX }{} {\ldots}
\right| ^{2}=\frac{\NX}{N'} \delta _{\NNX N'\left( s_{0}/\NX +c_{
2}\sum _{j=1}^{n}s_{j}/M\right) =0{\rm\hspace{.38ex}mod}\NNX }
\frac{1}{N'} \left| \sum _{k\in {\ErgoBbb Z}_{N'}}{} {\ldots}
\right| ^{2}\hspace{.6ex}.
\end{displaymath} 
The Kronecker $\delta $ assures that $ s_{0}'\equiv N'(s_{0}/\NX 
+c_{2}\sum _{j=1}^{n}s_{j}/M)$ is an integer. The sum over $k$ gives 
$ N'|\hat{g}|^{2}[N',M',X'_{k}=a^{k}](s_{0}',{\bf s}')$ and Theorem 
3.4 applies. With $ \delta _{s_{0}'{\rm\hspace{.38ex}integer\hspace{.38ex}}
}$ and Eq.\ (\ref{56}) the condition in Eq.\ (\ref{7}) becomes 
\begin{eqnarray*}
&&N'\left( \frac{s_{0}}{\NX} +\frac{c_{2}\sum _{j=1}^{n}s_{j}}{M
} \right) +\frac{M'c_{1}s_{a}}{Mm'} a^{k}\left( a^{N_{m'}}-1
\right) =\frac{m_{0}'}{2}\delta _{2|M_{0}'}{\rm\hspace{.38ex}mod\hspace{.38ex}}m
_{0}'\\ 
&\stackrel{{\rm (\ref{8})\hspace{.38ex}}}{=}&\frac{N'}{\NX} s_{0
}+\frac{M'}{Mm'} \left( c_{1}s_{a}a^{N_{m'}+k}+c_{2}N_{m'}\sum _{
j=1}^{n}s_{j}-c_{1}s_{a}a^{k}\right) =\frac{m_{0}'}{2}\delta _{2
|M'_{0}}{\rm\hspace{.38ex}mod\hspace{.38ex}}m_{0}'\hspace{.6ex}.
\end{eqnarray*} 
The expression in the brackets equals $ {\bf s}\cdot ({\bf X}_{N_
{m'}+k}-{\bf X}_{k})$. 
So we have confirmed the invariance of the condition in Eq.\ (\ref{7}) 
as claimed in Eq.\ (\ref{28}). Similarly 
\begin{eqnarray*}
\lefteqn{|\hat{g}|^{2}[N'/\Nn ,M,X'_{k+\Nn\bullet }]\left( s_{0}',
{\bf s}'\cdot {\bf X}_{0}'\right) }\\ 
&=&\left| \sum _{k'}\exp\left( 2\pi i\left( \frac{\Nn}{\NX} s_{0
}+\frac{\Nn}{M} c_{2}\sum _{j=1}^{n}s_{j}\right) k'+\frac{2\pi 
i}{M} c_{1}s_{a}a^{k+\Nn k'}\right) \right| ^{2}{}\\ 
&=&|\hat{g}|^{2}[N'/\Nn ,M',X_{k+\Nn \bullet }]\left( N'/\NX 
\cdot s_{0},M'/M \cdot {\bf s}\cdot {\bf X}_{0}\right) \\ 
&=&|\hat{g}|^{2}[\NX /\Nn ,M,X_{k+\Nn \bullet }]\left( s_{0},
{\bf s}\cdot {\bf X}_{0}\right) .
\end{eqnarray*} 
Altogether we have proven that $ |\hat{g}|^{2}[\NX ,M]$ is given 
by Eqs.\ (\ref{17}), (\ref{7}) and Eq.\ (\ref{28}). With Eqs.\ (\ref{26}), 
(\ref{8}), (\ref{18}), (\ref{7}) we find Eq.\ (\ref{34}).\hfill $\Box $
\pagebreak[3] 

\section{Generators\label{generators}} 
\subsection{$ X_{0}=c$, $ X_{k+1}=aX_{k}{\rm\hspace{.38ex}mod\hspace{.38ex}}M
$} 
We assume that $ 1\neq a,c\in {\ErgoBbb Z}_M^\times $. Generators 
of this type are called multiplicative. The recursion formula implies 
$ X_{k}=ca^{k}{\rm\hspace{.38ex}mod\hspace{.38ex}}M$. The harmonic 
analysis of many multiplicative generators is easily obtained by 
Eqs.\ (\ref{7}) and (\ref{34}).

\pagebreak[3]

\noindent {\bf Examples} 4.1. 
\begin{enumerate}\item{}$ N_{m}=1$ (Fig.\ 1). We find $ m={\rm\hspace
{.38ex}gcd}(a-1,M)$, $ N=M/m$, $ m_{0}={\rm\hspace{.38ex}gcd}(m
,M/m)={\rm\hspace{.38ex}gcd}((a-1)^{2},M)/{\rm\hspace{.38ex}gcd}
(a-1,M)$ and $ M_{0}=M/{\rm\hspace{.38ex}gcd}((a-1)^{2},M)$.

With $ M'=M/{\rm\hspace{.38ex}gcd}(s_{a},M)=M/s_{a,M}$ we get $ N_
{m'}=N_{m_{0}'}=1$, $ m'={\rm\hspace{.38ex}gcd}(a-1,M')$ and $ M
/M_{0}'={\rm\hspace{.38ex}gcd}(m'^{2}{\rm\hspace{.38ex}gcd}(s_{
a},M),M)={\rm\hspace{.38ex}gcd}({\rm\hspace{.38ex}gcd}((a-1)^{2
},M'^{2})s_{a},M)={\rm\hspace{.38ex}gcd}(s_{a}m^{2},M)=mm_{0}{\rm\hspace
{.38ex}gcd}(s_{a},M_{0})=mm_{0}s_{a,M_{0}}$. Thus 
\begin{equation}
\label{35}M_{0}'=M_{0}/s_{a,M_{0}}\hbox{ if }N_{m}=1
\hspace{.6ex}.
\end{equation} 
We obtain from Eq.\ (\ref{34}) 
\begin{equation}
\label{31}|\hat{g}|^{2}\left( s_{0},{\bf s}\right) =m_{0}s_{a,M_{
0}}\delta _{s_{0}+cs_{a}\left( a-1\right) /m=\frac{1}{2}m_{0}s_
{a,M_{0}}\delta _{2|\frac{M_{0}}{s_{a,M_{0}}{}} }{\rm\hspace
{.38ex}mod\hspace{.38ex}}m_{0}s_{a,M_{0}}}\hspace{.6ex}.
\end{equation} 
In the case of a power of two modulus $ M=2^d$ with $ m=4$ ($ m=2
$ is not possible for $ d>2$) we obtain for a proper choice of $a$ 
(cf.\ Sec.\ \ref{param}) 
\begin{equation}
\NX =M/4\hspace{.6ex},\hspace{2ex}Q_{1}=\sqrt{2} /4
\hspace{.6ex},\hspace{2ex}Q_{n\ge 2}\approx M^{1/n-1}
\hspace{.6ex}.
\end{equation} 
If we look at correlations of $k$ with $X$$_{k}$, disregarding higher 
$n$-tupels, we get $ \tilde{N}_{1}= \tilde{M}_{1}=NQ_{1}=\sqrt{2
} M/16$ reasonable random numbers with $ \log (\sqrt{2} M/16)$ digits.

A detailed discussion is postponed to the next section where we analyze 
the mixed multiplicative generator which is very similar but more 
popular than the multiplicative generator. 
\item{}$ M=P^d$, $P$ is odd, prime and $a$ is a primitive element 
of $ {\ErgoBbb Z}_{P^d}^\times $, $ \langle a\rangle _{P^d}=
{\ErgoBbb Z}_{P^d}^\times $, $ c\in {\ErgoBbb Z}_{P^d}^\times $ 
(Figs.\ 2, 3). We find $ N_{m}=P-1$, $ m=P$, $ N=(P-1)P^{d-1}$, 
$ m_{0}=1$ if $ d=1$ and $ m_{0}=P$ if $ d\ge 2$.

For the case $ d=1$ Eq.\ (\ref{7}) is useless. However, we saw in 
Sec.\ \ref{gentest} how to calculate $ |\hat{g}|^{2}$ from the very 
definition. In fact it is even simpler to observe that $ |\hat
{g}|^{2}$ is constant for all $ s_{a}\neq 0{\rm\hspace{.38ex}mod\hspace{.38ex}}P
$ (Eq.\ (\ref{20})). With Eqs.\ (\ref{29}), (\ref{30}) we obtain 
$ (P-1)\delta _{s_{0}=0}+(P-1)|\hat{g}|^{2}(s_{0},s_{a}\neq 0)=P
$ with the result given in Eq.\ (\ref{36}).

If $ d\ge 2$ we obtain for $ s_{a}\neq 0{\rm\hspace{.38ex}mod\hspace{.38ex}}P
$ from Eq.\ (\ref{7}) $ |\hat{g}|^{2}(s_{0},{\bf s})=P/(P-1)
\cdot \sum _{k\in {\ErgoBbb Z}_{P-1}}$ $ \delta _{s_{0}+s_{a}a^{
k}(a^{P-1}-1)/P{\rm\hspace{.38ex}mod\hspace{.38ex}}P}=P/(P-1)
\cdot \delta _{s_{0}\neq 0{\rm\hspace{.38ex}mod\hspace{.38ex}}P
}$. For general $s$$_{a}$ we use Eq.\ (\ref{28}) and obtain the 
result 
\begin{equation}
\label{36}
\begin{array}{c|cc}|\hat{g}|^{2}\left( s_{0},{\bf s}\right) &s_{
0}=0&s_{0}\neq 0\\\hline  
s_{a}=0&P^{d-1}\left( P-1\right) &0\\ 
\lambda =d-1&P^{d-1}/\left( P-1\right) &P^d\delta _{\mu =d-1}/
\left( P-1\right) \\ 
\lambda \le d-2&0&P^{\lambda +1}\delta _{\mu =\lambda }/\left( P
-1\right) \\ 
\multicolumn{3}{l}{\hbox{\hspace{.38ex}where }s_{a,P^d}\equiv {\rm\hspace
{.38ex}gcd}\left( s_{a},P^d\right) \equiv P^\lambda \hbox
{\hspace{.38ex}, }{\rm\hspace{.38ex}gcd}\left( s_{0},P^d
\right) \equiv P^{\mu }\hspace{.6ex}.}
\end{array}
\end{equation} 
We find that $ Q_{1}=Q_{1}(P^\lambda ,P^\lambda )=\sqrt{2} (P-1)
/P$ is independent of $a$ and even greater than 1.\ For $ n\ge 2
$ only the case $ (s_{0},s_{a})=(0,0)$ contributes to $Q$$_{n}$ 
and the discussion is equal to the case with power of two modulus 
presented in Sec.\ \ref{param}. We find 
\begin{equation}
\NX =\left( P-1\right) P^{d-1}\hspace{.6ex},\hspace{2ex}Q_{1}=
\sqrt{2} \left( P-1\right) /P\hspace{.6ex},\hspace{2ex}Q_{n\ge 
2}\approx P^{1/n-1}\hspace{.6ex}.
\end{equation} 
From a mathematical point of view odd prime number moduli give good 
random number generators. In particular $ \tilde{N}_{1}=P-1$, $ 
\tilde{M}_{1}=P$ whereas for a power of two modulus $M$ we had $ 
\tilde{N}_{1}= \tilde{M}_{1}=\sqrt{2} M/16$. So we need $ M>16P^d
/\sqrt{2}$ to obtain power of two multiplicative generators which 
behave better than generators with powers of odd prime numbers. 
However, one has to take into account that computers calculate automatically 
modulo powers of two. Moreover, the power of two generator will 
be improved in the next sections until we achieve $ Q_{1}=1$ (cf.\ 
Sec.\ \ref{int2}). As a byproduct a better behavior of $Q$$_{n}$ 
for $ n\ge 2$ is obtained, too.

Best performance allow prime numbers of the form $ P=2^{k}\pm 1$ 
\cite{Knu}. In this case $ a\cdot b=c_{1}2^{k}+c_{2}$ leads to $ a
\cdot b=c_{2}\mp c_{1}{\rm\hspace{.38ex}mod\hspace{.38ex}}P$. The 
extra effort, compared with a calculation $ {\rm\hspace
{.38ex}mod\hspace{.38ex}}2^{k}$ is one addition and, which is more 
important, the calculation of $c$$_{1}$. In Ex.\ 5.1 1 we discuss 
the improved generator with $ M=2^{256}$ which can most easily be 
changed to $ M=2^{128}$. Alternatively one may construct a generator 
$ {\rm\hspace{.38ex}mod\hspace{.38ex}}2^{127}-1$ which is a prime 
number. This generator will however be more time consuming and moreover 
it has worse quality $ \tilde{N}_{1}\approx 2^{127}$, $ \tilde
{N}_{2}\approx 2^{63.5}$, $ \tilde{N}_{3}\approx 2^{42.3}$, etc.\ 
vs.\ $ \tilde{N}_{1}=2^{129}$, $ \tilde{N}_{2}\approx 2^{86.3}$, 
$ \tilde{N}_{3}\approx 2^{65}$, etc.\ for the generator with power 
of two modulus (cf.\ Sec.\ \ref{int2}, Sec.\ \ref{param}). So, power 
of two generators are more efficient than multiplicative generators 
with prime number modulus. The situation is slightly different if 
one considers multiply recursive generators with prime number modulus, 
analyzed in Sec.\ \ref{multrec} and Ex.\ 5.1 2.

Multiplicative generators with prime number modulus and primitive 
$a$ produce every random number $ \neq 0{\rm\hspace{.38ex}mod\hspace{.38ex}}P
$ exactly once in a period. We recommend to use a prime number modulus 
only if one needs this quality. 
\item{}$ M=2^d$, $ c=1$, $ a=3{\rm\hspace{.38ex}mod\hspace{.38ex}}8
$ (the cases $ a=1$, $ a=5{\rm\hspace{.38ex}mod\hspace{.38ex}}8
$ were discussed in Ex.\ 1). This example and the following ones 
are less interesting from the random number point of view. They 
are discussed to demonstrate how Eq.\ (\ref{7}) applies in less 
trivial cases. We restrict ourselves to $ s_{a}\in {\ErgoBbb Z}_M^
\times $ and find 
\begin{eqnarray*}
&&\hspace{-.7cm} N_{m}=\left\{ 
\begin{array}{cl}2&\hbox{\hspace{.38ex}if }d\ge 3\\ 
1&\hbox{\hspace{.38ex}if }d=1,2
\end{array}\right. \hspace*{1ex},\hspace*{1ex}m=\left\{ 
\begin{array}{cl}8&\hbox{\hspace{.38ex}if }d\ge 3\\ 
2&\hbox{\hspace{.38ex}if }d=1,2
\end{array}\right. \hspace*{1ex},\hspace*{1ex}N=\left\{ 
\begin{array}{cl}2^{d-2}&\hbox{\hspace{.38ex}if }d\ge 3\\ 
2^{d-1}&\hbox{\hspace{.38ex}if }d=1,2
\end{array}\right. \hspace*{1ex},\\ 
&&\hspace{-.7cm} m_{0}=\left\{ 
\begin{array}{cl}8&\hbox{\hspace{.38ex}if }d\ge 6\\ 
2^{d-3}&\hbox{\hspace{.38ex}if }d=3,4,5\\ 
2&\hbox{\hspace{.38ex}if }d=2\\ 
1&\hbox{\hspace{.38ex}if }d=1
\end{array}\right. \hspace*{1ex},\hspace*{1ex}N_{m_{0}}=\left\{ 
\begin{array}{cl}2&\hbox{\hspace{.38ex}if }d\ge 5\\ 
1&\hbox{\hspace{.38ex}if }d\le 4
\end{array}\right. \hspace{.6ex}.
\end{eqnarray*} 
The most complicated case is $ d=4$ where we get $ |\hat{g}|^{2}
(s_{0},{\bf s})=\delta _{s_{0}+s_{a}(a^{2}-1)/8=0{\rm\hspace
{.38ex}mod\hspace{.38ex}}2}\cdot \left| \exp(\frac{2\pi i}{16}s_{
0})+\exp(\frac{2\pi i}{16}(4s_{0}+as_{a}))\right| ^{2}=2\delta _
{s_{0}=1{\rm\hspace{.38ex}mod\hspace{.38ex}}2}(1+\cos(\frac{
\pi }{8}(4s_{0}+(a-1)s_{a})))$. We obtain the following table 
\begin{equation}
\begin{array}{c|c}M=2^d&|\hat{g}|^{2}\left( s_{0},{\bf s}
\right) \\\hline  
d\ge 7&4\sum _{k=0,1}\delta _{s_{0}+s_{a}3^{k}\left( a^{2}-1
\right) /8=4{\rm\hspace{.38ex}mod\hspace{.38ex}}8}\\ 
d=6&4\sum _{k=0,1}\delta _{s_{0}+s_{a}3^{k}\left( a^{2}-1
\right) /8=0{\rm\hspace{.38ex}mod\hspace{.38ex}}8}\\ 
d=5&2\delta _{s_{0}=1{\rm\hspace{.38ex}mod\hspace{.38ex}}2}\\ 
d=4&
\begin{array}{c|cc}&s_{a}=1,3{\rm\hspace{.38ex}mod\hspace{.38ex}}8
&s_{a}=5,7{\rm\hspace{.38ex}mod\hspace{.38ex}}8\\\hline  
s_{0}=0,2{\rm\hspace{.38ex}mod\hspace{.38ex}}4&0&0\\ 
s_{0}=1{\rm\hspace{.38ex}mod\hspace{.38ex}}4&2\mp \sqrt{2}&2
\pm \sqrt{2}\\ 
s_{0}=3{\rm\hspace{.38ex}mod\hspace{.38ex}}4&2\pm \sqrt{2}&2\mp 
\sqrt{2}\\ 
\multicolumn{3}{l}{\hbox{\hspace{.38ex}upper sign: $ a=3{\rm\hspace
{.38ex}mod\hspace{.38ex}}16$, lower sign: }a=11{\rm\hspace
{.38ex}mod\hspace{.38ex}}16}
\end{array}\\ 
d=3&1\\ 
d=2&2\delta _{s_{0}=1{\rm\hspace{.38ex}mod\hspace{.38ex}}2}\\ 
d=0,1&1
\end{array}
\end{equation} 
and $ Q_{1}=\sqrt{2} /4$, $ Q_{n\ge 2}\approx M^{1/n-1}$ for $ d
\ge 6$ like in the case $ a=5{\rm\hspace{.38ex}mod\hspace{.38ex}}8
$. 
\item{}$ M=10^d$, $ c=1$, $ a=3$, $ {\rm\hspace{.38ex}gcd}(s_{a}
,10)=1$. We find 
\begin{equation}
\begin{array}{c|ccccc|c}M=10^d&N_{m}&m&N&m_{0}&N_{m_{0}}&|\hat
{g}|^{2}\left( s_{0},{\bf s}\right) \\\hline  
d\ge 9&4&80&5\cdot 10^{d-2}&80&4&20\sum _{k=0}^{3}\delta _{s_{0}
+3^{k}s_{a}=40{\rm\hspace{.38ex}mod\hspace{.38ex}}80}\\ 
5\le d\le 8&4&80&5\cdot 10^{d-2}&5\cdot 2^{d-4}&4&5\cdot 2^{d-6
}\sum _{k=0}^{3}\delta _{s_{0}+3^{k}s_{a}=0{\rm\hspace{.38ex}mod\hspace{.38ex}}5
\cdot 2^{d-4}}\\ 
2\le d\le 4&4&5\cdot 2^d&4\cdot 5^{d-1}&5&4&5/4\cdot \delta _{s_{
0}\neq 0{\rm\hspace{.38ex}mod\hspace{.38ex}}5}\\ 
d=1&4&10&4&1&1&1/4+\delta _{s_{0}\neq 0{\rm\hspace{.38ex}mod\hspace{.38ex}}4
}
\end{array}
\end{equation} 
and $ Q_{1}=\sqrt{2} /20$, $ Q_{n}\le Q_{n}(0,-3,1,0,{\ldots},0)
=2\cdot 10^{3/2-d}$, if $ d\ge 8$. 
\item{}$ M=2^{7}\cdot 7$, $ c=1$, $ a=3$, $ {\rm\hspace{.38ex}gcd}
(s_{a},14)=1$. This example illustrates all aspects of calculating 
$ |\hat{g}|^{2}$. We find $ N_{m}=6$, $ m=8\cdot 7$, $ N=2^{5}
\cdot 3$, $ \Mm =2^{4}$, $ \N =4$, $ m_{0}=8$, $ N_{m_{0}}=2$, $ 
(a^{N_{m}}-1)/m=-3{\rm\hspace{.38ex}mod\hspace{.38ex}}m_{0}$. Eq.\ 
(\ref{7}) gives $ |\hat{g}|^{2}(s_{0},{\bf s})=4/3\cdot \sum _{
k=0,1}\delta _{s_{0}-3^{k+1}s_{a}=4{\rm\hspace{.38ex}mod\hspace{.38ex}}8
}|\hat{g}|^{2}_{{\ErgoBbb Z}_{3}}[N\!=\!24,M\!=\!2^{7}\cdot 7,a
\!=\!3^{4}](s_{0},3^{k}s_{a})$.

Firstly, $ -3^{k+1}s_{a}=3^{k}(1-4)s_{a}=3^{k}s_{a}+4{\rm\hspace
{.38ex}mod\hspace{.38ex}}8$.

Secondly, in the $ |\hat{g}|^{2}_{{\ErgoBbb Z}_{3}}$ we sum over 
$ k'\in {\ErgoBbb Z}_{3}$. Since $ 8\in {\ErgoBbb Z}_{3}^
\times $ we can replace $ k'$ by $ 8k'$ yielding $ |\hat{g}|^{2
}[N\!=\!3,M\!=\!2^{7}\cdot 7,a\!=\!3^{32}](s_{0},3^{k}s_{a})$.

Finally, we notice that $ 3^{32}=1{\rm\hspace{.38ex}mod\hspace{.38ex}}2
^{7}$. With the Euclidean algorithm we find integers $ c_{1}$, $ c
_{2}$ with $ 1=2^{7}c_{1}+7c_{2}$. This plugged into the definition 
of $ \hat{g}$ gives $ \exp(\frac{2\pi i}{M} s_{a}3^{32k'}(2^{7}c
_{1}+7c_{2}))=\exp(\frac{2\pi i}{7}s_{a}3^{32k'}c_{1}+\frac{2
\pi i}{2^{7}{}} s_{a}c_{2})$. The second term is independent of 
$ k'$ and drops in $ |\hat{g}|^{2}$. Since $ 3^{32}=2{\rm\hspace
{.38ex}mod\hspace{.38ex}}7$ and $ c_{1}=2^{-7}=4{\rm\hspace
{.38ex}mod\hspace{.38ex}}7$ we obtain with Eq.\ (\ref{20}) 
\begin{displaymath}
|\hat{g}|^{2}\left( s_{0},{\bf s}\right) =\frac{4}{3}\sum _{k=0,
1}\delta _{s_{0}+3^{k}s_{a}=0{\rm\hspace{.38ex}mod\hspace{.38ex}}8
}|\hat{g}|^{2}[N\!=\!3,M\!=\!7,a\!=\!2]\left( s_{0},3^{k}s_{a}
\right) \hspace{.6ex}.
\end{displaymath} 
An explicit calculation (cf.\ Eq.\ (\ref{57})) is needed to obtain 
\begin{equation}
\begin{array}{lp{4cm}}
\begin{array}{c|cc}|\hat{g}|^{2}\left( s_{0},{\bf s}\right) &s_{
a}=1,9,11{\rm\hspace{.38ex}mod\hspace{.38ex}}14&s_{a}=3,5,13{\rm\hspace
{.38ex}mod\hspace{.38ex}}14\\\hline  
s_{0}=0{\rm\hspace{.38ex}mod\hspace{.38ex}}3&8/3&8/3\\ 
s_{0}=1{\rm\hspace{.38ex}mod\hspace{.38ex}}3&2/3\cdot \left( 7-
\sqrt{21}\right) &2/3\cdot \left( 7+\sqrt{21}\right) \\ 
s_{0}=2{\rm\hspace{.38ex}mod\hspace{.38ex}}3&2/3\cdot \left( 7+
\sqrt{21}\right) &2/3\cdot \left( 7-\sqrt{21}\right) 
\end{array}&\hbox{\hspace{.38ex}if $ \exists k\in 0,1$ with\hfill\hspace{.38ex}} 
\hbox{\hspace{.38ex}\hspace{ 1.4ex}$ s_{0}+3^{k}s_{a}=0{\rm\hspace
{.38ex}mod\hspace{.38ex}}8$\hspace{.38ex}}\\ 
\hspace{ 2.4ex}|\hat{g}|^{2}\left( s_{0},{\bf s}\right) 
\hspace*{1ex}=\hspace*{1ex}0&else .
\end{array}
\end{equation}
\end{enumerate}

\subsection{$ X_{0}=0$, $ X_{k+1}=aX_{k}+c{\rm\hspace{.38ex}mod\hspace{.38ex}}M
$} 
Let $ 1\neq a,c\in {\ErgoBbb Z}_M^\times $. The recursion formula 
implies $ X_{k}=c(a^{k}-1)/(a-1){\rm\hspace{.38ex}mod\hspace{.38ex}}M
$. Generators of this type are called mixed multiplicative.

There are two ways of deriving the mixed multiplicative generator 
from the multiplicative generator.

Firstly, we expect from a good generator to have random differences 
$ \Delta X_{k}\equiv X_{k+1}-X_{k}$. For the multiplicative generator 
we have $ \Delta X_{k}=(a-1)X_{k}{\rm\hspace{.38ex}mod\hspace{.38ex}}M
$ and in deed for the case $ M=P^d$, (Ex.\ 4.1 2) where the multiplicative 
generator was good, the difference generator has the same quality 
as the original one. However if $ M=2^d$ obviously $ a-1$ and $M$ 
have common divisors. If e.g.\ $ a=5{\rm\hspace{.38ex}mod\hspace{.38ex}}8
$ then $ (\Delta X_{k})_{k}$ produces only multiples of 4 which 
deteriorates the randomness of the sequence. This suggests to use 
the sum generator $ \Sigma X_{k}\equiv \sum _{j=0}^{k-1}X_{j}=c
(a^{k}-1)/(a-1)$ which is the mixed multiplicative generator.

Secondly, we observe that $ a-1$ divides $ X_{k}-c=c(a^{k}-1){\rm\hspace
{.38ex}mod\hspace{.38ex}}M$. Since $ X_{k}-c$ and $X$$_{k}$ differ 
only by a shift they have the same quality. The flaw that $ X_{
k}-c$ produces only multiples of $ {\rm\hspace{.38ex}gcd}(a-1,M
)$ can be corrected by a division by $ a-1$ which yields the mixed 
multiplicative generator.

\pagebreak[3]

\noindent {\bf Theorem} 4.2. Let $B$ be defined by 
\begin{equation}
\label{33}B\equiv {\rm\hspace{.38ex}max\hspace{.38ex}}_{k}{\rm\hspace
{.38ex}gcd}\left( a-1,M^{k}\right) 
\end{equation} 
then the Fourier transform of the mixed multiplicative generator 
$ X_{0}=0$, $ X_{k+1}=aX_{k}+c{\rm\hspace{.38ex}mod\hspace{.38ex}}M
$, $ 1\neq a,c\in {\ErgoBbb Z}_M^\times $ is given by Thm.\ 3.4 
and Cor.\ 3.5 with $ MB$ and $ XB$ instead of $M$ and $X$, respectively.

\pagebreak[3]

\noindent {\bf Proof}. From $ a,c,(a-1)/B\in {\ErgoBbb Z}_M^
\times $ we conclude $ a,c,(a-1)/B\in {\ErgoBbb Z}_{MB}^\times 
$. If we denote the inverse of $ (a-1)/B{\rm\hspace{.38ex}mod\hspace{.38ex}}MB
$ by $ \frac{B}{a-1}$ we have $ X_{k}/M=c\frac{B}{a-1} (a^{k}-1
)/MB{\rm\hspace{.38ex}mod\hspace{.38ex}}1$. Cor.\ 3.5 applies with 
$ XB$ and $ MB$ instead of $X$ and $M$.\hfill $\Box $
\pagebreak[3] 

\pagebreak[3]

\noindent {\bf Example} 4.3. 
\begin{enumerate}\item{}$ N_{m_{MB}}=1$ which is equivalent to 
\begin{equation}
\label{32}\hbox{\hspace{.38ex}1. }a-1\hbox{ contains every prime 
factor of $M$,\hspace*{1cm}2. }4|\left( a-1\right) \hbox{ if }4
|M.
\end{equation} 
(Figs.\ 4--8). With 
\begin{equation}
b={\rm\hspace{.38ex}gcd}\left( a-1,M\right) ={\rm\hspace
{.38ex}gcd}\left( B,M\right) 
\end{equation} 
we find $ m_{MB}={\rm\hspace{.38ex}gcd}(a-1,MB)=B$, $ {m_{0}}_{MB
}=b$, $ {M_{0}}_{MB}=M/b$, $ \NX \stackrel{{\rm (\ref{8})\hspace{.38ex}}
}{=} MB/m_{MB}=M$ and $ {\bf s}\cdot ({\bf X}_{1}-{\bf X}_{0})=s
_{a}c$. We obtain from Eqs.\ (\ref{34}), (\ref{35}) 
\begin{equation}
\label{46}|\hat{g}|^{2}\left( s_{0},{\bf s}\right) =bs_{a,M/b}
\delta _{s_{0}+cs_{a}=\frac{1}{2}{bs_{a,M/b}}\delta _{2|\frac{M
}{bs_{a,M/b}{}} }{\rm\hspace{.38ex}mod\hspace{.38ex}}bs_{a,M/b}
}\hspace{.6ex}.
\end{equation} 
In particular for $ s_{0}=0$ we observe that $ \hat{g}$ is only non-zero 
if $ M/bs_{a,M/b}$ is odd and $ s_{a}=0{\rm\hspace{.38ex}mod\hspace{.38ex}}bs_
{a,M/b}$. From the second identity we conclude $ s_{a}=0{\rm\hspace
{.38ex}mod\hspace{.38ex}}M$ and thus 
\begin{equation}
|\hat{g}|^{2}\left( 0,{\bf s}\right) =M\delta _{s_{a}=0{\rm\hspace
{.38ex}mod\hspace{.38ex}}M}
\end{equation} 
For $ n=1$ we get $ s_{a}=s_{1}$ and $ \hat{g}$ is the Fourier transform 
of the uniform distribution. So, $X$$_{k}$ has the property to produce 
every number in $ {\ErgoBbb Z}_M$ exactly once in a period. The 
generator is called to have a full period.

We proceed with a discussion of the parameters $a$ and $c$. 
\begin{enumerate}\item{}Choice of $c$. We can restrict ourselves 
to $ 1\le c\le b/2$ since every $c$ emerges from a $ c\in (0,b/2
)$ via translations ($ X_{k}\mapsto X_{k+\Delta k}-X_{\Delta k}
=a^{\Delta k}X_{k}$) or reflection ($ X_{k}\mapsto -X_{k}$). We 
can determine $c$ by the condition that there should be no small 
$ (s_{0},s_{a})$, $ {\rm\hspace{.38ex}gcd}(M,s_{a})=1$ with $ s_{
0}+s_{a}c=b/2\cdot \delta _{2|M/b}{\rm\hspace{.38ex}mod\hspace{.38ex}}b
$. If $ M/b$ is odd $ c\approx \sqrt{b}$ gives $ Q_{1}(s_{0}
\approx \sqrt{b} ,-1)\approx \sqrt{b+1} /b\approx b^{-1/2}$. In 
the case where $ M/b$ is even and $b$ is small the choice $ c=1
$ is best with the result $ Q_{1}(s_{0}\approx s_{1}\approx b/4
)\approx \sqrt{2} \cdot (b/4)/b=\sqrt{2} /4\approx 0.35$.

In the case of Fig.\ 5 with the 'wrong' choice $ c=3$ one has $ a
=9{\rm\hspace{.38ex}mod\hspace{.38ex}}16$, $ b=8$ and the smallest 
$ (s_{0},s_{1})$ with non-vanishing $ \hat{g}$ is (1,1). Since $ 
|\hat{g}|^{2}(1,1)=8$ we obtain $ Q_{1}(1,1)=\sqrt{2} /8
\approx 0.18$ (notice the correlations perpendicular to the (1,1)-direction 
in Fig.\ 5I). With the right choice $ c=1$ (Fig.\ 6) it takes an 
$ (s_{0},s_{1})=(1,3)$ (or (3,1)) to get $ |\hat{g}|^{2}=8$. Therefore 
$ Q_{1}(1,3)=\sqrt{10} /8\approx 0.40$ which means that the random 
numbers are more uniformly distributed in Fig.\ 6I. The large value 
of $ Q_{1}(1,3)$ is yet misleading since $ Q_{1}=Q_{1}(-M/8,M/8
)=\sqrt{2} /8$. However $ |\hat{g}|^{2}$ assumes the small value 
of $Q$$_{1}$ at much less sites as in the case of $ c=3$ which means 
that the choice $ c=1$ is better than $ c=3$.

Notice the similar pair distributions in Fig.\ 5II and Fig.\ 6II. 
In general the quality dependence on $c$ can not be obtained by 
the standard spectral test (corresponding to $ s_{0}=0$) since the 
$n$-tuple distributions are only shifted by changing $c$. 
\item{}Choice of $b$. In general $b$ should be as small as possible 
in order to prevent $ |\hat{g}|^{2}$ from being concentrated on 
too few points. If $ M/b$ is odd, $ c\approx \sqrt{b}$ then $Q$$
_{1}$ behaves like $ b^{-1/2}$. If $ M/b$ is even, $ c=1$ then $ Q
_{1}(s_{0},s_{1})=Q_{1}(s_{0}\approx s_{1}\approx b/4)\approx 
\sqrt{2} /4$ for small $ (s_{0},s_{1})$ (cf.\ (a)). However $ Q_{
1}(-M/b,M/b)=\sqrt{2} /b$ which forbids large values of $b$.

For a power of two modulus the smallest value possible is $ b=4$ 
which implies $ Q_{1}=\sqrt{2} /4$ (Fig.\ 4). In particular $ M
=10^d$ (Fig.\ 7) should be avoided since in this case $ b\ge 20
$.

These arguments require $ s_{0}\neq 0$. They are not obtained by 
the standard spectral test. 
\item{}Choice of $a$. Up to now we have evaluated $ |\hat{g}|^{2
}(s_{0},s_{1})$ for $ n=1$ which is given by $b$ and $c$. In order 
to determine $a$ more precisely we have to look at the distribution 
of $n$-tupels for $ n\ge 2$. In this case $ Q_{n}={\rm\hspace
{.38ex}min\hspace{.38ex}}_{\bf s}Q_{n}(s_{0}=0,s_{a}=0)={\rm\hspace
{.38ex}min\hspace{.38ex}}_{{\bf s}:s_{a}=0}|{\bf s}|/M$. A further 
discussion of the choice of $a$ is postponed to Sec.\ \ref{param}. 
We will see that a reasonable $a$ gives $ Q_{n}\approx M^{1/n-1
}$.

Since $ s_{0}=0$ this part of the choice of $a$ is identical with 
the standard spectral test.
\end{enumerate} 
Let us summarize the result for $ M=2^d$, 
\begin{eqnarray}
&&c=1\hspace{.6ex},\hspace{2ex}a=5{\rm\hspace{.38ex}mod\hspace{.38ex}}8
\hspace{.6ex},\hspace{2ex}{\rm\hspace{.38ex}max\hspace{.38ex}}_a
{\rm min\hspace{.38ex}}_{{\bf s}:s_{a}=0}|{\bf s}|,\hbox{ for }n
=2,3,{\ldots}\hbox{ gives\hspace{.38ex}}\\ 
&&\NX =M\hspace{.6ex},\hspace{2ex}Q_{1}=\sqrt{2} /4
\hspace{.6ex},\hspace{2ex}Q_{n}\approx M^{1/n-1}\hspace{.6ex}.
\end{eqnarray} 
A loss of randomness for $n$-tupels is avoided if one takes $ {\rm\hspace
{.38ex}gcd}(n,M)=1$. 
\item{}$ a=3{\rm\hspace{.38ex}mod\hspace{.38ex}}8$, $ M=2^d$. If 
$ s_{a,M}\le M/32$ then $ B=2$, $ m'_{MB}={m_{0}}_{MB}'=8$, $ {M
_{0}}'_{MB}=M/32s_{a,M}$, $ N_{m'_{MB}}=N_{{m_{0}}_{MB}'}=2$ and 
from Eq.\ (\ref{34}) we get 
\begin{equation}
|\hat{g}|^{2}\left( s_{0},{\bf s}\right) =\frac{8\NX s_{a,M}}{M} 
\sum _{k=0,1}\delta _{\frac{M}{2\NX} s_{0}+3^{k}c\frac{a+1}{4}s_{
a}=4s_{a,M}\delta _{64|\frac{M}{s_{a,M}{}} }{\rm\hspace
{.38ex}mod\hspace{.38ex}}8s_{a,M}}\hspace{.6ex}.
\end{equation} 
We obtain $ \NX =M/2$ (cf.\ Eq.\ (\ref{42})) and either $ Q_{1}(1
,1)=\sqrt{2} /4$ or $ Q_{1}(-1,1)=\sqrt{2} /4$. The period is doubled 
compared with the multiplicative case and the quality remains unchanged. 
Since $ a=3{\rm\hspace{.38ex}mod\hspace{.38ex}}8$ has only half 
the period of $ a=5{\rm\hspace{.38ex}mod\hspace{.38ex}}8$ the latter 
should be preferred.
\end{enumerate}

The only improvement we achieved was an increased period by a factor 
of $B$. The mixed multiplicative generator $ {\rm\hspace
{.38ex}mod\hspace{.38ex}}M$ equals the multiplicative generator 
$ {\rm\hspace{.38ex}mod\hspace{.38ex}}MB$. For a power of two modulus 
we still have $ Q_{1}=\sqrt{2} /4<1$.

This does not mean that one can not use the mixed multiplicative 
generator if the modulus is large enough. In Fig.\ 8 we see that 
all problems dissolve if the modulus is much larger than the range 
of random numbers used. Nevertheless it will be profitable to look 
for improvements in the next sections.

\vspace{1ex}
\noindent{}We close this section with a well known theorem which 
is a corollary of our general theory.

\pagebreak[3]

\noindent {\bf Corollary} 4.4 (Greenberger \cite{Greenberger}, 
Hull and Dobell \cite{Hull}). 
The mixed multiplicative generator has a full period if and only 
if (\ref{32}) holds.

\pagebreak[3]

\noindent {\bf Proof}. We saw in Ex.\ 4.3 1 that (\ref{32}) implies 
a full period. If on the other hand $ N_{m_{MB}}>1$, due to Eq.\ 
(\ref{27}) there either exists a prime factor $ P\neq 2$ of $M$ 
with $ N_P>1$ or $ 4|M$ and $ N_{4}>1$. In the first case we look 
at the generator $ {\rm\hspace{.38ex}mod\hspace{.38ex}}P$ and find 
$ B=1$. Hence $ N_{X{\rm\hspace{.38ex}mod\hspace{.38ex}}P}=N_P<P
$ and the generator has no full period. In the second case $ a=3
{\rm\hspace{.38ex}mod\hspace{.38ex}}4$ and $\forall $$k$: $ X_{
k}=0{\rm\hspace{.38ex}mod\hspace{.38ex}}4$ or $ X_{k}=c{\rm\hspace
{.38ex}mod\hspace{.38ex}}4$.\hfill $\Box $
\pagebreak[3] 

\subsection{$ X_{0}=0$, $ X_{k+1}=aX_{k}+ck{\rm\hspace{.38ex}mod\hspace{.38ex}}M
$\label{ck}} 
Let $ 1\neq a,c\in {\ErgoBbb Z}_M^\times $. The recursion formula 
gives $ X_{k}=\frac{c}{a-1} (\frac{a^{k}-1}{a-1} -k){\rm\hspace
{.38ex}mod\hspace{.38ex}}M$.

The generator of this section improves the mixed multiplicative generator 
according to both ways presented in the last section. Firstly, it 
is the sum generator $ \sum _{j=0}^{k-1}X_{j}^{\rm mm}=\sum _{j
=0}^{k-1}c(a^{k}-1)/(a-1)$ of the mixed multiplicative generator. 
Secondly, we observe that $ X_{k}^{\rm mm}-ck=c\sum _{j=0}^{k-1
}(a^{j}-1){\rm\hspace{.38ex}mod\hspace{.38ex}}M$ can be divided 
by $ a-1$. On the other hand the randomness of $ X_{k}$ should not 
differ much from that of $ X_{k}-ck$ and the division by $ a-1$ 
improves the generator.

\pagebreak[3]

\noindent {\bf Theorem} 4.5. Let $B$ be defined by Eq.\ (\ref{33}) 
then the Fourier transform of the generator $ X_{0}=0$, $ X_{k+
1}=aX_{k}+ck{\rm\hspace{.38ex}mod\hspace{.38ex}}M$, $ 1\neq a,c
\in {\ErgoBbb Z}_M^\times $ is given by Cor.\ 3.5 with 
\begin{equation}
\label{52}X\mapsto XB^{2}\hspace{.6ex},\hspace{2ex}M\mapsto MB^{
2}\hspace{.6ex}.\hspace*{1cm}\NX ={\rm\hspace{.38ex}lcm}\left( N
_{m},2^{\delta _{2|M}}M\right) 
\end{equation} 
is the period of $X$.

\pagebreak[3]

\noindent {\bf Proof}. Analogously to the proof of Thm.\ 4.2 we 
find that $ X_{k}/M=c((\frac{B}{a-1} )^{2}{} (a^{k}-1)-\frac{B}
{a-1} Bk)/MB^{2}{\rm\hspace{.38ex}mod\hspace{.38ex}}1$ where $ 
\frac{B}{a-1}$ is the inverse of $ (a-1)/B$ in $ {\ErgoBbb Z}_{MB
^{2}}$. With Cor.\ 3.5 it only remains to show that the period of 
$X$ is given by Eq.\ (\ref{52}).

Since $ 0=X_{1}=X_{\NX +1}=aX_\NX +c\NX =c\NX {\rm\hspace
{.38ex}mod\hspace{.38ex}}M$ we know that $ M|\NX$ and since $ a^
\NX -1=0{\rm\hspace{.38ex}mod\hspace{.38ex}}M$ implies $ N_{m}|
\NX$ we have $ {\rm\hspace{.38ex}lcm}(N_{m},M)|\NX$. If we split 
$M$ into its prime components $ M=2^{k}p_{1}^{k_{1}}\cdots p_{
\ell }^{k_{\ell }}$, it suffices to show that $ X_\NX =0{\rm\hspace
{.38ex}mod\hspace{.38ex}}p_{j}^{k_{j}}\hspace*{1ex}\forall j=1,
{\ldots},\ell $ and $ X_{2^{k}}=2^{k-1}{\rm\hspace{.38ex}mod\hspace{.38ex}}2
^{k}$. If $ N_{p_{j}}>1$ then $ a-1\in {\ErgoBbb Z}_{p_{j}}^
\times $ and one obtains immediately $ X_\NX =\frac{c}{a-1} (
\frac{a^\NX -1}{a-1} -\NX )=0{\rm\hspace{.38ex}mod\hspace{.38ex}}p
_{j}^{k_{j}}$. The case $ N_{p_{j}}=1$ is treated in Ex.\ 4.6 1.\ 
For $ M=2$ we get $ X_{2}=1$ and for $ M=2^{k}$ the result follows 
by induction since 
\begin{displaymath}
X_{2M}=\frac{c}{a-1} \left( \frac{a^{2M}-1}{a-1} -2M\right) =
\frac{2c}{a-1} \left( \frac{a^M-1}{a-1} -M\right) +c\left( 
\frac{a^M-1}{a-1}\right) ^{2}=2X_M{\rm mod\hspace{.38ex}}2M
\end{displaymath} 
for $ M\ge 2$.\hfill $\Box $
\pagebreak[3] 

\pagebreak[3]

\noindent {\bf Examples} 4.6.
\begin{enumerate}\item{}$ N_{m_{MB^{2}}}=1$ which means that $ a
-1$ contains every prime factor of $M$ and $ 4|(a-1)$ if $M$ is 
even (Fig.\ 9). We find $ m_{MB^{2}}={m_{0}}_{MB^{2}}=B$, $ {M_{
0}}_{MB^{2}}=M$ and for odd $M$ we obtain 
\begin{equation}
\label{38}|\hat{g}|^{2}\left( s_{0},{\bf s}\right) =\frac{\NX}{M
} s_{a,M}\delta _{\frac{M}{\NX} s_{0}+{\bf s}\cdot {\bf X}_{0}^{\rm mm}
=0{\rm\hspace{.38ex}mod\hspace{.38ex}}s_{a,M}}
\hspace{.6ex},\hspace{2ex}X_{k}^{\rm mm}=c\frac{a^{k}-1}{a-1} 
\hspace{.6ex}.
\end{equation} 
With Eq.\ (\ref{42}) we can conclude that $ \NX =M$ since any $ 
\NX =\lambda M$ leads to $ |\hat{g}|^{2}(s_{0},{\bf s})=0$ unless 
$ \lambda |s_{0}$. This completes the proof of Theorem 4.5.

If $M$ is even we have $ \NX =2M$ yielding 
\begin{equation}
\label{39}|\hat{g}|^{2}\left( s_{0},{\bf s}\right) =2s_{a,M}
\delta _{s_{0}+2{\bf s}\cdot {\bf X}_{0}^{\rm mm}=s_{a,M}
\delta _{2|\frac{M}{s_{a,M}{}} }{\rm\hspace{.38ex}mod\hspace{.38ex}}2s_
{a,M}}\hspace{.6ex}.
\end{equation} 
For $ n=1$ we have $ s_{a,M}=s_{1,M}={\rm\hspace{.38ex}gcd}(s_{1
},M)$ and $ {\bf s}\cdot {\bf X}_{0}^{\rm mm}=s_{1}X_{0}^{\rm mm}
=0$. Thus for odd $M$ we have $ Q_{1}=Q_{1}(0,s_{1,M})=1$. If $M$ 
is even then $ Q_{1}(1,1)=\sqrt{2} /2$ and if $ \exists k\in 
{\ErgoBbb N}$ so that $ 1\neq M/2^{k}$ is odd then $ Q_{1}(0,2^{
k})=1/2$. Therefore 
\begin{equation}
\NX =2^{\delta _{2|M}}M,\hspace*{1ex}Q_{1}=\left\{ 
\begin{array}{cl}1&\hbox{\hspace{.38ex}if $M$ is odd\hspace{.38ex}}
\\ 
\sqrt{2} /2&\hbox{\hspace{.38ex}if }M=2^d\\ 
1/2&\hbox{\hspace{.38ex}else\hspace{.38ex}}
\end{array}\right. ,\hspace*{1ex}Q_{2}\approx M^{-1/3},
\hspace*{1ex}Q_{n\ge 3}\approx M^{1/\left( n-1\right) -1}
\hspace{.6ex}.
\end{equation} 
In case of a power of two modulus we have increased $Q$$_{1}$ from 
$ \sqrt{2} /4$ for the mixed multiplicative generator to $ 
\sqrt{2} /2\approx 0.70$ and $ \NX$ from $M$ to $ 2M$. Notice also 
the better behavior of $Q$$_{n}$ for $ n\ge 2$ which we will derive 
in Sec.\ \ref{param}.

For $ s_{0}=0$ we obtain $ |\hat{g}|^{2}(0,s_{1})=2^{\delta _{2|M
}}s_{1,M}\delta _{M/s_{1,M}{\rm  \ odd\hspace{.38ex}}}$. Only for 
$ M=2^d$ this is $ \NX\delta _{s_{1}=0}$, the Fourier transform 
of the uniform distribution. In this case one obtains two full periods, 
every random number occurs twice in a period. Analogously to the 
proof of Cor.\ 4.4 one can show that for $ N_{m_{MB^{2}}}>1$ a full 
period is not possible.

Thus the generator $ X_{k+1}=aX_{k}+ck{\rm\hspace{.38ex}mod\hspace{.38ex}}M
$ produces exactly uniformly distributed random numbers if and only 
if $M$ is a power of two and $ a=1{\rm\hspace{.38ex}mod\hspace{.38ex}}4
$.

The distribution of $n$-tupels is no longer a lattice as can be seen 
from Fig.\ 9II. So even for the standard spectral test $ (s_{0}
=0)$ one would need Eq.\ (\ref{34}) to analyze the generator. We 
skip a further discussion here since in the next section we will 
present an improved generator for a power of two modulus. 
\item{}$ M=2^d$, $ a=3{\rm\hspace{.38ex}mod\hspace{.38ex}}8$. Let 
$ s_{a,M}\le M/16$ then $ B=2$, $ m'_{MB^{2}}={m_{0}'}_{MB^{2}}
=8$, $ {M_{0}'}_{MB^{2}}=M/16s_{a,M}$, $ N_{m'_{MB^{2}}}=N_{{m_{
0}'}_{MB^{2}}}=2$ and we obtain 
\begin{equation}
|\hat{g}|^{2}\left( s_{0},{\bf s}\right) =8s_{a,M}\sum _{k=0,1}
\delta _{s_{0}+{\bf s}\cdot \left( {\bf X}_{k+1}^{\rm mm}+
{\bf X}_{k}^{\rm mm}\right)  =8s_{a,M}\delta _{32|\frac{M}{s_{a
,M}{}} }{\rm\hspace{.38ex}mod16\hspace{.38ex}}s_{a,M}}
\hspace{.6ex}.
\end{equation} 
For e.g.\ $ c=1$ we obtain $ Q_{1}(3,1)=\sqrt{10} /8\approx 0.40
$ or $ Q_{1}(1,3)=\sqrt{10} /8$. The improvement in the quality 
is very little in comparison with the mixed multiplicative generator. 
The period increases from $ M/2$ to $ 2M$. The generator with $ a
=5{\rm\hspace{.38ex}mod\hspace{.38ex}}8$ behaves clearly better. 
\item{}$ M=P^d$, $P$ is an odd prime number and $a$ is primitive, 
$ \langle a\rangle _{P^d}={\ErgoBbb Z}_{P^d}^\times $ (Fig.\ 10). 
We find $ B=1$, $ \NX =(P-1)P^d$. In the following $ \frac{c}{a
-1} \in {\ErgoBbb Z}_{P^d}^\times $ is $c$ times the inverse of 
$ a-1{\rm\hspace{.38ex}mod\hspace{.38ex}}P^d$. We have to distinguish 
three cases:

1.\ $ s_{a,P^d}=0{\rm\hspace{.38ex}mod\hspace{.38ex}}P^d$, $ m'=m
_{0}'=M_{0}'=1$, $ N_{m'}=N_{m_{0}'}=1$. The condition in Eq.\ (\ref{34}) 
is $ s_{0}+(P-1){\bf s}\cdot ({\bf X}_{1}-{\bf X}_{0})=0{\rm\hspace
{.38ex}mod}(P-1)P^d\Leftrightarrow s_{0}=0{\rm\hspace{.38ex}mod\hspace{.38ex}}P
-1$ and $ s_{0}+(P-1)c(s_{a}-\sum _{j=1}^{n}s_{j})/(a-1)=s_{0}-
(P-1)\frac{c}{a-1} \sum _{j=1}^{n}s_{j}=0{\rm\hspace{.38ex}mod\hspace{.38ex}}P^d
$. With the pre-factor $ (P-1)P^d$ we obtain the first line in Eq.\ 
(\ref{37}).

2.\ $ s_{a,P^d}=P^{d-1}$, $ m'=P$, $ m_{0}'=1=M_{0}'$, $ N_{m'}=P
-1$, $ N_{m_{0}'}=1$. We obtain from Eq.\ (\ref{34}) $ |\hat{g}
|^{2}(s_{0},{\bf s})=P^d|\hat{g}|^{2}_{{\ErgoBbb Z}_{P-1}}[(P-1
)P^d,X_\bullet ](s_{0},{\bf s})\delta _{s_{0}+{\bf s}\cdot (
{\bf X}_{P-1}-{\bf X}_{0})=0{\rm\hspace{.38ex}mod\hspace{.38ex}}P^d
}$. Since $ P^d\in {\ErgoBbb Z}_{P-1}^\times $ we get $ P^d|
\hat{g}|^{2}[P-1,X_{P^d\bullet }](s_{0},{\bf s})\delta _{s_{0}+
\frac{c}{a-1} (s_{a}(a^{P-1}-1)/(a-1)-(P-1)\sum _{j=1}^{n}s_{j}
)=0{\rm\hspace{.38ex}mod\hspace{.38ex}}P^d}=P^d|\hat{g}|^{2}[N=P
-1,M=P,X_{k}=\frac{c}{(a-1)^{2}{}} a^{P^dk}](s_{0},{\bf s}/P^{d
-1})\delta _{s_{0}=(P-1)\frac{c}{a-1} \sum _{j=1}^{n}s_{j}{\rm\hspace
{.38ex}mod\hspace{.38ex}}P^d}$. Since $ a^{P^dk}=a^k{\rm mod\hspace{.38ex}}P
$ we can use the result of the multiplicative generator, Eq.\ (\ref{36}), 
$ d=1$, yielding the second line in Eq.\ (\ref{37}).

3.\ $ s_{a,P^d}\equiv P^\lambda \le P^{d-2}$, $ m'=m_{0}'=P$, $ M
_{0}'=P^{d-\lambda -2}$, $ N_{m'}=N_{m_{0}'}=P-1$. We obtain $ 
|\hat{g}|^{2}(s_{0},{\bf s})=P^{\lambda +2}/(P-1)\cdot \sum _{k
\in {\ErgoBbb Z}_{P-1}}\delta _{s_{0}+{\bf s}\cdot ({\bf X}_{P-1
+k}-{\bf X}_{k})=0{\rm\hspace{.38ex}mod\hspace{.38ex}}P^{
\lambda +2}}=P^{\lambda +2}/(P-1)\cdot \sum _{k\in {\ErgoBbb Z}_
{P-1}}\delta _{s_{0}+\frac{c}{a-1} (s_{a}a^{k}(a^{P-1}-1)/(a-1)
-(P-1)\sum _{j=1}^{n}s_{j})=0{\rm\hspace{.38ex}mod\hspace{.38ex}}P^
{\lambda +2}}=P^{\lambda +2}/(P-1)\cdot \sum _{k'\in 
{\ErgoBbb Z}_P^\times }$ $ \delta _{s_{0}-\frac{c}{a-1} (P-1)
\sum _{j=1}^{n}s_{j}=k's_{a}P{\rm\hspace{.38ex}mod\hspace{.38ex}}P^
{\lambda +2}}$. Altogether, 
\begin{equation}
\label{37}
\begin{array}{c|ccc}|\hat{g}|^{2}\left( s_{0},{\bf s}\right) &
\mu =d\hbox{\hspace{.38ex}, }s_{0}=0{\rm\hspace{.38ex}mod\hspace{.38ex}}P
\!-\!1&\mu =d\hbox{\hspace{.38ex}, }s_{0}\neq 0{\rm\hspace
{.38ex}mod\hspace{.38ex}}P\!-\!1&\mu <d\\\hline  
s_{a}=0&\left( P\!-\!1\right) P^d&0&0\\ 
\lambda =d\!-\!1&P^d/\!\left( P\!-\!1\right) &P^{d+1}\!/\left( P
\!-\!1\right) &0\\ 
\lambda \le d\!-\!2&0&0&\hspace{-2ex} P^{\lambda +2}\delta _{
\mu =\lambda +1}/\!\left( P\!-\!1\right) \\ 
\multicolumn{4}{l}{\hbox{\hspace{.38ex}with }s_{a,P^d}\equiv {\rm\hspace
{.38ex}gcd}\left( s_{a},P^d\right) \equiv P^\lambda \hbox
{\hspace{.38ex}, }{\rm\hspace{.38ex}gcd}\left( \left( a-1
\right) s_{0}-c\left( P-1\right) \sum _{j=1}^{n}s_{j},P^d
\right) \equiv P^{\mu }.}
\end{array}
\end{equation} 
We find that a proper choice of $a$ and $c$ leads to $ Q_{1}
\approx \sqrt{P} (P-1)/P^{2}\approx P^{-1/2}$. The by a factor of 
$P$ longer period compared with the multiplicative case (Ex.\ 4.1 
2) is to some extend compensated by a worse relative quality $Q$
$_{1}$.

If we set $ s_{0}=0$ and $ n=1$ we find $ \lambda =\mu $ and therefore 
$ |\hat{g}|^{2}=\NX \delta _{s_{1}=0}$. The random numbers are exactly 
uniformly distributed, the generator produces $ P-1$ full periods.

Although $ \NX Q_{n}$ is larger than for the multiplicative generator 
we can not recommend the generator. If one is not interested in 
having full periods, generators with a power of two modulus are 
more efficient. If one needs a full period the velocity of the generator 
is not important since $M$ has to be comparatively small. Then only 
the relative quality $Q$$_{n}$ is essential and the multiplicative 
generator is better. 
\item{}$ X_{k}=c(k^{2}-k)/2$. This quadratic polynomial is obtained 
with $ a=M+1$. Since, of course, $ N_{m_{MB^{2}}}=1$ Eqs.\ (\ref{38}) 
and (\ref{39}) of Ex.\ 1 for odd and even $M$ are valid, respectively. 
Although this generator has reasonable $Q$$_{1}$ it is too simple 
to have good pair and $n$-tupel distributions (cf.\ Sec.\ \ref{param}).
\end{enumerate} 
We close this section with a lemma which will play an essential role 
in Sec.\ \ref{int2}.

\pagebreak[3]

\noindent {\bf Lemma} 4.7. Let $ A\in {\ErgoBbb N}$, $ M'=MB^{
2}{} /s_{a,MB^{2}}$, $ m'=m_{a^A,M'}$, $ m_{0}'={m_{0}}_{a^A,M'}
$, $ M_{0}'=M'/m'm_{0}'$. Moreover $ \tilde{M}'= \tilde{M}B^{2}
{} /s_{a,\tilde{M}B^{2}}$, $ \tilde{m}'=m_{a^A,\tilde{M}'}$, $ 
\tilde{m}_{0}'={m_{0}}_{a^A,\tilde{M}'}$, $ \tilde{M}_{0}'= 
\tilde{M}'/\tilde{m}'\tilde{m}_{0}'$. If one of the following conditions 
holds there exists at most one $ C\in {\ErgoBbb Z}_A$ with $ 
\hat{g}[N,M,{\bf X}_\bullet =(X_{A\bullet +C+j-1})_{j=1{\ldots}n
}]\neq 0$. 
\begin{eqnarray}
\label{41}1.&&\exists M_A\in {\ErgoBbb N}\hbox{ with }N_{M_A}=A\hbox{ 
and }M_A|\left( M'/m'\right) \hspace{.6ex}.\\ 
2.&&\exists \tilde{M}|M\hbox{ with }a=1{\rm\hspace{.38ex}mod\hspace{.38ex}}
\tilde{m}_{0}'\hbox{ and\hspace{.38ex}}\nonumber \\ 
\label{40}&&\frac{MN_{a^A,m'}}{\NX} As_{0}+{\bf s}\cdot \left( 
{\bf X}_{AN_{a^A,m'}}-{\bf X}_{0}\right) =\frac{M}{2M_{0}'} 
\delta _{2|M_{0}'}{\rm\hspace{.38ex}mod\hspace{.38ex}}\frac{
\tilde{M}}{\tilde{M}_{0}'} \hspace{.6ex}.
\end{eqnarray}

\pagebreak[3]

\noindent {\bf Proof}. Since $ M_A|(M'/m')$ we get $ M_A|M'$. Thus 
$ M_A|{\rm\hspace{.38ex}gcd}(a^A-1,M')|{\rm\hspace{.38ex}gcd}(a^
{AN_{m'}}-1,M')=m'$, hence $ M_A|m_{0}'$. Because of the condition 
in Eq.\ (\ref{7}) and Eq.\ (\ref{28}) with $ XB^{2}$ and $ MB^{
2}$ instead of $X$ and $M$ we obtain as a necessary condition for 
$ \hat{g}[{\bf X}_\bullet =(X_{A\bullet +C_{1}+j-1})_{j}]\neq 0
\neq \hat{g}[{\bf X}_\bullet =(X_{A\bullet +C_{2}+j-1})_{j}]$: 
\begin{eqnarray*}
&&\hspace{-.6cm}\exists k_{1},k_{2}\hbox{\hspace{.38ex}, with }
\frac{M'}{MB^{2}m'} \sum _{i=1,2}\left( -1\right) ^{i}\sum _{j=
1}^{n}s_{j}B^{2}\left( X_{A\left( N_{m'}+k_{i}\right) +C_{i}+j-1
}-X_{Ak_{i}+C_{i}+j-1}\right) =0{\rm\hspace{.38ex}mod\hspace{.38ex}}m
_{0}'\\ 
&&\hspace{-.6cm}\Leftrightarrow \hspace*{1ex}\frac{s_{a}}{s_{a,MB
^{2}}{}} \frac{cB^{2}}{\left( a-1\right) ^{2}{}} \frac{a^{AN_{m'}
}-1}{m'} \left( a^{Ak_{1}+C_{1}}-a^{Ak_{2}+C_{2}}\right) =0{\rm\hspace
{.38ex}mod\hspace{.38ex}}m_{0}'\Rightarrow a^{C_{1}}=a^{C_{2}}{\rm\hspace
{.38ex}mod\hspace{.38ex}}M_A
\end{eqnarray*} 
and therefore $ C_{1}=C_{2}$.

In case of Eq.\ (\ref{40}) we may relax the condition in Eq.\ (\ref{34}) 
and replace (after multiplication with $M$) $ M/M_{0}'$ by $ 
\tilde{M}/\tilde{M}_{0}'$. Since $ a=1{\rm\hspace{.38ex}mod\hspace{.38ex}}
\tilde{m}_{0}'$ we see from the above calculation that in this case 
the condition is independent of $C$. Therefore all the $ \hat
{g}[{\bf X}_\bullet =(X_{A\bullet +C+j-1})_{j}]$ are zero if $ 
\hat{g}[{\bf X}_\bullet =(X_{A\bullet +j-1})_{j}]$ is zero. The 
dependence on $k$$_{1}$, $k$$_{2}$ drops, too, and we end up with 
the condition in Eq.\ (\ref{40}).\hfill $\Box $
\pagebreak[3] 

\vspace{1ex}
\noindent{}If $ \hat{g}$ is decomposed $ \sqrt{\NX} \hat{g}[\NX 
]=\sum _{C\in {\ErgoBbb Z}_A}\exp(2\pi iCs_{0}/\NX )\sqrt{\NX /A
}\hat{g}[\NX /A,(X_{A\bullet +C+j-1})_{j}]$ only one of the terms 
on the right hand side can be non-zero. Therefore $ \NX |\hat
{g}|^{2}[\NX ]=\sum _{C\in {\ErgoBbb Z}_A}{} \NX /A |\hat{g}|^{
2}[\NX /A,(X_{A\bullet +C+j-1})_{j}]$.

This shows explicitly that the quality of $n$-tupels deteriorates 
if $ A|n$ and $ A|\NX$ because the period reduces by $A$ and $ {\rm\hspace
{.38ex}max\hspace{.38ex}}_{C\in {\ErgoBbb Z}_A}$ $ |\hat{g}|^{2
}[\NX /A,(X_{A\bullet +C+j-1})_{j}]=A|\hat{g}|^{2}[\NX ,X_
\bullet ]$.

\subsection{$ X_{0}=0$, $ X_{k+1}=aX_{k}+c{\rm\hspace{.38ex}int}
(k/t){\rm\hspace{.38ex}mod\hspace{.38ex}}M$\label{int2}} 
Let $ 1\neq a,c\in {\ErgoBbb Z}_M^\times $, $ 2|t|M$ and $ N_{MB
^{2}}=1$ (Fig.\ 11).

This section is restricted to even $M$ since for odd $M$ the random 
number generator of Sec.\ \ref{ck} was already satisfactory. We 
only discuss $ N_{MB^{2}}=1$ since this suffices to find a power 
of two generator with $ Q_{1}=1$.

If one tries to improve the generator of the last section by building 
sum generators one would obtain the recursion relation $ X_{k+1
}=aX_{k}+P(k){\rm\hspace{.38ex}mod\hspace{.38ex}}M$, where $P$ is 
a polynomial of degree greater than one. Such generators may be 
interesting from a mathematical point of view, however they are 
not easy to analyze and the calculation of the random numbers becomes 
more time consuming.

We follow the second way to improve the generator of the last section. 
Looking at this generator $ {\rm\hspace{.38ex}mod\hspace{.38ex}}2
$ one obtains the sequence $ 0,0,1,1,0,0,1,1{\ldots}$ and to make 
all random numbers even one could perform the map $ X_{k}
\mapsto X_{k}+{\rm\hspace{.38ex}int}(k/2)$ which does not seriously 
affect the quality of the generator. If we afterwards divide the 
random numbers by 2 we should end up with an improved generator. 
A slightly more general framework leads to the recursion formula 
of the generator of this section (cf.\ Prop.\ 4.9). It is perhaps 
surprising that even this generator can be Fourier transformed algebraically 
at almost all points $ (s_{0},{\bf s})$.

It would be possible to analyze this generator for $ N_{MB^{2}}>1
$. However this would cause some extra effort and for practical 
purposes this is not needed.

The Fourier transform of the generator is again derived from Eq.\ 
(\ref{34}).

\pagebreak[3]

\noindent {\bf Theorem} 4.8. Let $ X_{0}=0$, $ X_{k+1}=aX_{k}+c
{\rm\hspace{.38ex}int}(k/t){\rm\hspace{.38ex}mod\hspace{.38ex}}M
$, $ 1\neq a,c\in {\ErgoBbb Z}_M^\times $, $ 2|t|M$ and $ N_{MB^{
2}}=1$ then 
\begin{equation}
\label{43}|\hat{g}|^{2}\left( s_{0},{\bf s}\right) =S\delta _{s_{
0}+{\bf s}\cdot {\bf X}_{0}^{\rm mm}=0{\rm\hspace{.38ex}mod\hspace{.38ex}}s_
{a,M}}\hbox{ with }S=s_{a,M}\hbox{ if }s_{a,M}|M/t
\hspace{.6ex}.
\end{equation} 
$ X_{k}^{\rm mm}=c(a^{k}-1)/(a-1)$ is the mixed multiplicative generator 
related to $X$. The period is $ \NX =Mt$.

\pagebreak[3]

\noindent {\bf Proposition} 4.9. There exists a $ c'\in 
{\ErgoBbb Z}_M^\times $, a $ d'\in {\ErgoBbb Z}$, odd, and a periodic 
function $R$, $ R(k)=R(k+t)\in {\ErgoBbb Q}$, $ R(0)=0$, so that 
we get with $ X_{k}'=\frac{c'}{a-1} (\frac{a^{k}-1}{a-1} -k)$, 
\begin{equation}
\label{44}X_{k}=\frac{1}{t} \left( X'_{k}-\frac{d'}{2}k+R\left( k
\right) \right) {\rm\hspace{.38ex}mod\hspace{.38ex}}M
\hspace{.6ex}.
\end{equation}

\pagebreak[3]

\noindent {\bf Proof}. The equation is trivial for $ k=0$. We can 
assume by induction that the equation holds for $k$ and prove it 
for $ k+1$. With $ k=tk'+k_{0}$, $ 0\le k_{0}<t$ we get $ X_{k+
1}=\frac{1}{t} (aX'_{k}-ad'k/2+aR(k)+tc{\rm\hspace{.38ex}int}(k
/t))=\frac{1}{t} (X'_{k+1}-c'k-ad'k/2+aR(k)+c(k-k_{0}))$. If we 
compare this with $ \frac{1}{t} (X'_{k+1}-d'(k+1)/2+R(k+1)){\rm\hspace
{.38ex}mod\hspace{.38ex}}M$ we find (a) $ c'+ad'/2-c=d'/2$. Further 
on, $ R(k+1)$ is determined by $ R(k)$ and $k$$_{0}$. Therefore 
the periodicity of $R$ is equivalent to $ R(t)=R(0)=0$. This means 
(b) $ X_t=\frac{c'}{a-1} (\frac{a^t-1}{t(a-1)} -1)-d'/2$. To solve 
Eqs.\ (a) and (b) we need the following observation. Due to Eq.\ 
(\ref{14}) (replace $M$ by $ MBt^{2}$, $ B=m={\rm\hspace
{.38ex}gcd}(a-1,MB)$) we know that there exists a $ c_t\in 
{\ErgoBbb Z}_{Mt^{2}}$ with $ a^t-1=c_tBt{\rm\hspace{.38ex}mod\hspace{.38ex}}MBt
^{2}$. Thus $ \frac{a^t-1}{t(a-1)} \in {\ErgoBbb Z}_{Mt}^
\times $ and it has an inverse $ {\rm\hspace{.38ex}mod\hspace{.38ex}}Mt
$ which we will denote by $ \frac{t(a-1)}{a^t-1}$. Now we can solve 
Eqs.\ (a) and (b) for $ c'$ and $ d'$ $ {\rm\hspace{.38ex}mod\hspace{.38ex}}Mt
$, 
\begin{displaymath}
c'=\frac{t\left( a-1\right) }{a^t-1} \left( \left( a-1\right) X_t
+c\right) \hspace{.6ex},\hspace{2ex}d'=\frac{2c}{a-1} \left( 1-
\frac{t\left( a-1\right) }{a^t-1}\right) -2\frac{t\left( a-1
\right) }{a^t-1} X_t\hspace{.6ex}.
\end{displaymath} 
Since every prime factor of $M$ is contained in $ a-1$ and $ c
\in {\ErgoBbb Z}_M^\times $ we have $ (a-1)X_t+c\in 
{\ErgoBbb Z}_M^\times $ and therefore $ c'\in {\ErgoBbb Z}_M^
\times $. It remains to show that $ d'$ is odd. We obtain $ {\rm\hspace
{.38ex}mod\hspace{.38ex}}2$: 
\begin{displaymath}
d'=2\frac{a-1}{a^t-1} \frac{c}{a-1} \left( \frac{a^t-1}{a-1} -{t
}\right) =2\frac{a-1}{a^t-1} X_t^{\rm Sec.\ \ref{ck}}=2\frac{a-1
}{a^t-1} \frac{t}{2}=1{\rm\hspace{.38ex}mod\hspace{.38ex}}2
\hspace{.6ex}.
\end{displaymath}\hfill $\Box $
\pagebreak[3] 

\pagebreak[3]

\noindent {\bf Proof} of Thm.\ 4.8. We start with Prop.\ 4.9. Let 
$ k=tk'+k_{0}$ and $ |P(k_{0})|^{2}=1$ then $ |\hat{g}|^{2}$ equals 
\begin{displaymath}
\frac{1}{\NX} \left| \sum _{k_{0}\in {\ErgoBbb Z}_t}\!P\left( k_{
0}\right) \!\sum _{k'\in {\ErgoBbb Z}_{\NX /t}}\hspace{-1ex}\exp
\left( \frac{2\pi i}{\NX} s_{0}\left( tk'\!+\!k_{0}\right) +
\frac{2\pi i}{Mt} \sum _{j=1}^{n}s_{j}\left( X'_{tk'+k_{0}+j-1}
-\frac{d'}{2}\left( tk'\!+\!k_{0}\right) \right) \!\right) 
\right| ^{2}.
\end{displaymath} 
Now we combine the $ d'tk'/2$-term with the $ s_{0}tk'$-term yielding 
$ (s_{0}-\NX /2Mt\cdot d'\sum _{j=1}^{n}s_{j})\cdot (tk'+k_{0})
$. For the moment we set $ \NX =2Mt$ since we want to apply Lemma 
4.7 on the generator $ X'$. Later we will show that actually we 
have $ \NX =Mt$.

We assume that one of the conditions (\ref{41}), (\ref{40}) is valid 
(which we will show later). Then only one term in the sum over $k$
$_{0}$ can be non-zero. Therefore $ |\hat{g}|^{2}=1/\NX \cdot 
\sum _{k_{0}}|P(k_{0})\sum _{k'{\ldots}}|^{2}$ and $ P(k_{0})$ drops. 
Afterwards we interchange the sum over $k$$_{0}$ and taking the 
square once more and recombine the $k$$_{0}$-sum with the $ k'$-sum. 
We end up with the squared Fourier transform of $ X'$ which was 
already analyzed in Ex.\ 4.6 1.\ The result (Eq.\ (\ref{39})) is 
\begin{displaymath}
|\hat{g}|^{2}[\NX =2Mt]\left( s_{0},{\bf s}\right) =2s_{a,Mt}
\delta _{s_{0}-d'\sum _{j=1}^{n}s_{j}+2{\bf s}\cdot {\bf X}_{0}'^
{\rm mm}=s_{a,Mt}\delta _{2|\frac{Mt}{s_{a,Mt}{}} }{\rm\hspace
{.38ex}mod\hspace{.38ex}}2s_{a,Mt}}\hspace{.6ex}.
\end{displaymath} 
We find $ -d'\sum _{j=1}^{n}s_{j}+2{\bf s}\cdot {\bf X}_{0}'^{\rm mm}
=-d'\sum _{j=1}^{n}s_{j}+2c'(s_{a}-\sum _{j=1}^{n}s_{j})/(a-1)=
(d'(a-1)+2c')(s_{a}-\sum _{j=1}^{n}s_{j})/(a-1)-d's_{a}$ and with 
Eq.\ (a) in the proof of Prop.\ 4.9 this equals $ 2{\bf s}
\cdot {\bf X}_{0}^{\rm mm}-d's_{a}$. Now we assume that $ Mt/s_
{a,Mt}$ is even. Then $ s_{a}/s_{a,Mt}$ is odd and since $ d'$ is 
odd $ d's_{a}=s_{a,Mt}{\rm\hspace{.38ex}mod\hspace{.38ex}}2s_{a
,Mt}$. Altogether the equation in the Kronecker $\delta $ in $ 
|\hat{g}|^{2}$ gives $ s_{0}+2{\bf s}\cdot {\bf X}_{0}^{\rm mm}
=d's_{a}\delta _{Mt/s_{a,Mt}{\rm\hspace{.38ex}odd\hspace{.38ex}}
}{\rm\hspace{.38ex}mod\hspace{.38ex}}2s_{a,Mt}$. We recognize that 
$ Mt/s_{a,Mt}$ can only be odd if $s$$_{a}$ is even so that $ |
\hat{g}|^{2}[\NX =2Mt](s_{0},{\bf s})$ vanishes for odd $s$$_{0
}$. From Eq.\ (\ref{42}) for $ c_{1}=c_{2}=2$ we see that $X$ has 
actually the period $ \NX =Mt$ and since $ d'$ is odd 
\begin{displaymath}
|\hat{g}|^{2}[\NX =Mt]\left( s_{0},{\bf s}\right) =s_{a,Mt}
\delta _{s_{0}+{\bf s}\cdot {\bf X}_{0}^{\rm mm}=\frac{s_{a}}{2
}\delta _{Mt/s_{a,Mt}{\rm\hspace{.38ex}odd\hspace{.38ex}}}{\rm\hspace
{.38ex}mod\hspace{.38ex}}s_{a,Mt}}\hspace{.6ex}.
\end{displaymath} 
Now we check the conditions of Lemma 4.7. We set $ M_t={\rm\hspace
{.38ex}max\hspace{.38ex}}_{k\in {\ErgoBbb N}}{\rm\hspace
{.38ex}gcd}(tB,t^{k})$. Since $ m_{M_t}={\rm\hspace{.38ex}gcd}(a
-1,M_t)=M_t/t$ there exists a $ c_{k}\in {\ErgoBbb Z}_t^\times 
$ (Eq.\ (\ref{14})) with $ a^{k}-1=c_{k}kM_t/t{\rm\hspace
{.38ex}mod\hspace{.38ex}}M_t$, hence $ N_{M_t}=t$. Assume $ s_{a
,M}|M/t$. Since $ M_t$ consists only of prime factors of $t$ and 
$ {\rm\hspace{.38ex}gcd}(M_t,t^{k})={\rm\hspace{.38ex}gcd}(Bt,t^{
k})|{\rm\hspace{.38ex}gcd}(MB/s_{a,M},t^{k})$ it suffices to show 
that $ {\rm\hspace{.38ex}gcd}(MB/s_{a,M},t^{k})|{\rm\hspace
{.38ex}gcd}(M'/m',t^{k})$ for all $k$ ($ M'=MB^{2}t/s_{a,MB^{2}t
}$). Since $ s_{a,M}|M/t$ we get $ {\rm\hspace{.38ex}gcd}(s_{a,M
},t^{k})={\rm\hspace{.38ex}gcd}(s_{a,MB^{2}t},t^{k})$ and therefore 
$ {\rm\hspace{.38ex}gcd}(MB/s_{a,M},t^{k})={\rm\hspace{.38ex}gcd}
(M'/Bt,t^{k})$. Finally from Eq.\ (\ref{14}) we get $ {\rm\hspace
{.38ex}gcd}(a^t-1,MB^{2}t)=Bt$, hence $ m'={\rm\hspace{.38ex}gcd}
(a^t-1,M')|Bt$. So Eq.\ (\ref{41}) holds if $ s_{a,M}|M/t$. Moreover, 
in this case $ s_{a,Mt}=s_{a,M}$ and $ 2|t|Mt/s_{a,M}$ which proves 
Eq.\ (\ref{43}) if $ s_{a,M}|M/t$.

If $ s_{a,M}$ does not divide $ M/t$ we take $ \tilde{M}=ts_{a,M
}$ in Eq.\ (\ref{40}). So $ \tilde{M}|Mt$ (notice that $M$ has to 
be replaced by {\em Mt }in Lemma 4.7). Since $ \tilde{M}'=s_{a,M
}B^{2}t/s_{a,MB^{2}t}|B^{2}t$ and $ \tilde{m}'={\rm\hspace
{.38ex}gcd}(a^t-1,\tilde{M}')={\rm\hspace{.38ex}gcd}(Bt,\tilde
{M}')$ we have $ \tilde{m}_{0}'|(\tilde{M}'/\tilde{m}')|B$. Therefore 
$ a=1{\rm\hspace{.38ex}mod\hspace{.38ex}}\tilde{m}_{0}'$ and $ 
\tilde{M}_{0}'= \tilde{M}'/\tilde{m}'\tilde{m}_{0}'=1$. We can use 
Lemma 4.7 if $ \frac{Mt}{2Mt} t(s_{0}-d'\sum _{j=1}^{n}s_{j})+
{\bf s}\cdot ({\bf X}'_t-{\bf X}'_{0})\neq \frac{Mt}{2M_{0}'} 
\delta _{2|M_{0}'}{\rm\hspace{.38ex}mod\hspace{.38ex}}\tilde{M}
$. Now $ {\bf s}\cdot ({\bf X}'_t-{\bf X}'_{0})=\frac{c'}{a-1} 
(s_{a}\frac{a^t-1}{a-1} -t\sum _{j=1}^{n}s_{j})=s_{a}X'_t+t
\frac{c'}{a-1} (s_{a}-\sum _{j=1}^{n}s_{j})$ and with Eq.\ (a) in 
the proof of Prop.\ 4.9 we get $ \frac{c'}{a-1} (s_{a}-\sum _{j
=1}^{n}s_{j})={\bf s}\cdot {\bf X}^{\rm mm\hspace{.38ex}}_{0}-
\frac{d'}{2}(s_{a}-\sum _{j=1}^{n}s_{j})$. Collecting all pieces 
we obtain the condition $ ts_{0}/2+t{\bf s}\cdot {\bf X}^{\rm mm\hspace{.38ex}}
_{0}+s_{a}(X'_t-d't/2)\neq \frac{Mt}{2M_{0}'} \delta _{2|M_{0}'}
{\rm\hspace{.38ex}mod\hspace{.38ex}}\tilde{M}$.

Since $ m_{a^t,MB^{2}t}={m_{0}}_{a^t,MB^{2}t}=Bt$ we get with Eq.\ 
(\ref{35}) $ M_{0}'=M/{\rm\hspace{.38ex}gcd}(ts_{a},M)$. Therefore 
$ \frac{Mt}{2M_{0}'} =t{\rm\hspace{.38ex}gcd}(ts_{a},M)/2=0{\rm\hspace
{.38ex}mod\hspace{.38ex}}\tilde{M}$ if $ 2|M_{0}'$. With Eq.\ (\ref{44}) 
we get $ s_{a}(X'_t-d't/2)=s_{a}tX_t=0{\rm\hspace{.38ex}mod\hspace{.38ex}}
\tilde{M}$. We can divide the condition by $t$ and get $ s_{0}/2
+{\bf s}\cdot {\bf X}^{\rm mm\hspace{.38ex}}_{0}\neq 0{\rm\hspace
{.38ex}mod\hspace{.38ex}}s_{a,M}$.

This condition holds for odd $s$$_{0}$ which completes the proof 
that $ \hat{g}[\NX =2Mt](s_{0},{\bf s})$ vanishes for odd $s$$
_{0}$. The period is $ \NX =Mt$ and due to Eq.\ (\ref{42}) we may 
replace $ s_{0}/2$ by $ s_{0}$ if we calculate $ \hat{g}[\NX =Mt
](s_{0},{\bf s})$. This means that Eq.\ (\ref{43}) is valid if the 
Kronecker $\delta $ vanishes. Otherwise the equation holds trivially 
with $ S\equiv |\hat{g}|^{2}(s_{0},{\bf s})$.\hfill $\Box $
\pagebreak[3] 

\vspace{1ex}
\noindent{}If $ s_{a,M}$ fails to be a divisor of $ M/t$ then $ 
|\hat{g}|^{2}$ remains undetermined for $ Mt/s_{a,M}$ values of 
$s$$_{0}$. At these points we only know that the average of $ |
\hat{g}|^{2}$ is $ s_{a,M}$ (Eq.\ (\ref{30})).

Therefore one should choose $t$ as small as possible, namely $ t
=2$. In this case $ |\hat{g}|^{2}$ is known if $ s_{a,M}|M/2$ and 
an explicit calculation using Eq.\ (\ref{44}) (note that $ a^{k
}-1=(a^{2}-1)k/2-(a-1)^{2}\delta _{k{\rm  \ odd\hspace{.38ex}}}
/2{\rm\hspace{.38ex}mod\hspace{.38ex}}2B^{2}$) gives 
\begin{equation}
\label{53}S=M\left( 1+\cos\left( \frac{\pi }{M} \left( s_{0}+2c
\sum _{j=3}^{n}\frac{a^{j-1}-a^{\delta _{2|j}}}{a^{2}-1} s_{j}
\right) \right) \right) \hspace*{1cm}\hbox{\hspace{.38ex}if }s_{
a}=0{\rm\hspace{.38ex}mod\hspace{.38ex}}M
\hspace{.6ex},\hspace{2ex}t=2\hspace{.6ex}.
\end{equation} 
In particular for $ n=1$ we find $ |\hat{g}|^{2}(s_{0},0)=2M
\delta _{s_{0}=0}$ (cf.\ Eq.\ (\ref{29})), and $ n=2$ gives $ |
\hat{g}|^{2}(s_{0},s_{a}=0)=M(1+\cos(\pi s_{0}/M))\approx 2M$ for 
small $s$$_{0}$.

If moreover $M$ is a power of two then $ |\hat{g}|^{2}$ is completely 
determined by Eqs.\ (\ref{43}), (\ref{53}) and the generator can 
also best be implemented (cf.\ Ex.\ 5.1 1). We obtain 
\begin{equation}
\NX =2M\hspace{.6ex},\hspace{2ex}Q_{1}=1
\hspace{.6ex},\hspace{2ex}Q_{2}\approx M^{-1/3}
\hspace{.6ex},\hspace{2ex}Q_{n}\approx M^{1/\left( n-1\right) -1
}\hbox{ for }n\ge 3
\end{equation} 
as will be shown in Sec.\ \ref{param}. It is advantageous to have 
two parameters $a$ and $c$ at hand to optimize the quality of higher 
$n$-tuples and not only $a$ as in the case of the mixed multiplicative 
generator.

The generator does not provide full periods since $ |\hat{g}|^{2
}(0,s_{1})=s_{1,M}\neq 2M\delta _{s_{1}=0}$. However the deviation 
from an exact uniform distribution is not larger than in a finite 
true random sequence. If one does not use the entire period of the 
generator (and this is not recommended because of the $n$-tupel 
distribution) the feature of having a full period is anyway irrelevant. 
If, for some reasons, one insists in a full period we recommend 
to use a multiplicative generator with prime number modulus or the 
multiply recursive generator which will be analyzed next.

The generator of this section behaves in every aspect better than 
the widely used mixed multiplicative generator. This is also confirmed 
by the figures (cf.\ Fig.\ 4 vs.\ Fig.\ 11). The extra effort in 
calculating random numbers is little (cf.\ Ex.\ 5.1 1). If $n$-tuples 
are used one should take odd $n$ and occasionally omit one random 
number.

Nevertheless the most essential step for producing good random numbers 
is to use large moduli (cf.\ Fig.\ 8 and Sec.\ \ref{param}).

\subsection{$ X_{0}\!=\!1,X_{-1}\!=\!. ..\!=\!X_{-r+1}\!=\!0$, $ X
_{k+1}\!=\!a_{r-1}X_{k}+. ..+a_{0}X_{k-r+1}{\rm\hspace{.38ex}mod\hspace{.38ex}}P
$\label{multrec}} 
Let $P$ be a prime number and $X$$_{k}$ have the maximum period of 
$ P^r-1$ (Fig.\ 12). In this case the random number generator has 
a full period in the sense that every $r$-tuple $ (X_{0},
{\ldots},X_{r-1})\neq (0,{\ldots},0)$ occurs exactly once in a period.

\pagebreak[3]

\noindent {\bf Theorem} 4.10 (Grube \cite{Grube}). Let 
\begin{equation}
P\left( \lambda \right) =\lambda ^r-a_{r-1}\lambda ^{r-1}-
{\ldots}-a_{0}.
\end{equation} 
The corresponding generator has maximum period if and only if $P$ 
is a primitive polynomial over $ {\ErgoBbb Z}_{P^r}$.

The proof is found in \cite[Satz 2.1]{Grube}.

\pagebreak[3]

\noindent {\bf Theorem} 4.11. Let $X$$_{k}$ be defined as above 
and $ {\bf s}_{a}\equiv ({\bf s}\cdot {\bf X}_{k})_{0\le k<r}=(
\sum _{j=1}^{n}s_{j}X_{k+j-1}\hspace{-1pt})_{0\le k<r}$ then $ 
|\hat{g}|^{2}$ is given by the following table. 
\begin{equation}
\label{59}
\begin{array}{c|cc}|\hat{g}|^{2}\left( s_{0},{\bf s}\right) &s_{
0}=0&s_{0}\neq 0\\\hline  
{\bf s}_{a}={\bf 0}&P^r-1&0\\ 
{\bf s}_{a}\neq {\bf 0}&1/\left( P^r-1\right) &P^r/\left( P^r-1
\right) 
\end{array}
\end{equation}

\pagebreak[3]

\noindent {\bf Proof}. First we notice that $ ({\bf s}\cdot 
{\bf X}_{k})_{k}$ obeys the same recursion relation as $ (X_{k}
)_{k}$ since $ {\bf s}\cdot {\bf X}_{k+1}=\sum _{j=1}^{n}s_{j}X_{
k+j}=\sum _{\ell =1}^ra_{r-\ell }\sum _{j=1}^{n}s_{j}X_{k+j-
\ell }=\sum _{\ell =1}^ra_{r-\ell }\,{\bf s}\cdot {\bf X}_{k+1-
\ell }$. So, $ ({\bf s}\cdot {\bf X}_{k})_{k}$ is either identically 
zero or it has maximum period. In the latter case every number $ 
\in {\ErgoBbb Z}_P^\times $ is produced $ P^{r-1}$ times in a period 
and the zero is generated $ P^{r-1}-1$ times. Since the same holds 
for $ ({\bf s}\cdot ({\bf X}_{k+\Delta k}-{\bf X}_{k}))_{k}$ we 
get 
\begin{displaymath}
\sum _{k\in {\ErgoBbb Z}_{P^r-1}}\exp\left( \frac{2\pi i}{P} 
{\bf s}\cdot \left( {\bf X}_{k+\Delta k}-{\bf X}_{k}\right) 
\right) =P^r\delta _{{\bf s}\cdot \left( {\bf X}_{k+\Delta k}-
{\bf X}_{k}\right) =0\hspace*{1ex}\forall 0\le k<r}-1
\hspace{.6ex}.
\end{displaymath} 
If $ {\bf s}_{a}={\bf 0}$ then $ {\bf s}\cdot {\bf X}_{k}=0$ $ 
\forall k$, the Kronecker $\delta $ gives 1 and from Eq.\ (\ref{15}) 
we obtain $ |\hat{g}|^{2}(s_{0},{\bf s}_{a}={\bf 0})=(P^r-1)
\delta _{s_{0}=0}$. If on the other hand $ {\bf s}_{a}\neq 
{\bf 0}$ then $ ({\bf s}\cdot {\bf X}_{k})_{k}$ has maximum period 
and the Kronecker $\delta $ vanishes unless $ \Delta k=0$. In this 
case Eq.\ (\ref{15}) yields $ P^r/(P^r-1)-\delta _{s_{0}=0}$.\hfill $\Box $
\pagebreak[3] 

The choice of parameters is determined by avoiding small $s$ with 
$ {\bf s}_{a}={\bf 0}$. For practical purposes it is more convenient 
to replace the condition $ {\bf s}_{a}=({\bf s}\cdot {\bf X}_{k
})_{0\le k<r}={\bf 0}$ by the equivalent requirement $ {\bf 0}=
({\bf s}\cdot {\bf X}_{1-k})_{1\le k\le r}\Leftrightarrow \sum _{
j=k}^{n}s_{j}X_{j-k}=0{\rm\hspace{.38ex}mod\hspace{.38ex}}P\hbox
{\hspace{.38ex}, }k=1,2,{\ldots},{\rm\hspace{.38ex}min}(r,n)$. If 
$ n\le r$ the only solution is $ {\bf s}={\bf 0}{\rm\hspace
{.38ex}mod\hspace{.38ex}}P$. For $ n>k$ one has the problem of finding 
the smallest lattice vector of an $n$-dimensional lattice. The unit 
cell of this lattice has the volume $ P^r$ (cf.\ Sec.\ \ref{param}). 
Thus for proper parameters the quality of the generator is 
\begin{equation}
\NX =P^r-1\hspace{.6ex},\hspace{2ex}Q_{1}=Q_{2}={\ldots}=Q_r=
\sqrt{2} /\left( 1-P^{-r}\right) \hspace{.6ex},\hspace{2ex}Q_{n
>r}\approx P^{r/n-r}\hspace{.6ex}.
\end{equation} 
This is the first generator which has $ Q_{n}\ge 1$ for $ 1\le n
\le r>1$. For $ 2\le n\le r$ the generator has higher $ \tilde
{N}_{n}={\rm\hspace{.38ex}max}(\NX Q_{n},\NX )=P^r-1$ but lower 
$ \tilde{M}_{n}={\rm\hspace{.38ex}max}(\NX Q_{n},M)=P$ than the 
multiplicative generator $ {\rm\hspace{.38ex}mod\hspace{.38ex}}P^r
$ (Ex.\ 4.1 2, with $ \tilde{N}_{n}\approx  \tilde{M}_{n}
\approx P^{r/n}$).

In particular if one needs the full periods this generator may be 
recommended. For prime numbers of the form $ P=2^{k}\pm 1$ the generator 
has good performance, too. If the prime factors of $ P^r-1$ are 
known it is no problem to find multipliers which lead to a full 
period. A short discussion of the choice of parameters for large 
$P$ and $r$ is given in the next section and an implementation is 
presented in Ex.\ 5.2 2.

\section{Choice of parameters\label{param}} 
We start with a discussion of the mixed multiplicative generator 
(the multiplicative generator is analogous). For practical purposes 
we can restrict ourselves to $ M=2^d$ and $ a=5{\rm\hspace
{.38ex}mod\hspace{.38ex}}8$. We set $ c=1$ which is equivalent to 
any other odd $c$ and assume $ n\ge 2$ since the case $ n=1$ depends 
only on $b$ which is 4 for $ a=5{\rm\hspace{.38ex}mod\hspace{.38ex}}8
$.

From Eq.\ (\ref{46}) we obtain, as long as $ bs_{a,M/b}<M$, that 
$ \hat{g}$ vanishes unless $ s_{a,M/b}|s_{0}\neq 0$ and therefore 
(A) $ Q_{n}(s_{0},{\bf s})=\sqrt{1+{\bf s}^{2}{} /s_{a,M/b}^{2}
} /4>1/4$. However if $ s_{a,M/b}=M/b$ we get $ |\hat{g}|^{2}=M
$ for (B) $ s_{0}=s_{a}=0{\rm\hspace{.38ex}mod\hspace{.38ex}}M$. 
This leads to $ Q_{n}=|{\bf s}|/M$ which for some $ {\bf s}$ is 
much smaller than $ 1/4$.

So Eq.\ (B) is more important. We solve it for $s$$_{1}$ yielding 
$ s_{1}=kM-as_{2}-a^{2}s_{3}-{\ldots}-a^{n-1}s_{n}$ depending on 
the free integer constants $k$, $s$$_{2}$, $s$$_{3}$, {\dots}, $s$
$_{n}$ which give rise to an $n$-dimensional lattice (cf.\ \cite{Knu}). 
The lattice is given by an $n$ by $n$ matrix $A$ according to $ 
{\bf s}=A\cdot (k,s_{2},{\ldots},s_{n})^T$ and we read off 
\begin{equation}
\label{48}A=\left( 
\begin{array} {ccccc}M&-a&-a^{2}&{\ldots}&-a^{n-1}{}\\ 
&1&&&\\ 
&&1&&\\ 
&&&\raisebox{ 3mm}{$\ddots$}&1
\end{array}\right) \sim \left( 
\begin{array} {ccccc}M&-a&&&\\ 
&1&-a&&\\ 
&&1&\ddots &-a\\ 
&&&&1
\end{array}\right) \hspace{.6ex},
\end{equation} 
where zeros have been omitted and both matrices define the same lattice 
since they differ only by $ SL(n,{\ErgoBbb Z})$ lattice transformations.

We denote the length of the smallest non-vanishing lattice vector 
by $\nu _{n}$. Since the quality of the random numbers is determined 
by $ \nu _{n}=MQ_{n}$ we search for an $a$ which large $\nu _{n
}$. Most important are small $n$, in particular the pair correlation 
$ n=2$. In the best case the lattice has a cubic unit-cell and $
\nu _{n}$ is determined by the dimension of the lattice and the 
volume of the unit-cell. Since the volume is given by the determinant 
of $A$ we get as an approximate upper bound $ \nu _{n}
\lessapprox M^{1/n}$. The calculation of $\nu _{n}$ is a standard 
problem in mathematics for which efficient algorithms exist \cite{Lenstra}.

To simplify the search for reasonable multipliers it is useful to 
have also a lower bound for $\nu _{n}$. Due to the specific form 
of $A$ it is easy to see that $\nu _{n}$ has to be larger than the 
smallest ratio $ >1$ between two elements of then set $ \{1,a,a^{
2}{\rm\hspace{.38ex}mod\hspace{.38ex}}M,{\ldots},a^{n-1}{\rm\hspace
{.38ex}mod\hspace{.38ex}}M,M\}$. If we take e.g.\ $ a\approx M^{
1/2}$ we find $ \nu _{2}\gtrapprox M^{1/2}$ which is identical with 
the upper bound.

Similarly we obtain $ \nu _{3}\approx M^{1/3}$ if we take $ a
\approx M^{1/3}$ or $ a\approx M^{2/3}$. However this is not compatible 
with $ a\approx M^{1/2}$ and we only get $ \nu _{2}\gtrapprox M^{
1/3}$. On the other hand we can take $ a\approx M^{1/2}{} +
\frac{1}{2}M^{1/4}$ which differs little from $M$$^{1/2}$. Therefore 
$ \nu _{2}\approx M^{1/2}$ and since $ a^{2}\approx M+M^{3/4}{} 
+\frac{1}{4}M^{1/4}\approx M^{3/4}{\rm\hspace{.38ex}mod\hspace{.38ex}}M
$ we have $ \nu _{3}\gtrapprox M^{1/4}$. Generally, with $ a
\approx M^{1/2}{} +\frac{1}{2}M^{1/4}{} +{\ldots}+\frac{1}{k-1}M^
{1/2^{k-1}}$ (the plus signs may as well be replaced by minus signs) 
we get $ \nu _{n}\gtrapprox M^{1/2^{n-1}}$ as long as $ k\ge n$ 
and $ M^{1/2^{n-1}}\gg 1$. Note that $ M^{1/2^{n-1}}$ is only a 
lower bound for $\nu _{n}$. In the generic case $\nu _{n}$ will 
be close to $M$$^{1/n}$ (cf.\ Ex.\ 5.1 1).

Obviously $\nu _{n}$ increases with $M$. For all practical purposes 
the magnitude of $M$ is only limited by the performance of the generator. 
In practice one has to split $M$ into groups of digits (16 or 32 
bit) that can be treated on a computer. The multiplication by $a$ 
performs best if the pre-factors $ 1/j$ are omitted. This should 
be done even though for $ a\approx M^{1/2}+M^{1/4}+{\ldots}+M^{1
/2^{k-1}}$ the lower bounds for $\nu _{n}$ decrease, $ \nu _{n}
\gtrapprox M^{1/2^{n-1}}/(n-1)!$. Note that the number of digits 
of $M$ is much more important for randomness than the fine-tuning 
of $a$.

Finally, we have to add not too small a constant $ a_{0}=5{\rm\hspace
{.38ex}mod\hspace{.38ex}}8$ (16 or 32 bit) to the sum of powers 
of $M$. This constant can be fixed by explicit calculation of the 
$\nu _{n}$ or by looking at (A) from the beginning of this section 
which implies that $a$$_{0}$ should have large $ |{\bf s}|$ for 
all $ 16<m=s_{a,M}|M^{1/2^{k-1}}$. A suitable choice is e.g.\ $ a
_{0}=3~580~621~541=62~181{\rm\hspace{.38ex}mod\hspace{.38ex}}2^{
16}$. With this value of $a$$_{0}$ we find $ |{\bf s}|\approx m^{
1/n}$ for $ n=2,3$.

We summarize the result for the parameters of the mixed multiplicative 
generator: 
\begin{eqnarray}
&&M=2^{2^{k}d_{0}}\hspace{.6ex},\hspace{2ex}c=1
\hspace{.6ex},\hspace{2ex}a=2^{2^{k-1}d_{0}}+2^{2^{k-2}d_{0}}+
{\ldots}+2^{2d_{0}}+a_{0}\hspace{.6ex},\hspace{2ex}\hbox
{\hspace{.38ex}with\hspace{.38ex}}\nonumber \\ 
&&a_{0}=5{\rm\hspace{.38ex}mod\hspace{.38ex}}8
\hspace{.6ex},\hspace{2ex}a_{0}\approx 2^{d_{0}}
\hspace{.6ex},\hspace{2ex}\hbox{\hspace{.38ex}e.g. }a_{0}=3~580~621
~541{\rm\hspace{.38ex}mod\hspace{.38ex}}2^{d_{0}}\hbox
{\hspace{.38ex}, leads to\hspace{.38ex}}\nonumber \\ 
&&\label{49}\NX =2^{256}\hspace{.6ex},\hspace{2ex}Q_{1}=\sqrt{2
} /4\hspace{.6ex},\hspace{2ex}Q_{n}\approx M^{1/n-1}
\hspace{.6ex}.
\end{eqnarray}

Now we turn to the improved generator of Sec.\ \ref{int2}. The discussion 
of the generator of Sec.\ \ref{ck} is analogous.

The Fourier transform of the generator is given by Eq.\ (\ref{43}). 
We set $ n\ge 2$ since independently of the parameters $ Q_{1}=1
$. Further on, we fix an $ m|M$ and find that $ |\hat{g}|^{2}=m
$ if and only if (C) $ s_{a}=km$, $k$ odd if $ m<M$, and (D) $ s
_{0}+\frac{c}{a-1} \sum _{j=2}^{n}(a^{j-1}-1)s_{j}=\ell m$. (We 
neglect here that $|$\^{g}$|^{2}$ may even be $ 2M$ for $ m=M$, 
cf Eq.\ (\ref{53}).) Eq.\ (C) can be solved for $s$$_{1}$ and Eq.\ 
(D) for $s$$_{0}$ depending on the integer parameters $k$, $ 
\ell $, $s$$_{2}$, {\dots}, $s$$_{n}$. This gives rise to an ($ n
+1$)-dimensional lattice (for $ m<M$ we actually obtain an affine 
sub-lattice since $k$ has to be odd) determined by the matrix $B$ 
via $ (s_{0},{\bf s})=B\cdot (\ell ,k,s_{2},{\ldots},s_{n})^T$, 
\begin{eqnarray}
\hspace{-1ex}B&\hspace{-1ex}=&\hspace{-1ex}\left( 
\begin{array} {ccccc}m&&-c&{\ldots}&-c\left( a^{n-2}+{\ldots}+1
\right) \\ 
&m&-a&{\ldots}&-a^{n-1}{}\\ 
&&1&&\\ 
&&&\raisebox{ 1mm}{$\ddots$}&\\ 
&&&&1
\end{array}\right) \sim \left( 
\begin{array} {cccccc}m&&-c&-c&{\ldots}&-c\\ 
&m&-a&&&\\ 
&&1&-a&&\\ 
&&&1&\ddots &-a\\ 
&&&&&1
\end{array}\right) \\ 
\label{47}&\hspace{-1ex}\sim &\hspace{-1ex}\left( 
\begin{array} {ccccccc}m&&-c&&&&\\ 
&m&-a&a&a^{2}+a&{\ldots}&a^{n-2}+{\ldots}+a\\ 
&&1&-a-1&-a^{2}-a-1&{\ldots}&-a^{n-2}-{\ldots}-1\\ 
&&&1&&&\\ 
&&&&1&&\\ 
&&&&&\raisebox{ 3mm}{$\ddots$}&1
\end{array}\right) \hspace{.6ex},
\end{eqnarray} 
where again zeros have been omitted.

\noindent{}B describes an ($ n+1$)-dimensional lattice which has 
a unit-cell with volume $m$$^{2}$. However this does not imply that 
the smallest lattice vector $\nu _{n}$ has length of about $ m^
{2/(n+1)}$. We see from (\ref{47}) that there exists an ($ n-1$)-dimensional 
sub-lattice with $ s_{0}=0$ and $ s_{2}=-s_{1}-s_{3}-{\ldots}-s_{
n}$ (delete the first and the third row and column in (\ref{47})). 
The unit-cell of the sub-lattice has volume $m$ and $ \nu _{n}
\approx m^{1/(n-1)}$ which, for $ n\ge 4$, is smaller than $ m^
{2/(n+1)}$. The smallest lattice vector for $ n\ge 4$ will have 
the form $ (0,s_{1},-s_{1}-s_{3}-{\ldots}-s_{n},s_{3},{\ldots},s
_{n})$ with the length $ (s_{1}^{2}+s_{3}^{2}+{\ldots}+s_{n}^{2
}+(s_{1}+s_{3}+{\ldots}+s_{n})^{2})^{1/2}$. Since this is of about 
the same magnitude as $ (s_{1}^{2}+s_{3}^{2}+{\ldots}+s_{n}^{2}
)^{1/2}$ we may simply omit $s$$_{2}$ and reduce the problem to 
the ($ n-1$) dimensions given by ($ s_{1},s_{3},{\ldots},s_{n}$). 
Geometrically this means that the lattice corresponding to $B$ for 
$ n\ge 4$ never has an approximately cubic unit-cell. Note moreover 
that the sub-lattice is independent of $c$ which means that $c$ 
can not be fixed by looking at the $n$-tupel distributions for $ n
\ge 4$.

The smallest value of $ Q_{n}=\nu _{n}/m$ is obtained for $ m=M$ 
which is thus the most important case. For $ m=M$ we are not restricted 
to odd $k$. The situation is similar to the ($ n-1$)-dimensional 
case of the mixed multiplicative generator, Eq.\ (\ref{48}), with 
$ -a^{j}$ replaced by $ a^{j}+a^{j-1}+{\ldots}+a$. This allows us 
to use $ a=M^{1/2}+M^{1/4}+{\ldots}+M^{1/2^{k-1}}+a_{0}$ again. 
Since $ 1\ll a\approx M^{1/2}\ll a^{2}{\rm\hspace{.38ex}mod\hspace{.38ex}}M
\approx 2M^{3/4}\ll {\ldots}\ll a^{n-2}{\rm\hspace{.38ex}mod\hspace{.38ex}}M
\approx (n-2)!M^{1-2^{2-n}}\ll M$ we have $ a^{j}+a^{j-1}+
{\ldots}+a\approx a^{j}{\rm\hspace{.38ex}mod\hspace{.38ex}}M$. The 
minus sign is irrelevant, thus we can copy the corresponding lower 
bounds from the mixed multiplicative generator: $ \nu _{n}
\gtrapprox M^{1/2^{n-2}}/(n-2)!$ for $ k+1\ge n\ge 4$. The constant 
$a$$_{0}$ is given by the case $ m<M$ as will be discussed below.

The constant $c$ can be fixed by the case $ n=2$. We have to meet 
two equations (E) $ s_{1}+as_{2}=0{\rm\hspace{.38ex}mod\hspace{.38ex}}M
$ and (F) $ s_{0}+cs_{2}=0{\rm\hspace{.38ex}mod\hspace{.38ex}}M
$ to get $ |\hat{g}|^{2}=M$. Both equations are solved by e.g.\ 
$ s_{0}=-c$, $ s_{1}=-a\approx -M^{1/2}$, $ s_{2}=1$ with $ |(s_{
0},{\bf s})|\approx (c^{2}+M)^{1/2}$. In order to reach the theoretical 
limit $ \nu _{2}\approx M^{2/3}$ one needs $ c\gtrapprox M^{2/3
}$. So, the simplest ansatz for $c$ is $ c=M^\lambda +1$ for $ 
\lambda \ge 2/3$, $ M^\lambda \in {\ErgoBbb N}$. On the other hand, 
if $ s_{1}=M^{1-\lambda }s_{1}'$, $ s_{2}=M^{1-\lambda }s_{2}'$ 
then $ s_{1}'+as_{2}'=0{\rm\hspace{.38ex}mod\hspace{.38ex}}M^
\lambda $ has a solution with $ |{\bf s}'|\lessapprox M^{
\lambda /2}$. Since (F) is solved by $ s_{0}=-M^{1-\lambda }s_{
2}'$ we find $ |(s_{0},{\bf s})|\approx |s_{0}|\lessapprox M^{1
-\lambda }M^{\lambda /2}=M^{1-\lambda /2}$. To allow for the maximum 
value $ M^{2/3}$ one needs $ \lambda \le 2/3$. In general, $c$ should 
not have more successive zero digits than $ M^{2/3}$ has. The simplest 
reasonable choice is therefore $ c=M^{2/3}+1$. We can generalize 
this slightly to $ c=(2^{d_{1}}+1)c_{0}$, where $c$$_{0}$ is a 16 
or 32 bit number and $ 2^{d_{1}}\le M^{2/3}\le c_{0}2^{d_{1}}$. 
This choice of $c$ leads to best performance among all reasonable 
$c$. We will see in Ex.\ 5.1 1 that it actually gives $ \nu _{2
}\approx M^{2/3}$ and $ \nu _{3}\approx M^{1/2}$. As a lower bound 
for $\nu _{2}$, $\nu _{3}$ one has only the values $M$$^{1/2}$, 
$M$$^{1/4}$ that are obtained from Eqs.\ (E), (C) alone.

Now we determine $c$$_{0}$ and $a$$_{0}$ by looking at $ s_{a,M}
=m<M$. The case $ m<M$ is more important than for the mixed multiplicative 
generator since $Q$$_{n}$ is not limited by $ 1/4$. To some extent 
the smaller $Q$$_{n}$ for $ m<M$ is compensated by the fact that 
for small $m$ there are more points with $ s_{a}={\rm\hspace
{.38ex}odd\hspace{.38ex}}\cdot m$. We use $ a_{0}=3~580~621~541
$ as for the mixed multiplicative generator and find with $ c_{
0}=3~370~134~727=11~463{\rm\hspace{.38ex}mod\hspace{.38ex}}2^{1
6}$ that $ Q_{2}(s_{0},{\bf s})\approx m^{2/3-1}$ and $ Q_{3}(s_{
0},{\bf s})\approx m^{1/2-1}$ if $ s_{a,M}=m$.

We summarize the result for the generator of Sec.\ \ref{int2}: 
\begin{eqnarray}
&&M=2^{2^{k}d_{0}},\hspace*{1ex}a=2^{2^{k-1}d_{0}}+2^{2^{k-2}d_{
0}}+{\ldots}+2^{2d_{0}}+a_{0},\hspace*{1ex}c=\left( 2^{{\rm int}
\left( 2^{k+1}{} /3\right) d_{0}}+1\right) c_{0},\hbox{ with\hspace{.38ex}}
\nonumber \\ 
&&a_{0}=5{\rm\hspace{.38ex}mod\hspace{.38ex}}8
\hspace{.6ex},\hspace{2ex}a_{0}\approx 2^{d_{0}}
\hspace{.6ex},\hspace{2ex}c_{0}{\rm\hspace{.38ex}odd\hspace{.38ex}}
\hspace{.6ex},\hspace{2ex}c_{0}\ge 2^{2/3\cdot d_{0}}
\hspace{.6ex},\nonumber \\ 
&&\hbox{\hspace{.38ex}e.g. }a_{0}=3~580~621~541{\rm\hspace
{.38ex}mod\hspace{.38ex}}2^{d_{0}}\hspace{.6ex},\hspace{2ex}c_{
0}=3~370~134~727{\rm\hspace{.38ex}mod\hspace{.38ex}}2^{d_{0}}\hbox
{\hspace{.38ex}\hspace*{2ex}leads to\hspace{.38ex}}\nonumber \\
&&\label{50}\NX =2^{257}\hspace{.6ex},\hspace{2ex}Q_{1}=1
\hspace{.6ex},\hspace{2ex}Q_{2}\approx M^{2/3-1}
\hspace{.6ex},\hspace{2ex}Q_{n\ge 3}\approx M^{1/\left( n-1
\right) -1}\hspace{.6ex}.
\end{eqnarray}

Finally we give a short discussion of the multiply recursive generator 
of Sec.\ \ref{multrec} (cf.\ Ex.\ 5.1 2).

\noindent{}P should not be taken too small to provide enough digits 
for the random numbers. To optimize the performance one should use 
a prime number of the form $ P=2^d\pm 1$, e.g.\ $ P=2^{31}-1$. Moreover 
we set $ a_{r-1}=1$, $ a_{r-2}={\ldots}=a_{1}=0$.

The most severe problem is to find the prime factors of $ P^r-1$. 
To this end it is useful to take $ r=2^k$ since in this case $ P^
{2^{k}}-1=(P^{2^{k-1}}+1)\cdot {\ldots}\cdot (P+1)\cdot (P-1)$ and 
one is basically left with the problem to determine the prime factors 
of $ P^{2^{k-1}}+1$.

Afterwards it is easy to find an $ a_{0}\in {\ErgoBbb Z}_P^
\times $ that makes the polynomial $ P(\lambda )=\lambda ^r-
\lambda ^{r-1}-a_{0}$ primitive over $ {\ErgoBbb Z}_{P^r}$. Since 
\begin{equation}
\label{60}\left( 
\begin{array} {c}X_{k}{}\\ 
X_{k-1}{}\\ 
\vdots \\ 
X_{k-r+1}
\end{array}\right) =X^{k}\cdot \left( 
\begin{array}{c}1\\ 
0\\ 
\vdots \\ 
0
\end{array}\right) \hspace{.6ex},\hspace{2ex}\hbox
{\hspace{.38ex}with }X\equiv \left( 
\begin{array} {ccccc}a_{r-1}&a_{r-2}&{\ldots}&a_{1}&a_{0}{}\\ 
1&0&&&\\ 
&&\ddots &&\\ 
&&&1&0
\end{array}\right) \hspace{.6ex},
\end{equation} 
a necessary and sufficient condition for a maximum period is $ X^
{(P^r-1)}=1\hspace{-.6ex}{\rm l} {\rm\hspace{.38ex}mod\hspace{.38ex}}P
$ and $ X^{(P^r-1)/p}\neq 1\hspace{-.6ex}{\rm l} {\rm\hspace
{.38ex}mod\hspace{.38ex}}P$ for all prime factors $p$ of $ P^r-1
$. High powers of $X$ are easily computed. If $ N=\sum b_{i}2^{
i}$, $ b_{i}\in \{0,1\}$ then $ X^N=\prod _{\{i:b_{i}=1\}}X^{2^{
i}}$ and $ X^{2^{i}}=(X^{2^{i-1}})^{2}$.

Now one has to check the $n$-tupel distributions for $ n>r$. We found 
(Eq.\ (\ref{59})) that $ |\hat{g}|^{2}=P^r-1$ if and only if $ s
_{0}=0$ and $ {\bf s}_{a}={\bf 0}$. The latter equation is equivalent 
to $ 0=\sum _{j=k}^{n}s_{j}X_{j-k}=\ell _{k}P$, $ k=1,2,
{\ldots},r$, $ \ell _{k}\in {\ErgoBbb Z}$ (cf.\ Sec.\ \ref{multrec}) 
and gives rise to an $n$-dimensional lattice determined by $C$ via 
$ {\bf s}=C\cdot (\ell _{1},{\ldots},\ell _r,s_{r+1},{\ldots},s_{
n})^T$, $ C=C_{1}\cdots C_r$, 
\begin{equation}
C_{k}=\left( 
\begin{array}{cccccc}1&&&&&\\ 
&\raisebox{ 3mm}{$\ddots$}\hspace*{1ex}1&&&&\\ 
&&P&-X_{1}&{\ldots}&-X_{n-k}{}\\ 
&&&1&&\\ 
&&&&\raisebox{ 3mm}{$\ddots$}&1
\end{array}\right) .\hspace*{1ex}C\sim C_{0}\equiv \left( 
\begin{array}{cccccc} 
P&&&-a_{0}&&\\ 
&\raisebox{ 3mm}{$\ddots$}&P&-1&\raisebox{ 3mm}{$\ddots$}&-a_{0}
{}\\ 
&&&1&\raisebox{ 3mm}{$\ddots$}&-1\\ 
&&&&\raisebox{ 3mm}{$\ddots$}&1
\end{array}\right) ,
\end{equation} 
(after some lattice transformations) if $ r<n\le 2r$ and $ a_{r-1
}=1$, $ a_{r-2}={\ldots}=a_{1}=0$. The determinant of $C$$_{0}$ 
is $ P^r$, however the symmetry of $C$$_{0}$ leads to $ \nu _{r
+1}={\ldots}=\nu _{2r}\equiv \nu $ which is given by the shortest 
lattice vector of the 2 by 3 matrix $ 
\renewcommand{\arraystretch}{1}\left( 
\begin{array} {ccc}P&0&0\\ 
-a_{0}&-1&1
\end{array}\right) ^T\renewcommand{\arraystretch}{1.2}$. Since the 
second and third row are identical (up to a minus sign) the problem 
is analogous to the calculation of $\nu _{2}$ in the case of the 
mixed multiplicative generator. We obtain $ \nu \lessapprox 2^{
1/4}P^{1/2}$ with $ a_{0}\approx 2^{1/4}P^{1/2}\approx 55109$ for 
$ P=2^{31}-1$.

We summarize the result for the generator of Sec.\ \ref{multrec}: 
\begin{eqnarray}
&&P=2^d-1,\hbox{ prime\hspace{.38ex}},\hspace*{1ex}r=2^{k},
\hspace*{1ex}a_{r-1}=1,\hspace*{1ex}a_{r-2}={\ldots}=a_{1}=0,
\hspace*{1ex}a_{0}\approx 2^{1/4}P^{1/2},\hbox{ with\hspace{.38ex}}
\nonumber \\ 
&&X^{\left( P^r-1\right) }=1\hspace{-.6ex}{\rm l} {\rm\hspace
{.38ex}mod\hspace{.38ex}}P\hbox{ and }X^{\left( P^r-1\right) /p
}\neq 1\hspace{-.6ex}{\rm l} {\rm\hspace{.38ex}mod\hspace{.38ex}}P
\hspace*{1ex}\forall p|\left( P^r-1\right) ,\hspace*{1ex}p\hbox{ 
prime, leads to\hspace{.38ex}}\nonumber \\ 
&&\NX =\left( P^r-1\right) \hspace{.6ex},\hspace{2ex}Q_{1}=
{\ldots}=Q_r\approx \sqrt{2} \hspace{.6ex},\hspace{2ex}Q_{r+1}=
{\ldots}=Q_{2r}\approx 2^{1/4}P^{1/2-r}\hspace{.6ex}.
\end{eqnarray} 
Notice that the effort for calculating random numbers does not increase 
with $r$.

Let us finally mention that the quality of the $n$-tupel $ 
{\bf X}_{\ell }$ of the (non-successive) random numbers $ X_{
\ell },X_{k_{2}+\ell },{\ldots},X_{k_{n}+\ell }$ deteriorates to 
$ Q_{n}\lessapprox P^{(r-d)/n-r}$ if there exist $ d>r-n$ values 
of $ j\in \{-1,{\ldots},-r\}$ with $ {\bf X}_{j}={\bf 0}$ (see the 
remark at the end of Sec.\ \ref{gentest}). In particular if $ X_{
k-1}={\ldots}=X_{k-r+1}=0$ the pair $ (X_{0},X_{k})$ has quality 
of less than $ P^{1/2-r}$ because $ aX_{\ell }=bX_{k+\ell }$ $ 
\forall \ell $ if $ a=bX_{k}{\rm\hspace{.38ex}mod\hspace{.38ex}}P
$. From Eq.\ (\ref{60}) we see immediately that this happens for 
multiples of $ k=(P^r-1)/(P-1)$ (notice the equidistant zeros in 
Fig.\ 12I). This makes it not desirable to use more than $ (P^r
-1)/(P-1)$ multiply recursive random numbers.

\pagebreak[3]

\noindent {\bf Example} 5.1. 
\begin{enumerate}\item{}We set $ M=2^{256}=2^{2^{4}\cdot 16}$, $ a
=2^{128}+2^{64}+2^{32}+62~181$ and in case of the generator of Sec.\ 
\ref{int2} $ c=(2^{160}+1)\cdot 11~463$. In the following table 
we compare the mixed multiplicative generator with the generator 
of Sec.\ \ref{int2}. The results can easily be obtained with a computer 
algebra program and Eq.\ (\ref{53}). 
\begin{equation}
\begin{array}{c|cc|cc} Q_{n}\equiv M^{\alpha _{n}-1}&X_{k+1}=aX_{
k}+1&\hbox{\hspace{.38ex}Eq.\ (\ref{49})\hspace{.38ex}}&X_{k+1}
=aX_{k}+c{\rm\hspace{.38ex}ink}\left( k/2\right) &\hbox
{\hspace{.38ex}Eq.\ (\ref{50})\hspace{.38ex}}\\ 
\alpha _{1}&0.99414&0.99414&1.00000&1.00000\\ 
\alpha _{2}&0.50000&0.50000&0.65658&0.66667\\ 
\alpha _{3}&0.33203&0.33333&0.49783&0.50000\\ 
\alpha _{4}&0.24859&0.25000&0.33436&0.33333\\ 
\alpha _{5}&0.19721&0.20000&0.24636&0.25000\\ 
\alpha _{6}&0.16335&0.16667&0.19882&0.20000
\end{array}
\end{equation} 
We see a good agreement of the quality parameters with the approximate 
upper bounds. This means that our choice of parameters is satisfactory. 
Moreover the table confirms that the quality parameter of the generator 
of Sec.\ \ref{int2} lies above the quality of the mixed multiplicative 
generator.

Finally we present an implementation of the generator in Pascal. 
We group the digits of $X$$_{k}$ to 16 blocks of 16 digits {\tt 
X[1], {\dots}, X[16]} starting from the highest digits.

\vspace{1ex}
\noindent{}\tt unit random1;\newline 
interface\newline 
const n=16; n0=(n+2) div 3; a0=62181; c0=11463;\newline 
var X:array[1..n] of longint;\newline 
procedure nextrandom;\newline 
implementation\newline 
var even:boolean; i:word; c:longint;\newline 
procedure nextrandom;\newline 
var j,k:word;\newline 
begin\newline 
if even then inc(c,c0); even:=not even;\newline 
for j:=1 to n do begin\newline 
\hspace*{1cm}X[j]:=X[j]*a0;\newline 
\hspace*{1cm}k:=2;while j+k$<$=n do begin inc(X[j],X[j+k]);k:=k shl 
1 end end;\newline 
inc(X[n-1],X[n] shr 16); X[n]:=(X[n] and \$FFFF)+c;\newline 
inc(X[n0-1],X[n0] shr 16); X[n0]:=(X[n0] and \$FFFF)+c;\newline 
for j:=n downto 2 do begin\newline 
\hspace*{1cm}inc(X[j-1],X[j] shr 16); X[j]:=X[j] and\hspace*{1ex}\$FFFF 
end;\newline 
X[1]:=X[1] and \$FFFF\newline 
end;\newline 
begin for i:=1 to n do X[i]:=0; c:=0; even:=true end.

\vspace{1ex}
\noindent{}\rm The corresponding mixed multiplicative generator is 
obtained by omitting or changing the lines containing {\tt c}. On 
a 100MHz Pentium computer this (not optimized) program produces 
19~563 random numbers per second whereas 20~938 mixed multiplicative 
random numbers can be produced. A loss of speed of about 6.6{\%} 
seems us worth the gain of better random numbers. Note that the 
number {\tt c} suffers an overflow every about $ 750~000$th random 
number. This does not affect randomness and it is not worth the 
effort to correct this flaw. 
\item{}We set $ P=2^{31}-1$, $ r=8$ which leads to $ P^r-1=2^{34
}\cdot 3^{2}\cdot 5\cdot 7\cdot 11\cdot 17\cdot 31\cdot 41
\cdot 151\cdot 331\cdot 733\cdot 1709\cdot 21529\cdot 368140581013
\cdot 708651694622727115232673724657$. Moreover we take $ a_{r-1
}=1$, $ a_{r-2}={\ldots}=a_{1}=0$, $ a_{0}=60~045$ yielding 
\begin{eqnarray}
\label{58}\hspace{-.7cm}&&X_{0}=1, X_{-1}={\ldots}=X_{-7}=0
\hspace{.6ex},\hspace{2ex}X_{k+1}=X_{k}+60~045X_{k-7}{\rm\hspace
{.38ex}mod\hspace{.38ex}}2^{31}-1\hspace{.6ex},\\ 
\hspace{-.7cm}&&\NX =P^{8}\!-\!1\approx 2^{248},\hspace{ 3pt} Q_{
1}={\ldots}=Q_{8}=\left( \!P^{8}\!\right) \!^{1.00202-1},
\hspace{ 3pt} Q_{9}={\ldots}=Q_{16}=\left( \!P^{8}\!\right) \!^{
0.06368-1}.\nonumber 
\end{eqnarray} 
The following program gives on a 100MHz Pentium 74~473 random numbers 
({\tt X[k]}) per second.

\tt unit random2;\newline 
interface\newline 
var X:array[0..7] of longint; k:integer;\newline 
procedure nextrandom;\newline 
implementation\newline 
const a0=60045;\newline 
var i:integer; x0,x1,x2:longint;\newline 
procedure nextrandom;\newline 
begin\newline 
x0:=X[(k+1) and 7];\newline 
x2:=(x0 and \$FFFF)*a0; x1:=(x0 shr 16)*a0+(x2 shr 16);\newline 
x2:=(x2 and \$FFFF)+(x1 shr 15)+((x1 and \$7FFF) shl 16);\newline 
if (x2 shr 31)=1 then x2:=(x2 xor \$80000000)+1;\newline 
inc(x2,X[k]);\newline 
while (x2 shr 31)=1 do x2:=(x2 xor \$80000000)+1;\newline 
k:=(k+1) and 7;\newline 
if x2=\$7FFFFFFF then X[k]:=0 else X[k]:=x2\newline 
end;\newline 
begin k:=0; X[0]:=1; for i:=1 to 7 do X[i]:=0 end.
\end{enumerate}

\section{Results and outlook} 
We have generalized the spectral test. As the new feature we analyze 
the sequence of random numbers (I in the figures) not only the distribution 
of $n$-tupels (II in the figures).

We saw that the mixed multiplicative generator did not pass the test 
with an ideal result. We were able to construct an improved generator 
which has the recursion formula 
\begin{equation}
\label{54}X_{0}=0\hspace{.6ex},\hspace{2ex}X_{k+1}=aX_{k}+c{\rm\hspace
{.38ex}int}\left( k/2\right) {\rm\hspace{.38ex}mod\hspace{.38ex}}2^d
\hspace{.6ex}.
\end{equation} 
For the choice of the parameters $a$, $c$, $d$ we made suggestions 
in Eq.\ (\ref{50}). This generator (or the multiply recursive generator 
given in Eq.\ (\ref{58})) seems us to be the best choice in quality 
and performance. An implementation of a generator of this type with 
modulus $ 2^d=2^{256}\approx 10^{77}$ was presented in Ex.\ 5.1 
1.\ The calculation of random numbers is fast even though the modulus 
is that large. We think that for all practical purposes pseudo random 
numbers generated with this generator can not be distinguished from 
a true random sequence.

We were able to analyze this generator and several others by virtue 
of a remarkable formula on the harmonic analysis of multiplicative 
rings of remainder class rings (cf.\ Thm.\ 3.4 and Eq.\ (\ref{34})). 
The choice of parameters was discussed in Sec.\ \ref{param}.

\vspace{1ex}
\noindent{}For practical purposes there is essentially no need for 
further improvements. From a purely mathematical point of view however 
there are lots of open questions.

First of all one could be interested in the cases where Eq.\ (\ref{34}) 
does not provide the result (take e.g.\ $ M=P$ prime and $a$ no 
primitive element of $ {\ErgoBbb Z}_P^\times $, cf.\ Ex.\ 4.1 5). 
In these cases $ N|\hat{g}|^{2}(s_{0},s_{1})$ is given as zero of 
the polynomial 
\begin{equation}
\label{57}P_{s_{0}}\left( Y\right) \equiv \prod _{s_{1}\in 
{\ErgoBbb Z}_M}\left( Y-N|\hat{g}|^{2}\left( s_{0},s_{1}
\right) \right) \hspace{.6ex}.
\end{equation} 
For multiplicative generators $ P_{s_{0}}(Y)=Y^M-MNY^{M-1}+
{\ldots}$. Numerical calculations show that $ P_{s_{0}}$ has integer 
coefficients. We were not able to prove this for $ s_{0}\neq 0$ 
nor to analytically determine the coefficients for non-trivial examples.

Further on, generators involving polynomials may be of interest. 
An example with a quadratic polynomial was presented in Ex.\ 4.6 
4.

Finally we would be interested in multiply recursive generators. 
Those generators are given by a matrix-valued multiplier. The simplest 
example with a prime number modulus was presented in Sec.\ \ref{multrec}. 
In this section we saw that multiply recursive generators are also 
the best candidates for being even more efficient than the generator 
given in (\ref{54}). In this connection multiply recursive generators 
with power of two modulus may be of special interest.

\section*{Aknowledgement} 
I am grateful to Manfred H{\"u}ck who motivated me to this work by 
showing me some figures of random number generators.

\section*{Figures} 
Some graphs of random number generators are presented to give a visual 
impression of what the generator looks like. There are two possibilities 
to draw a two-dimensional plot: first (I), to plot the $k$-th random 
number $X$$_{k}$ over $k$ and second (II), to plot $X$$_{k+1}$ over 
$X$$_{k}$ presenting the pair correlation. The third part of the 
figures give the absolute of the Fourier transform of I. $ |
\hat{g}|^{2}(s_{0},s_{1})$ is a measure for the correlations along 
a line perpendicular to $ (s_{0},s_{1})$ in I (cf.\ Eq.\ (\ref{55})). 
For ideal generators $ |\hat{g}|^{2}$ should be $ \le 1$ and Figs.\ 
I and II should look like first rain drops on a dry road.\vspace{1cm}

\noindent{} 
\unitlength0.027mm 
\begin{tabular}{crcrcr}\multicolumn{5}{l}{Fig.\ 1: $ X_{0}=1$, $ X
_{k+1}=37X_{k}{\rm\hspace{.38ex}mod\hspace{.38ex}}1024$}&\\ 
\raisebox{ 2.5cm}{\hspace{-3mm}$X$$_{k}$}&\hspace{-3ex}\input{a37a0.10
}&\hspace{-1.5ex}\raisebox{ 2.5cm}{$X$$_{k+1}$}&\hspace{-3ex}\input
{b37a0.10}&\hspace{-3ex}\raisebox{ 2.5cm}{$s$$_{1}$}\hspace{-3ex} 
0&\hspace{-5ex}
\begin{tabular}[b]{ccccc}0&0&0&0&16\\ 
0&4&0&0&0\\ 
0&0&8&0&0\\ 
0&4&0&0&0\\ 
2$^{10}$&0&0&0&0
\end{tabular}\\ 
\hspace{ 5ex} I&$k$&\hspace{ 3.5ex} II&$X$$_{k}$&$ |\hat{g}|^{2}
$\hspace{ 1mm}0&$s$$_{0}$\hspace*{ 2mm}
\end{tabular}\vspace{1cm}

\noindent{}
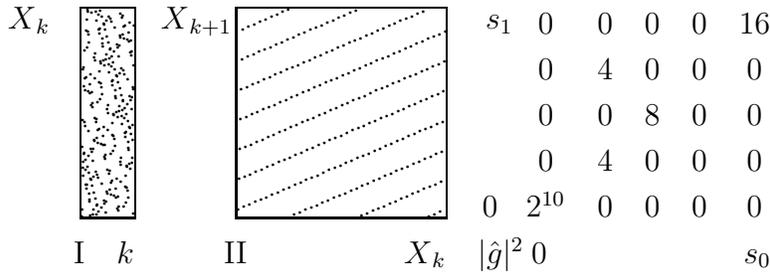
\begin{tabular}{crcrcr}\multicolumn{5}{l}{Fig.\ 2: $ X_{0}=1$, $ X
_{k+1}=195X_{k}{\rm\hspace{.38ex}mod\hspace{.38ex}}1009$}&\\ 
\raisebox{ 2.43cm}{\hspace{-3mm}$X$$_{k}$}&\hspace{-3ex}\input{a195a0.100
}&\hspace{-1.5ex}\raisebox{ 2.43cm}{$X$$_{k+1}$}&\hspace{-3ex}\input
{b195a0.100}&\hspace{-5ex}\raisebox{ 2.43cm}{$s$$_{1}$}\hspace{-2.7ex} 
0&\hspace{-6.4ex}
\begin{tabular}[b]{ccccc}$\frac{1}{1008}$&\hspace{-1ex}$\frac{1009}{1008}$
&\hspace{-1ex}$\frac{1009}{1008}$&\hspace{-1ex}$\frac{1009}{1008}$
&\hspace{-1ex}$\frac{1009}{1008}$\\ 
$\frac{1}{1008}$&\hspace{-1ex}$\frac{1009}{1008}$&\hspace{-1ex}$\frac{1009}{1008}$
&\hspace{-1ex}$\frac{1009}{1008}$&\hspace{-1ex}$\frac{1009}{1008}$\\ 
$\frac{1}{1008}$&\hspace{-1ex}$\frac{1009}{1008}$&\hspace{-1ex}$\frac{1009}{1008}$
&\hspace{-1ex}$\frac{1009}{1008}$&\hspace{-1ex}$\frac{1009}{1008}$\\ 
$\frac{1}{1008}$&\hspace{-1ex}$\frac{1009}{1008}$&\hspace{-1ex}$\frac{1009}{1008}$
&\hspace{-1ex}$\frac{1009}{1008}$&\hspace{-1ex}$\frac{1009}{1008}$\\ 
1008&0\hspace*{ 1ex}&0\hspace*{ 1ex}&0\hspace*{ 1ex}&0\hspace*{ 1ex}
\end{tabular}\\ 
\hspace{ 5ex} I&$k$&\hspace{ 3.5ex} II&$X$$_{k}$&$ |\hat{g}|^{2}
$\hspace{ 2.5mm}0&$s$$_{0}$\hspace*{ 2.5mm}
\end{tabular}\vspace{1cm}

\noindent{}
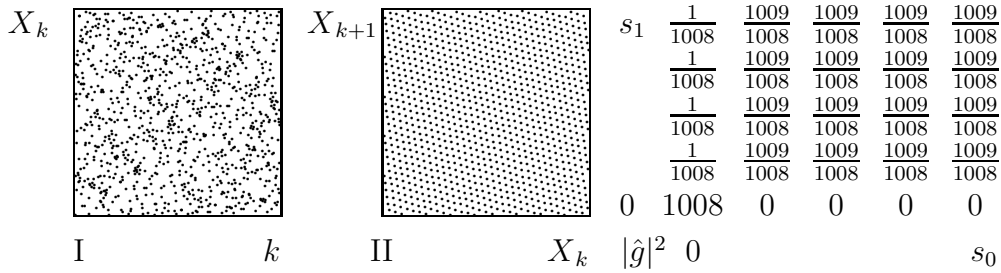
\begin{tabular}{crcrcr}\multicolumn{5}{l}{Fig.\ 3: $ X_{0}=1$, $ X
_{k+1}=44X_{k}{\rm\hspace{.38ex}mod\hspace{.38ex}}31^{2}$}&\\ 
\raisebox{ 2.3cm}{\hspace{-3mm}$X$$_{k}$}&\hspace{-3ex}\input{a44a0.961
}&\hspace{-1.5ex}\raisebox{ 2.3cm}{$X$$_{k+1}$}&\hspace{-3ex}\input
{b44a0.961}&\hspace{-40mm}\raisebox{ 2.3cm}{$s$$_{1}$\hspace*{1ex}31}\hspace{-3ex} 
0\hspace*{ 3mm}&\hspace{-5cm}
\begin{tabular}[b]{ccccc} \raisebox{-1mm}{$\frac{31}{30}$}&\raisebox{-1mm}{0}
&\raisebox{-1mm}{0}&\hspace{-1ex}{}{\dots}&\raisebox{-1mm}{\hspace{-1ex}$\frac{961}{30}$}\\ 
\raisebox{-1mm}{\vdots}&\raisebox{-1mm}{\vdots}&\raisebox{-1mm}{\vdots}
&&\raisebox{-1mm}{\vdots}\\ 
0&$\frac{31}{30}$&$\frac{31}{30}$&\hspace{-1ex}{}{\dots}&\hspace{-1ex}{}0\\ 
0&$\frac{31}{30}$&$\frac{31}{30}$&\hspace{-1ex}{}{\dots}&\hspace{-1ex}{}0\\ 
930&0&0&\hspace{-1ex}{}{\dots}&\hspace{-1ex}{}0
\end{tabular}\hspace*{ 5mm}\\ 
\hspace{ 5ex} I&$k$&\hspace{ 3.5ex} II&$X$$_{k}$&\hspace{ 3.5mm}$ 
|\hat{g}|^{2}$\hspace{ 4.5mm}0\hspace{ 29mm}31&$s$$_{0}$
\end{tabular}\vspace{1cm}

\noindent{}
\begin{tabular}{crcrcr}\multicolumn{5}{l}{Fig.\ 4: $ X_{0}=0$, $ X
_{k+1}=37X_{k}+1{\rm\hspace{.38ex}mod\hspace{.38ex}}1024$}&\\ 
\raisebox{ 2.5cm}{\hspace{-3mm}$X$$_{k}$}&\hspace{-3ex}\input{a37b1.10
}&\hspace{-1.5ex}\raisebox{ 2.5cm}{$X$$_{k+1}$}&\hspace{-3ex}\input
{b37b1.10}&\hspace{-3ex}\raisebox{ 2.5cm}{$s$$_{1}$}\hspace{-2.7ex} 
0&\hspace{-5ex}
\begin{tabular}[b]{ccccc}0&0&0&0&16\\ 
0&4&0&0&0\\ 
0&0&8&0&0\\ 
0&4&0&0&0\\ 
2$^{10}$&0&0&0&0
\end{tabular}\\ 
\hspace{ 5ex} I&$k$&\hspace{ 3.5ex} II&$X$$_{k}$&$ |\hat{g}|^{2}
$\hspace{ 1mm}0&$s$$_{0}$\hspace*{ 1.5mm}
\end{tabular}\vspace{1cm}

\noindent{}
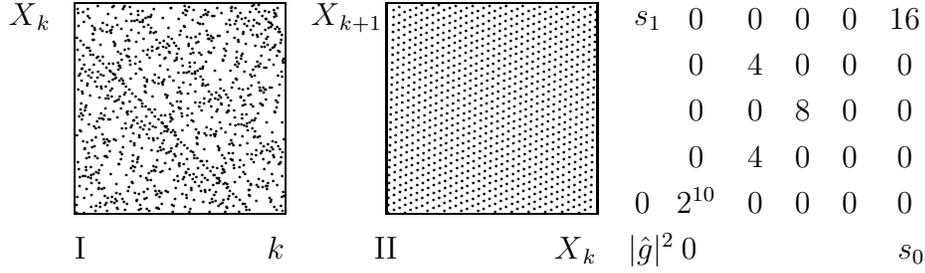
\begin{tabular}{crcrcr}\multicolumn{5}{l}{Fig.\ 5: $ X_{0}=0$, $ X
_{k+1}=41X_{k}+3{\rm\hspace{.38ex}mod\hspace{.38ex}}1024$}&\\ 
\raisebox{ 2.5cm}{\hspace{-3mm}$X$$_{k}$}&\hspace{-3ex}\input{a41b3.10
}&\hspace{-1.5ex}\raisebox{ 2.5cm}{$X$$_{k+1}$}&\hspace{-3ex}\input
{b41b3.10}&\hspace{-3ex}\raisebox{ 2.5cm}{$s$$_{1}$}\hspace{-2.7ex} 
0&\hspace{-5ex}
\begin{tabular}[b]{ccccc}0&0&0&0&32\\ 
0&8&0&0&0\\ 
0&0&16&0&0\\ 
0&8&0&0&0\\ 
2$^{10}$&0&0&0&0
\end{tabular}\\ 
\hspace{ 5ex} I&$k$&\hspace{ 3.5ex} II&$X$$_{k}$&$ |\hat{g}|^{2}
$\hspace{ 1mm}0&$s$$_{0}$\hspace*{ 1.5mm}
\end{tabular}\vspace{1cm}

\noindent{}
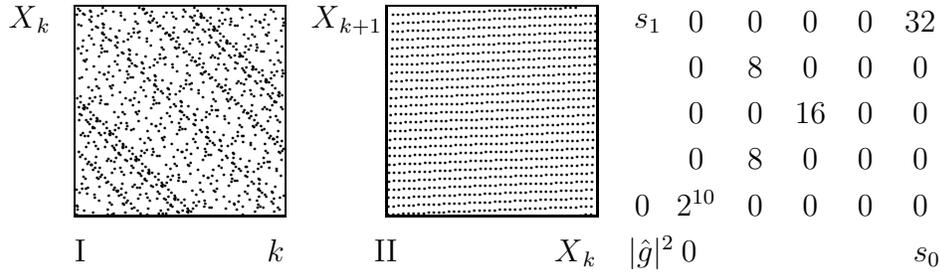
\begin{tabular}{crcrcr}\multicolumn{5}{l}{Fig.\ 6: $ X_{0}=0$, $ X
_{k+1}=41X_{k}+1{\rm\hspace{.38ex}mod\hspace{.38ex}}1024$}&\\ 
\raisebox{ 2.5cm}{\hspace{-3mm}$X$$_{k}$}&\hspace{-3ex}\input{a41b1.10
}&\hspace{-1.5ex}\raisebox{ 2.5cm}{$X$$_{k+1}$}&\hspace{-3ex}\input
{b41b1.10}&\hspace{-3ex}\raisebox{ 2.5cm}{$s$$_{1}$}\hspace{-2.7ex} 
0&\hspace{-5ex}
\begin{tabular}[b]{ccccc}0&0&0&0&0\\ 
0&8&0&0&0\\ 
0&0&0&0&0\\ 
0&0&0&8&0\\ 
2$^{10}$&0&0&0&0
\end{tabular}\\ 
\hspace{ 5ex} I&$k$&\hspace{ 3.5ex} II&$X$$_{k}$&$ |\hat{g}|^{2}
$\hspace{ 1mm}0&$s$$_{0}$\hspace{ 1.5mm}
\end{tabular}\vspace{1cm}

\noindent{}
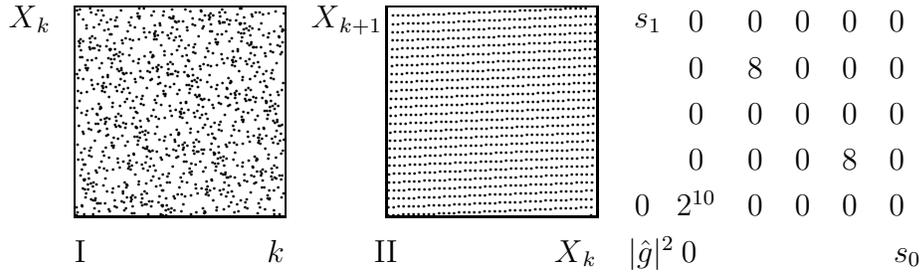
\begin{tabular}{crcrcr}\multicolumn{5}{l}{Fig.\ 7: $ X_{0}=0$, $ X
_{k+1}=21X_{k}+1{\rm\hspace{.38ex}mod\hspace{.38ex}}1000$}&\\ 
\raisebox{ 2.4cm}{\hspace{-3mm}$X$$_{k}$}&\hspace{-3ex}\input{a21b1.100
}&\hspace{-1.5ex}\raisebox{ 2.4cm}{$X$$_{k+1}$}&\hspace{-3ex}\input
{b21b1.100}&\hspace{-4ex}\raisebox{ 2.4cm}{$s$$_{1}$ 0}\hspace{-2.7ex}
$ -4$&\hspace{-6.3ex}
\begin{tabular}[b]{ccccc}1000&0&0&0&0\\ 
0&0&0&0&0\\ 
0&0&40&0&0\\ 
0&0&0&0&0\\ 
0&0&0&0&40
\end{tabular}\\ 
\hspace{ 5ex} I&$k$&\hspace{ 3.5ex} II&$X$$_{k}$&$ \hspace{ 1.5mm
}|\hat{g}|^{2}$\hspace{ 4.5mm}0&$s$$_{0}$\hspace*{ 1.5mm}
\end{tabular}\vspace{1cm}

\noindent{}
\begin{tabular}{crcrcr}\multicolumn{5}{l}{Fig.\ 8: $ X_{0}=0$, $ X
_{k+1}=(37+1024)X_{k}+1{\rm\hspace{.38ex}mod\hspace{.38ex}}1024^{
2}$\vspace{ 2mm}}&\\ 
\raisebox{ 2.5cm}{\hspace{-3mm}$X$$_{k}$}&\hspace{-3ex}\input{a1061e1.10
}&\hspace{-1.5ex}\raisebox{ 2.5cm}{$X$$_{k+1}$}&\hspace{-3ex}\input
{b1061e1.10}&\hspace{-5.5ex} &\hspace{-4ex}
\begin{tabular}[b]{c}cf. Fig.\ 4\\ 
with $ \NX =10^{20}$
\end{tabular}\\ 
\hspace{ 5ex} I&$k$&\hspace{ 3.5ex} II&$X$$_{k}$&&
\end{tabular}\vspace{1cm}

\noindent{}
\begin{tabular}{crcrcr}\multicolumn{5}{l}{Fig.\ 9: $ X_{0}=0$, $ X
_{k+1}=37X_{k}+129k{\rm\hspace{.38ex}mod\hspace{.38ex}}1024$}&\\ 
\raisebox{ 2.5cm}{\hspace{-3mm}$X$$_{k}$}&\hspace{-3ex}\input{a37c129.10
}&\hspace{-1.5ex}\raisebox{ 2.5cm}{$X$$_{k+1}$}&\hspace{-3ex}\input
{b37c129.10}&\hspace{-3ex}\raisebox{ 2.5cm}{$s$$_{1}$}\hspace{-2.7ex} 
0&\hspace{-5.5ex}
\begin{tabular}[b]{ccccc}0&0&0&0&8\\ 
0&2&0&2&0\\ 
0&0&4&0&0\\ 
0&2&0&2&0\\ 
2$^{11}$&0&0&0&0
\end{tabular}\\ 
\hspace{ 5ex} I&$k$&\hspace{ 3.5ex} II&$X$$_{k}$&$ |\hat{g}|^{2}
$\hspace{ 1.8mm}0&$s$$_{0}$\hspace*{ 1mm}
\end{tabular}\vspace{1cm}

\noindent{}
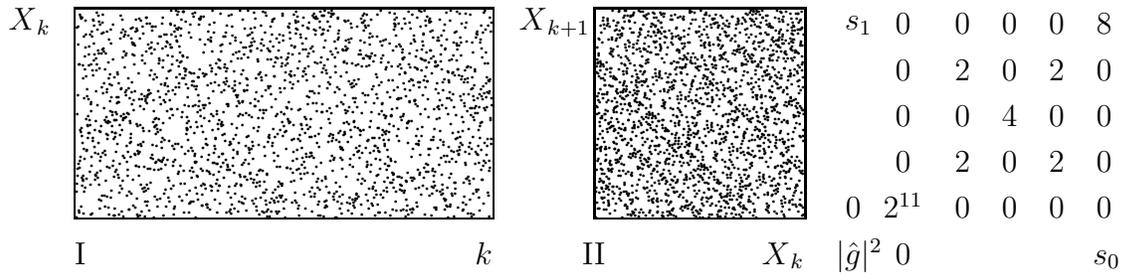
\begin{tabular}{crcrcr}\multicolumn{5}{l}{Fig.\ 10: $ X_{0}=0$, $ X
_{k+1}=30X_{k}+25k{\rm\hspace{.38ex}mod\hspace{.38ex}}11^{2}$}&\\ 
\raisebox{ 2.3cm}{\hspace{-3mm}$X$$_{k}$}&\hspace{-3ex}\input{a30c200.968
}&\hspace{-1.5ex}\raisebox{ 2.3cm}{$X$$_{k+1}$}&\hspace{-3ex}\input
{b30c200.968}&\hspace{-4ex}\raisebox{ 2.4cm}{$s$$_{1}$ 0}\hspace{-2.7ex}
$ -4$&\hspace{-6.3ex}
\begin{tabular}[b]{ccccc}1210&0&0&0&0\\ 
0&0&$\frac{121}{10}$&0&0\\ 
0&0&0&0&$\frac{121}{10}$\\ 
0&0&0&0&0\\ 
0&0&0&0&0
\end{tabular}\\ 
\hspace{ 5ex} I&$k$&\hspace{ 3.5ex} II&$X$$_{k}$&$ \hspace{ 1.5mm
}|\hat{g}|^{2}$\hspace{ 4.5mm}0&$s$$_{0}$\hspace*{ 2mm}
\end{tabular}\vspace{1cm}

\noindent{}
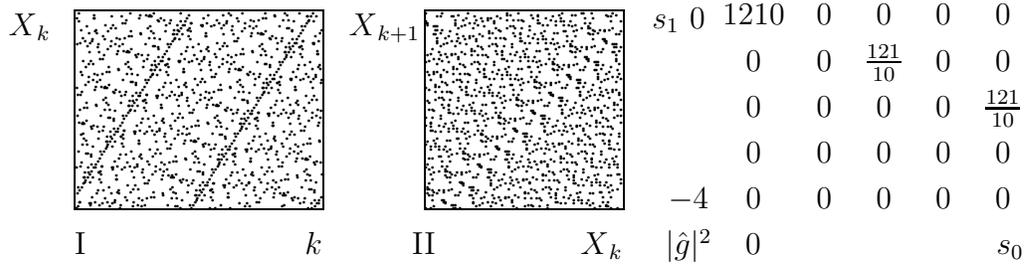
\begin{tabular}{crcrcr}\multicolumn{5}{l}{Fig.\ 11: $ X_{0}=0$, $ X
_{k+1}=37X_{k}+129{\rm\hspace{.38ex}int}(k/2){\rm\hspace
{.38ex}mod\hspace{.38ex}} 1024$}&\\ 
\raisebox{ 2.5cm}{\hspace{-3mm}$X$$_{k}$}&\hspace{-3ex}\input{a37d129.10
}&\hspace{-1.5ex}\raisebox{ 2.5cm}{$X$$_{k+1}$}&\hspace{-3ex}\input
{b37d129.10}&\hspace{-3ex}\raisebox{ 2.5cm}{$s$$_{1}$}\hspace{-2.7ex} 
0&\hspace{-5.5ex}
\begin{tabular}[b]{ccccc}4&0&0&0&4\\ 
1&1&1&1&1\\ 
2&0&2&0&2\\ 
1&1&1&1&1\\ 
2$^{11}$&0&0&0&0
\end{tabular}\\ 
\hspace{ 5ex} I&$k$&\hspace{ 3.5ex} II&$X$$_{k}$&$ |\hat{g}|^{2}
$\hspace{ 1.8mm}0&$s$$_{0}$\hspace*{ 1mm}
\end{tabular}\vspace{1cm}

\noindent{}
\begin{tabular}{crcrcr}\multicolumn{5}{l}{Fig.\ 12: $ X_{0}=1$, $ X
_{-1}=0$, $ X_{k+1}=X_{k}+7X_{k-1}{\rm\hspace{.38ex}mod\hspace{.38ex}}31
$}&\\ 
\raisebox{ 2.2cm}{\hspace{-3mm}$X$$_{k}$}&\hspace{-3ex}\input{a1f7.31
}&\hspace{-1.5ex}\raisebox{ 2.2cm}{$X$$_{k+1}$}&\hspace{-3ex}\input
{b1f7.31}&\hspace{-3ex}\raisebox{ 2.5cm}{$s$$_{1}$}\hspace{-2.7ex} 
0&\hspace{-7.5ex}
\begin{tabular}[b]{ccccc}$\frac{1}{960}$&\hspace{-1ex}$\frac{961}{960}$
&\hspace{-1ex}$\frac{961}{960}$&\hspace{-1ex}$\frac{961}{960}$&\hspace{-1ex}$\frac{961}{960}$\\ 
$\frac{1}{960}$&\hspace{-1ex}$\frac{961}{960}$&\hspace{-1ex}$\frac{961}{960}$
&\hspace{-1ex}$\frac{961}{960}$&\hspace{-1ex}$\frac{961}{960}$\\ 
$\frac{1}{960}$&\hspace{-1ex}$\frac{961}{960}$&\hspace{-1ex}$\frac{961}{960}$
&\hspace{-1ex}$\frac{961}{960}$&\hspace{-1ex}$\frac{961}{960}$\\ 
$\frac{1}{960}$&\hspace{-1ex}$\frac{961}{960}$&\hspace{-1ex}$\frac{961}{960}$
&\hspace{-1ex}$\frac{961}{960}$&\hspace{-1ex}$\frac{961}{960}$\\ 
960&0\hspace*{ 1ex}&0\hspace*{ 1ex}&0\hspace*{ 1ex}&0\hspace*{ 1ex}
\end{tabular}\\ 
\hspace{ 5ex} I&$k$&\hspace{ 3.5ex} II&$X$$_{k}$&$ |\hat{g}|^{2}
$\hspace{ 3.5mm}0&$s$$_{0}$\hspace*{ 1.5mm}
\end{tabular}

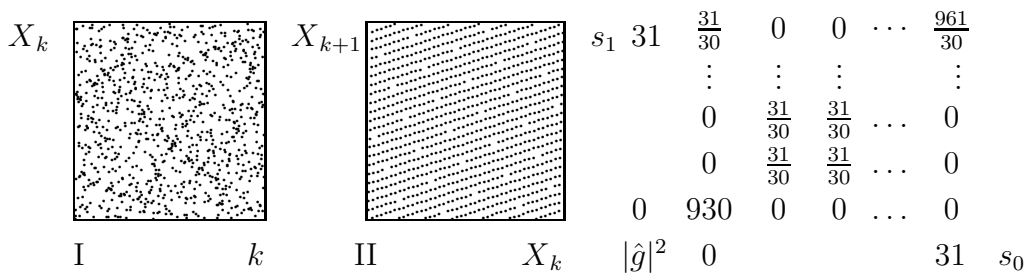
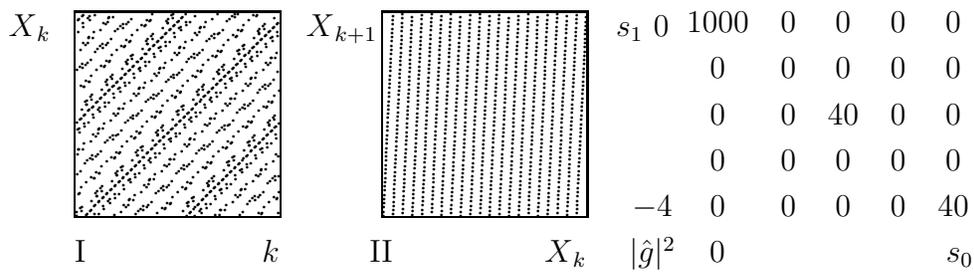
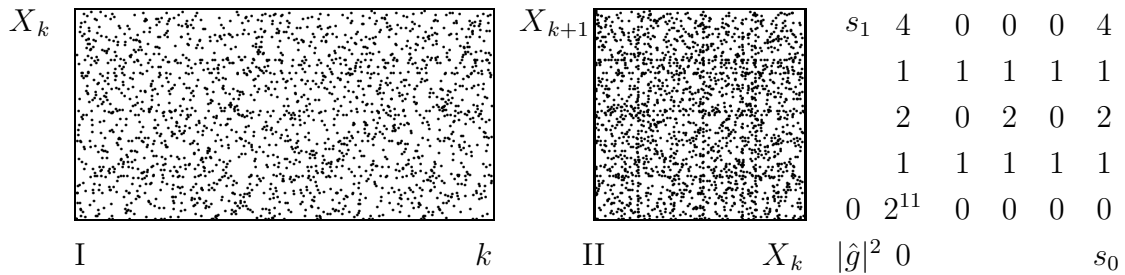
\end{document}